\documentclass[12pt]{article}
\usepackage{a4wide,epsfig}
\newcommand{\ice}[1]{\relax}

\renewcommand{\thefootnote}{\fnsymbol{footnote}}

\newcommand{\gsim}{\;\rlap{\lower 3.5 pt \hbox{$\mathchar \sim$}} \raise 1pt
 \hbox {$>$}\;}
\newcommand{\lsim}{\;\rlap{\lower 3.5 pt \hbox{$\mathchar \sim$}} \raise 1pt
 \hbox {$<$}\;}

\newcommand{\Lw}{L_\omega}
\newcommand{\lz}{L_z}
\newcommand{\cl}{C_l}
\newcommand{\sqrtthree}{\sqrt{3}}
\newcommand{\lifour}{l_4}

\begin{document}    

\title{\vskip-3cm{\baselineskip14pt
\centerline{\normalsize\hfill DESY 01--090}
\centerline{\normalsize\hfill Freiburg-THEP 01/07}
\centerline{\normalsize\hfill TTP01--14}
\centerline{\normalsize\hfill hep-ph/0108017}
\centerline{\normalsize\hfill June 2001}
}
\vskip.7cm
Heavy-light current correlators at order $\alpha_s^2$ in QCD and HQET
}

\author{
{K.G. Chetyrkin}$^{a,b,}$\thanks{Permanent address:
Institute for Nuclear Research, Russian Academy of Sciences,
60th October Anniversary Prospect 7a, Moscow 117312, Russia.}
\,\,
and
{M. Steinhauser}$^c$
  \\[3em]
  {\normalsize (a) Fakult{\"a}t f{\"u}r Physik,}\\
  {\normalsize Universit{\"a}t Freiburg, D-79104 Freiburg, Germany }
  \\[.5em]
  {\normalsize (b) Institut f\"ur Theoretische Teilchenphysik,}\\
  {\normalsize Universit\"at Karlsruhe, D-76128 Karlsruhe, Germany}
  \\[.5em]
  {\normalsize (c) II. Institut f\"ur Theoretische Physik,}\\ 
  {\normalsize Universit\"at Hamburg, D-22761 Hamburg, Germany}
}
\date{}
\maketitle

\begin{abstract}
\noindent
The non-diagonal correlators of vector and scalar currents are
considered at three-loop order in QCD. The full mass dependence is 
computed in the case where one of the quarks is massless and the other one
carries mass $M$. 
We exploit the decoupling relations between the full
theory and the heavy quark effective theory
(HQET) in order to obtain the
logarithmic parts of the leading threshold terms. 
With the help of conformal mapping and 
Pad\'e approximation numerical estimates for the 
non-logarithmic terms are extracted which in turn lead to a prediction
of the correlator in HQET at order $\alpha_s^2$.
As applications of the vector and scalar correlator we
consider the single-top-quark production via the process $q\bar{q}\to
t\bar{b}$ and the decay rate of a charged Higgs boson into hadrons,
respectively.
In both cases the computed NLO corrections are shown to be numerically 
much less important than the leading ones.
On the contrary, the NLO order QCD corrections to the HQET
sum rule for the leptonic decay rate of a heavy-light meson 
proves to be comparable to the leading one.

\vspace{.2cm}


\end{abstract}

\thispagestyle{empty}
\newpage
\setcounter{page}{1}

\renewcommand{\thefootnote}{\arabic{footnote}}
\setcounter{footnote}{0}


\section{Introduction}

In QCD the correlator of two currents 
is very often a central object from which important physical
consequences can be deduced. 
In particular, in case the coupling of the currents is diagonal
physical observables like
$e^+ e^-$ annihilation into hadrons and the decay of the $Z$ boson are
obtained from the vector and axial-vector current correlators.
Furthermore, total decay rates of CP even or CP odd Higgs bosons can
be obtained by considering 
the scalar and pseudo-scalar current densities, respectively.
For these cases the results are available up to order $\alpha_s^2$
taking into account the full quark mass dependence both for the
non-singlet~\cite{CheKueSte96} and singlet~\cite{CheHarSte98}
correlators.

In this work we want to extend the techniques developed
in~\cite{CheKueSte96,CheHarSte98} to the situation where the coupling
of the external currents to quarks is non-diagonal.
In particular we compute the three-loop correlators of
currents which couple to two different quark flavours. To be precise
we want to consider the case where one of the quarks is massless and
the other carries mass $M$. In this limit the vector (scalar) and
axial-vector (pseudo-scalar) correlators coincide.
A short account of our results has been presented in~\cite{CheSte01}.
In this paper the details of the calculation are given and 
explicit expressions for intermediate results are provided which might
be useful for other applications.

The main aim of this work is to obtain results which are valid for 
arbitray values of the quark mass, $M$, respectively the ratio 
$q^2/M^2$ where $q$ is the external momentum of the correlator.
Very often the knowledge of the full mass dependence is not
necessary. E.g., 
in high energy experiments where the center-of-mass energy,
$\sqrt{s}$, is much larger than the mass of the quarks, 
the latter can often be neglected or an expansion in $M/\sqrt{s}$
is sufficient to describe the data.
However, there are situations where the full dependence on $M$
and $\sqrt{s}$ is required.
One can, e.g., think on the high precision which meanwhile has been
reached at LEP (CERN), SLC (SLAC) or TEVATRON (Fermilab)
or on situations where the center-of-mass energy is
of the same order of magnitude as the quark masses.
In particular for threshold phenomena the masses
are important and the full dependence is desirable.

Concerning the physical applications of 
the non-diagonal vector and axial-vector correlators 
we have in mind properties connected to
the $W$ boson. With the results of this paper at hand
a certain (gauge invariant) class of corrections to the Drell-Yan
process becomes available. In particular we have in mind the production of
a quark pair through the decay of a virtual $W$ boson generated in
$p\bar{p}$ collisions.  The absorptive part of the considered
correlator is directly related to the decay width of the (highly virtual)
$W$ bosons into quark pairs and gluons.
Of particular interest in this connection is the single-top-quark
production via the process $q\bar{q}\to t\bar{b}$. The imaginary part
of the transversal $W$ boson polarization function constitutes a gauge
invariant and finite contribution of ${\cal O}(\alpha_s^2)$.

As an application of the scalar and pseudo-scalar
current correlator we want to
mention the decay of a charged Higgs boson which occurs in extensions
of the Standard Model (SM). The corrections provided in this paper describe
the total decay rate into a massive and a massless quark.

Another important application of the non-diagonal vector and scalar
current correlator is connected to the corresponding meson decay
constant. Within HQET it is related
to the imaginary part of the effective current correlator, i.e. the
spectral density, evaluated near threshold.  As the current in the
effective theory does not depend on the $\gamma$-matrix structure it
can be derived both from the vector and the scalar correlator of the
full theory which are evaluated in this work.  The corresponding
analysis at order $\alpha_s$ has been performed in
Ref.~\cite{BroGro92,Bagan92} with the conclusion that the determination of the
$B$ meson decay constant suffers from large pertubative
corrections. We show in this paper
that the corrections of order $\alpha_s^2$ are
also sizeable.

The paper is organized as follows. In Section~\ref{sec:pade} we
provide useful definitions and describe the method we use for the
computation. The connection between the effective and full theory is
established in Section~\ref{sec:eff}. In Section~\ref{sec:res}
we present explicit results for the moments of the vector and scalar
current correlator and discuss the results for the corresponding 
spectral densities in the full theory.
Afterwards, in Section~\ref{sec:reseff},
the procedure is discussed which allows for the extraction of 
threshold information from our results.
This can be translated to the effective theory which
provides a result for the spectral density in the effective theory
up to order $\alpha_s^2$.
Finally, in Section~\ref{sec:con}, we discuss some physical
applications and present our conclusions.
The Appendix contains the analytical results of the low- and
high-energy moments for the vector and scalar correlator.


\section{\label{sec:pade}Conformal mapping and Pad\'e approximation}

In this Section we want to describe the method we use for the
computation of the current correlators in full QCD. Let us start with
some definitions. 

In the vector case the polarization function is defined through
\begin{eqnarray}
  \left(-q^2g_{\mu\nu}+q_\mu q_\nu\right)\,\Pi^v(q^2)
  +q_\mu q_\nu\,\Pi^v_L(q^2)
  &=&
  i\int {\rm d}x\,e^{iqx}\langle 0|Tj^v_\mu(x) j^{v\dagger}_\nu(0)|0 \rangle,
  \label{eq:pivadef}
\end{eqnarray}
with
$j_\mu^v = \bar{\psi}_1\gamma_\mu\psi_2$.
Only the transversal part $\Pi^v(q^2)$ will be considered in the following.
The definition of the scalar polarization function reads
\begin{eqnarray}
  q^2\,\Pi^s(q^2)
  &=&
  i\int {\rm d}x\,e^{iqx} \langle 0|Tj^s(x)j^{s\dagger}(0)|0 \rangle
  \,,
  \label{eq:pispdef}
\end{eqnarray}
with $j^s = (m(\mu)/M) \bar{\psi}_1\psi_2$ where $m(\mu)$ is the
$\overline{\rm MS}$ and $M$ the on-shell mass of $\psi_2$.
$\psi_1$ we consider to be massless.
Thus we will identify $\psi_1$ with $q$, the massless quark,
and $\psi_2$ with $Q$ which is supposed to be a heavy quark of mass $M$.
This will become relevant later on when we consider the effective theory.
Throughout this paper we consider anti-commuting $\gamma_5$
which is justified as for $\psi_1\not=\psi_2$ only non-singlet diagrams
contribute.
As a consequence the axial-vector (pseudo-scalar) correlator coincides
with the vector (scalar) one.

It is convenient to introduce the dimensionless variable 
\begin{eqnarray}
  z &=& \frac{q^2}{M^2}
  \,,
\end{eqnarray}
where $M$ refers to the pole mass. 
For the overall normalization of $\Pi^\delta(q^2)$ ($\delta=v,s$)
we adopt the QED-like conditions $\Pi^\delta(0)=0$.

The physical observables $R(s)$ are related to $\Pi(q^2)$ through
\begin{eqnarray}
  R^v(s) &=& 12\pi \,\,\,\mbox{Im}\left[ \Pi^v(q^2=s+i\epsilon) \right]
  \,,
  \\
  R^s(s) &=&  8\pi \,\,\,\mbox{Im}\left[ \Pi^s(q^2=s+i\epsilon) \right]
  \,,
\end{eqnarray}
where the use of the variables
\begin{eqnarray}
  x\,\,=\,\,\frac{M}{\sqrt{s}}\,, && v\,\,=\,\, \frac{1-x^2}{1+x^2}\,,
\end{eqnarray}
turns out to be useful to describe the high energy and threshold
region, respectively.

The expansion of $\Pi^\delta$ in terms of $\alpha_s$ reads
($\delta=v,s$)
\begin{eqnarray}
  \Pi^\delta &=& \Pi^{(0),\delta} 
  + \frac{\alpha_s^{(n_f)}(\mu)}{\pi} C_F \Pi^{(1),\delta}
  + \left(\frac{\alpha_s^{(n_f)}(\mu)}{\pi}\right)^2 \Pi^{(2),\delta}
  + {\cal O}(\alpha_s^3)
  \,.
\label{eq:pidef}
\end{eqnarray}
It is convenient to further decompose the three-loop term according to the
colour structure
\begin{eqnarray}
  \Pi^{(2),\delta} &=& C_F^2 \Pi_{FF}^{(2),\delta}
                    +  C_AC_F \Pi_{FA}^{(2),\delta}
                    +  C_FTn_l \Pi_{FL}^{(2),\delta}
                    +  C_FT \Pi_{FH}^{(2),\delta}
  \label{eq:pidef2}
  \,,
\end{eqnarray}
where analogous formulae hold for $R(s)$ and also for the
corresponding quantities in the effective theory which we will define below.
$C_F=(N_c^2-1)/(2N_c)$ and
$C_A=N_c$ are the eigenvalues of the quadratic
Casimir operator of the fundamental and adjoint representation,
respectively, and $T=1/2$ is the index of the fundamental representation.
In Eq.~(\ref{eq:pidef2}) $\Pi_{FF}^{(2),\delta}$ corresponds to the
abelian part already present in QED whereas $\Pi_{FA}^{(2),\delta}$
represents the non-abelian structure. The remaining two structures
correspond to the fermionic contributions where $n_l$ counts the
number of massless quarks and $n_f=n_l+1$ is the total number of active
quark flavours.

For later convenience we want to list the exact
expressions for the Born results and the corrections of order
$\alpha_s$~\cite{Schilcher:1981kr,Bro81,Chang:1982qq,
  ReiRubYaz81,Djouadi:1994ss}.
In the vector case we have
\begin{eqnarray}
  R^{(0),v}(s) &=& \frac{N_c}{2}
                   \left(1-x^2\right)^2
                   \left(2+x^2\right)
  \,,
  \nonumber\\
 R^{(1),v}(s) &=& 
\frac{3\,N_c}{4}
\left\{
1 - \frac{5x^2}{2} + \frac{2x^4}{3} + \frac{5 x^6}{6}
  +\frac{1}{3}x^2(-5-4x^2 +5x^4) \,\ln(x^2)
\right.\nonumber\\&&\mbox{}
-\frac{1}{3} (1-x^2)^2 (4+5x^2)\,\ln(1-x^2)
\nonumber\\&&\left.\mbox{}
-\frac{2}{3}(1-x^2)^2(2+x^2)
\left[
- \,\ln \left(\frac{x^2}{1-x^2} \right)\,\ln(1-x^2)
 +       2 \,\mbox{Li}_2 \left(-\frac{x^2}{1-x^2} \right)
\right]
\right\}
\,.
\nonumber\\
\end{eqnarray}
Correspondingly, the scalar current correlators read
\begin{eqnarray}
  R^{(0),s}(s) &=& N_c
                   \left(1-x^2\right)^2
  \,,
  \nonumber\\
  R^{(1),s}(s) &=& 
  -\frac{N_c}{2}(1-x^2)
  \Bigg\{
  -(3-7 x^2 + 2 x^4) \,\ln(x^2) 
\nonumber\\&&\mbox{}
  +(1-x^2)\left[ -\frac{9}{2} + (5 -2x^2)\,\ln(1-x^2) \right]
\nonumber\\&&\mbox{}
  +2 (1-x^2) \left[
         -\,\ln\left(\frac{x^2}{1-x^2} \right)\,\ln(1-x^2)
 +       2 \,\mbox{Li}_2\left(-\frac{x^2}{1-x^2} \right)
           \right]
\Bigg\}
  \,.
\end{eqnarray}

A complete analytical computation of $\Pi^\delta(q^2)$ 
at three-loop order or its imaginary part is currently not feasible.
The method we use for the computation of the diagrams
has successfully been applied to several physical quantities
(see, e.g., Refs.~\cite{BaiBro95,CheKueSte96,CheHarSte98,Har01}).
It allows for the computation of a semi-numerical approximation
for $\Pi(q^2)$.
The aim is the reconstruction of the function $\Pi(q^2)$ from the
knowledge of some moments for $z\to0$ and $z\to-\infty$ and 
additional partial information about the behaviour of
$R(s)$ for $s\to M^2$.
For convenience we briefly summarize the different steps
which have to be performed for the individual pieces of
Eq.~(\ref{eq:pidef}) and~(\ref{eq:pidef2}). In the following we
generically write $\Pi(q^2)$.

\begin{enumerate}
\item
Compute as many moments as possible for small and large $z$.
In our case the expansion for $z\to0$ reduces to a simple Taylor
series of the Feynman diagrams in the external momentum.
For $z\to-\infty$, however, the rules of asymptotic 
expansion~\cite{asympexp} have to be applied. Thus in this limit 
besides the power corrections
there are logarithmic terms in $z$.

\item
\label{item:thr}
The information known about the imaginary part $R(s)$ for $s\to M^2$ 
has to be transformed to a function $\Pi^{thr}(q^2)$.
Afterwards $\Pi^{thr}(q^2)$ has to be expanded in the limits
$z\to0$ and $z\to-\infty$ and the moments have to be subtracted from
the ones of $\Pi(q^2)$.
It is important to construct $\Pi^{thr}(q^2)$ 
in such a way that its expansion for $z\to0$ is analytical.
In this way the information about the logarithmic part of $\Pi^{thr}(q^2)$
can be incorporated.

\item
Construct a function $\Pi^{log}(q^2)$ in such a way that the
combination
\begin{eqnarray}
  \tilde{\Pi}(q^2)\equiv\Pi(q^2) - \Pi^{thr}(q^2) - \Pi^{log}(q^2)
\end{eqnarray}
is polynomial in $z$ and $1/z$, i.e. in the small and high energy region.
Furthermore no logarithmic singularities may be introduced
at threshold.

\item
Perform a conformal mapping. The change of variables~\cite{FleTar94}
\begin{eqnarray}
  z &=& {4\omega\over (1+\omega)^2}
  \,,
\end{eqnarray}
maps the $z$ plane into the interior of the unit circle of the $w$
plane. Thereby the cut $[1,\infty)$ is mapped to the perimeter.

\item
Define~\cite{CheHarSte98}
\begin{eqnarray}
P_n(\omega) &=&  {(4\omega)^{n-1}\over (1+\omega)^{2n}}\left(
  \tilde\Pi(q^2) - 
  \sum_{j=0}^{n-1}{1\over j!}\left(
  {d^j\over d(1/z)^j}\tilde\Pi(q^2)\bigg|_{z =
  -\infty}\right) {(1+\omega)^{2j}\over (4\omega)^j}\right)
  \,,
\end{eqnarray}
where for $\tilde{\Pi}(q^2)$ the terms up to order
$1/z^n$ must be known.
The available information transforms into
$P_n(-1)$ and $P_n^{(k)}(0), (k = 0,1,\ldots,n+n_0-1)$, 
where $n_0$ is the number of moments for $z\to0$.

\item
In the last step a Pad\'e approximation is performed 
for the function $P_n(\omega)$. This means that $P_n(\omega)$ is
represented through a function
\begin{eqnarray}
  [i/j](\omega) &=& {a_0 + a_1\omega +\cdots + a_i\omega^i\over
  1 + b_1\omega + \cdots + b_j\omega^j}
  \,,
  \label{eq:padedef}
\end{eqnarray}
where the number of coefficients on the r.h.s. depend on the amount of
information available for $P_n(\omega)$.

\end{enumerate}

Once an approximation for $P(\omega)$ is known the above steps have to
be inverted in order to obtain the function $\Pi(q^2)$. 

Due to the structure of Eq.~(\ref{eq:padedef}) 
some Pad\'e approximants develop poles inside the unit circle
($|\omega|\le1$). In general we will
discard such results in the following.
In some cases, however, the
pole coincides with a zero of the numerator up to several digits
accuracy. These Pad\'e approximations 
will be taken into account in constructing our results.
To be precise: in addition
to the Pad\'e results without any poles inside the unit circle, we
will use the ones where the poles are accompanied by zeros within a
circle of radius 0.01, and the distance between the pole and the
physically relevant point $q^2/M^2=1$ is larger than 0.1.


\section{\label{sec:eff}Current correlator in the effective theory}

The computation of the current correlators is performed in the full
theory. In this Section the connection to the correlator in the HQET
is described. 
This connection is exploited to obtain the leading logarithmic terms
of $R^v(s)$ and $R^s(s)$ at threshold.

In the effective theory we denote the $\overline{\rm MS}$ renormalized
current which couples to a massive and a massless quark by 
\begin{eqnarray}
  \tilde{j}_\Gamma &=& \bar{q} \Gamma \tilde{Q}
  \,,
\end{eqnarray}
with $\Gamma\in\{1,\gamma^\mu,i\gamma_5,\gamma^\mu\gamma_5\}$.
The light quark flavour $q$ is identical in the full and the effective
theory which is not true for the heavy one as indicated by the tilde.
In the effective theory
$\tilde{Q}$ is considered as a static quark which fulfills the
relation
\begin{eqnarray}
  \tilde{Q} &=& \gamma_0 \tilde{Q}
  \,.
\end{eqnarray}
Due to this condition the set of $\gamma$ matrices can be divided into
two groups: one which commutes ($\{i\gamma_5,\gamma_j\}, j=1,2,3$)
and one which anti-commutes ($\{1,\gamma_j\gamma_5\}$) with $\gamma_0$.

Let us next consider the corresponding correlators which are defined
in analogy to Eqs.~(\ref{eq:pivadef}) and~(\ref{eq:pispdef})
\begin{eqnarray}
  \tilde{\Pi}_{\Gamma_1,\Gamma_2}(q_0) &=& i \int {\rm d}x e^{iqx} 
  \langle 0| T \tilde{j}_{\Gamma_1}(x) \tilde{j}^\dagger_{\Gamma_2}(0) 
  |0\rangle
  \,.
  \label{eq:tilpidef}
\end{eqnarray}
They only depend on the zeroth component $q_0$ as due to the
Feynman rules of HQET in coordinate space one has 
$\langle 0| T \tilde{j}_\Gamma(x) \tilde{j}^\dagger_\Gamma(0) |0 \rangle 
\sim \delta(\vec{x})$.
Furthermore, in our case where only one of the quarks is massive,
one can write all correlators in terms of 
$\tilde{\Pi}(q_0) \equiv \tilde{\Pi}_{1,1}(q_0)$.
It is possible to show that the following equations hold\footnote{
This is true only if no power suppressed non-perturbative
corrections are taken into account, what is implicitly understood throughout
the present paper.}
\begin{eqnarray}
  \tilde{\Pi}_{i\gamma_5,i\gamma_5}(q_0) &=& \tilde{\Pi}(q_0)
  \,,
  \nonumber \\
  \tilde{\Pi}_{\gamma_i,\gamma_j}(q_0) &=& \delta_{ij}\tilde{\Pi}(q_0)
  \,,
  \nonumber \\
  \tilde{\Pi}_{\gamma_i\gamma_5,\gamma_j\gamma_5}(q_0) &=& 
  \delta_{ij}\tilde{\Pi}(q_0)
  \,.
  \label{eq:tilpidef2}
\end{eqnarray}
Thus in HQET there is only one independent correlator,
$\tilde{\Pi}(q_0)$, which will be considered in the following. 
The corresponding current will be denoted by $\tilde{j}$.

The relation between $\tilde{j}$ and $j^\delta$ ($\delta=v,s$)
has been computed in~\cite{BroGro95} up to order $\alpha_s^2$.
For $\mu=M$ it can be written in the form
\begin{eqnarray}
  j^\delta &=& C_\delta(M) \tilde{j}
  \,,
  \label{eq:jdec}
\end{eqnarray}
where the decoupling constants $C_\delta(M)$ are given by
\begin{eqnarray}
  C_v(M) &=& 1 - C_F \frac{\alpha_s^{(n_f)}(M)}{\pi} +
  \left(\frac{\alpha_s^{(n_f)}(M)}{\pi}\right)^2
  \Bigg[
    C_F^2\left(
       \frac{1453}{768} 
     - \frac{173}{48}\zeta_2
     + \frac{7}{2}\zeta_2\ln2
     - \frac{11}{8}\zeta_3
     \right)
  \nonumber\\&&\mbox{}
     + C_F C_A\left(
     - \frac{6821}{2304} 
     + \frac{21}{16}\zeta_2
     - \frac{7}{4}\zeta_2\ln2
     + \frac{9}{16}\zeta_3
     \right)
     + C_F T n_l \left(\frac{445}{576} + \frac{1}{4}\zeta_2\right)
  \nonumber\\&&\mbox{}
     + C_F T     \left(\frac{709}{576} - \frac{5}{6}\zeta_2\right)
  \Bigg]
  \,,
  \label{eq:Cv}
  \\
  C_s(M) &=& 1 + \frac{C_F}{2} \frac{\alpha_s^{(n_f)}(M)}{\pi} +
  \left(\frac{\alpha_s^{(n_f)}(M)}{\pi}\right)^2
  \Bigg[
    C_F^2\left(
       \frac{369}{256} 
     + \frac{15}{16}\zeta_2 
     - \frac{3}{2}\zeta_2\ln2
     - \frac{1}{8}\zeta_3 
     \right)
  \nonumber\\&&\mbox{}
     + C_F C_A\left(
       \frac{1351}{768} 
     - \frac{3}{16}\zeta_2 
     + \frac{3}{4}\zeta_2\ln2
     - \frac{1}{16}\zeta_3 
     \right)
     + C_F T n_l \left( -\frac{95}{192} - \frac{1}{4}\zeta_2 \right)
  \nonumber\\&&\mbox{}
     + C_F T     \left( \frac{169}{192} - \frac{1}{2}\zeta_2 \right)
  \Bigg]
  \label{eq:Cs}
  \,,
\end{eqnarray}
with $\zeta_2=\pi^2/6$ and $\zeta_3\approx1.202\,057$.
The superscript attached to $\alpha_s$ defines the number of active
flavours.
$\tilde{j}$ is still defined with $n_f$ active flavours.
The transition to a theory with only $n_l$ active flavours is achieved
through
\begin{eqnarray}
  \tilde{j} = \tilde{C}(M) \tilde{j}^\prime
  \,,
\end{eqnarray}
with~\cite{Gro98}
\begin{eqnarray}
  \tilde{C}(M) &=& 1+\frac{89}{576}C_FT 
                      \left(\frac{\alpha_s^{(n_l)}(M)}{\pi}\right)^2
  \,,
  \label{eq:Ctil}
\end{eqnarray}
where again $\mu=M$ has been adopted.
In analogy to Eqs.~(\ref{eq:tilpidef}) and~(\ref{eq:tilpidef2})
one can define polarization functions also in the primed theory.
Again there is only one which is independent. It will be denoted by 
$\tilde{\Pi}^\prime(q_0)$ which only depends on the massless quark flavours.
Thus besides the renormalization scale $\mu$ 
there is only one more dimensionful quantity which 
in the framework of HQET is usually chosen to be $\omega=\sqrt{s}-M$.
As a consequence the renormalization group equation for
$\tilde{j}^\prime$ takes the simple form
\begin{eqnarray}
  \mu^2\frac{{\rm d}}{{\rm d}\mu^2} \tilde{j}^\prime &=& 
  \tilde{\gamma}^\prime \tilde{j}^\prime
  \,,
  \label{eq:rgeq}
\end{eqnarray}
with~\cite{JiMus91}
\begin{eqnarray}
  \tilde{\gamma}^\prime &=& C_F\frac{3}{8}\frac{\alpha_s^{(n_l)}}{\pi}
  + \left(\frac{\alpha_s^{(n_l)}}{\pi}\right)^2 \left[
    -C_F^2\left( \frac{5}{64}-\frac{1}{2}\zeta_2 \right)
    -C_AC_F\left( -\frac{49}{192}+\frac{1}{8}\zeta_2 \right)
    \right.\nonumber\\&&\left.\mbox{}
    -C_F T n_l \frac{5}{48}
    \right] 
  + {\cal O}(\alpha_s^3)
  \,.
  \label{eq:gam}
\end{eqnarray}
Note that the primed quantities only depend on $\alpha_s^{(n_l)}$.
This allows for a simple reconstruction of the logarithms
$\ln(\mu^2/\omega^2)$ at ${\cal O}(\alpha_s^2)$
of $\tilde{R}^\prime$ defined through
\begin{eqnarray}
  \tilde{R}^\prime(\omega) &=& 2\pi\,  \mbox{Im}\,\,
  \left[ \tilde{\Pi}^\prime(q_0) \right]\Bigg|_{q^2=s+i\epsilon,
                                                \omega=\sqrt{s}-M} 
  \,.
\end{eqnarray}
Once they are at hand Eqs.~(\ref{eq:Cv}),~(\ref{eq:Cs})
and~(\ref{eq:Ctil})
can be used to predict the logarithms 
of $R^v$ and $R^s$ at threshold via the relations
\begin{eqnarray}
  R^v(s) &=& 
  6 \left[C_v(M)\tilde{C}(M)\right]^2
  \frac{v^2}{\omega^2} \tilde{R}^\prime
  + {\cal O}(v^3)
  \,,
  \nonumber\\
  R^s(s) &=&
  4 \left[\frac{m(M)}{M}C_s(M)\tilde{C}(M)\right]^2
  \frac{v^2}{\omega^2} \tilde{R}^\prime
  + {\cal O}(v^3)
  \,,
  \label{eq:Rthr}
\end{eqnarray}
where it is understood that $\omega$ is expressed in terms of $v$ via
the relation
\begin{eqnarray}
1 - \frac{\omega}{M} &=& \sqrt{\frac{1-v}{1+v}}
  \,,
\end{eqnarray}
an expansion for $v\to0$ is performed and only the leading term is kept.

At this point we want to mention that the developed formalism --- in
particular the decoupling relations~(\ref{eq:jdec})
and~(\ref{eq:Ctil}) --- only applies to the imaginary part. For the
polarization function one would have to take into account 
additional contributions.

The procedure described above determines the logarithmic parts of the
leading 
threshold term. They are incorporated into the Pad\'e procedure as
described in Section~\ref{sec:pade} (cf. point \ref{item:thr}).
Note, that the non-logarithmic part for $R^\delta$ are not fixed via
this procedure. In Section~\ref{sec:reseff} we
will extract numerical approximations
from our Pad\'e results.

At the end of this Section we want to list the available information
for $\tilde{R}^\prime(\omega)$.
Using Eqs.~(\ref{eq:rgeq}) and~(\ref{eq:gam}) and the ${\cal O}(\alpha_s)$
result for $\tilde{R}^\prime$~\cite{BroGro92,Bagan92} one obtains
\begin{eqnarray}
  \tilde{R}^\prime(\omega) &=& N_c \omega^2 \Bigg\{ 1 +
  \frac{\alpha_s^{(n_l)}(\mu)}{\pi}
  C_F\left(
      \frac{17}{4} 
    - \frac{3}{2}\ln2
    + 2\zeta_2
    + \frac{3}{4}\Lw
  \right)
  \nonumber\\&&\mbox{}
  +\left(\frac{\alpha_s^{(n_l)}(\mu)}{\pi}\right)^2
  \Bigg[
  C_F^2\left(
      \tilde{c}_{FF} 
    + \left(
         \frac{97}{32}
       - \frac{9}{8}\ln2
       + \frac{5}{2}\zeta_2
      \right)\Lw
    + \frac{9}{32}\Lw^2
  \right)
  \nonumber\\&&\mbox{}
  +C_AC_F\left(
    \tilde{c}_{FA} 
    + \left( 
        \frac{141}{32}
      - \frac{11}{8}\ln2
      + \frac{19}{12}\zeta_2
      \right)\Lw
    + \frac{11}{32}\Lw^2 
  \right)
  \nonumber\\&&\mbox{}
  +C_FTn_l\left(
    \tilde{c}_{FL} 
    + \left(
      - \frac{13}{8}
      + \frac{1}{2}\ln2
      - \frac{2}{3}\zeta_2
      \right)\Lw
    - \frac{1}{8}\Lw^2
  \right)
  \Bigg]
  \Bigg\}
  \,,
  \label{eq:Rtil}
\end{eqnarray}
with $\Lw=\ln(\mu^2/\omega^2)$. The coefficients 
$\tilde{c}_{FF}$, $\tilde{c}_{FA}$ and $\tilde{c}_{FL}$ are not
known. In Section~\ref{sec:reseff} we will provide numerical approximations.

With the help of Eq.~(\ref{eq:Rthr}) one obtains the leading terms of
$R^v$ and $R^s$ for $v\to0$.
Separated according to the colour factors
they read ($\mu=M$).
\begin{eqnarray}
R^{v,thr} &=& N_c v^2 \Bigg\{ 6 +
  \frac{\alpha_s^{(n_f)}(M)}{\pi}
  C_F\left(
    \frac{27}{2} 
    + 12\zeta_2 
    - 9\ln2
    - 9\ln v
  \right)
  \nonumber\\&&\mbox{}
  +\left(\frac{\alpha_s^{(n_f)}(M)}{\pi}\right)^2
  \Bigg[
  C_F^2\left(
    c^v_{FF}
   +
   \left(
    -\frac{147}{8} 
    - 30\zeta_2 
    + \frac{27}{2}\ln2
   \right)\ln v 
   + \frac{27}{4}\ln^2 v
  \right)
  \nonumber\\&&\mbox{}
  +C_AC_F\left(
    c^v_{FA}
   + 
   \left(
    -\frac{423}{8} 
    - 19\zeta_2 
    + \frac{33}{2}\ln2
   \right)\ln v 
   + \frac{33}{4} \ln^2 v
  \right)
  \nonumber\\&&\mbox{}
  +C_FTn_l\left(
    c^v_{FL}
   + 
   \left(
    \frac{39}{2} 
    + 8\zeta_2 
    - 6\ln2
    \right)\ln v 
    - 3 \ln^2 v
  \right)
  \nonumber\\&&\mbox{}
  +C_FT\left( 
   \frac{133}{8}
   -10\zeta_2
  \right)
  \Bigg]
  \Bigg\}
  \,,
  \label{eq:Rvthr}
  \\
R^{s,thr} &=& N_c v^2 \Bigg\{ 4 +
  \frac{\alpha_s^{(n_f)}(M)}{\pi}
  C_F\left(
    13 + 8\zeta_2 - 6\ln2 - 6\ln v
  \right)
  \nonumber\\&&\mbox{}
  +\left(\frac{\alpha_s^{(n_f)}(M)}{\pi}\right)^2
  \Bigg[
  C_F^2\left(
    c^s_{FF}
    + 
    \left(
    -\frac{73}{4} 
    - 20\zeta_2
    + 9\ln2
    \right) \ln v 
    + \frac{9}{2}\ln^2 v
  \right)
  \nonumber\\&&\mbox{}
  +C_AC_F\left(
    c^s_{FA}
    + 
   \left(
   -\frac{141}{4} 
   - \frac{38}{3}\zeta_2 
   + 11\ln2
   \right)\ln v 
   + \frac{11}{2}\ln^2 v
  \right)
  \nonumber\\&&\mbox{}
  +C_FTn_l\left(
  c^s_{FL}+
    \left(
     13 
     + \frac{16}{3}\zeta_2 
     - 4\ln 2
    \right) \ln v
    - 2\ln^2 v
  \right)
  \nonumber\\&&\mbox{}
  +C_FT\left(
    \frac{727}{36}
    - 12 \zeta_2
  \right)
  \Bigg]
  \Bigg\}
  \,,
  \label{eq:Rsthr}
\end{eqnarray}
with $\zeta_2=\pi^2/6$.
The smooth behaviour proportional to $v^2$ is a 
consequence from the fact that $\tilde{R}^\prime$ is, for dimensional
reasons, proportional to $\omega^2$. This is in contrast to the
diagonal correlators where 
at order $\alpha_s^2$ even $1/v$ singularities appear~\cite{CheKueSte96}. 
Note that in 
Eqs.~(\ref{eq:Rvthr}) and~(\ref{eq:Rsthr})
the terms proportional to the colour factor $C_FT$ are
completely fixed.
For later use it is convenient to display explicitly the relations
between $\tilde{c}$ and $c^\delta$:
\begin{eqnarray}
  c^v_{FF} 
  &=& 
    6 \tilde{c}_{FF} 
    - \frac{1427}{64} 
    + 18\ln2  
    - \frac{269}{4}\zeta_2 
    + 42\zeta_2\ln2 
    - \frac{33}{2}\zeta_3
\,\,\approx\,\,
  -92.3884 + {6} \tilde{c}_{FF}
  \,,\nonumber\\
  c^v_{FA} 
  &=&   6 \tilde{c}_{FA} 
    -\frac{6821}{192}  + \frac{63}{4}\zeta_2 
    - 21\zeta_2\ln2 + \frac{27}{4}\zeta_3
\,\,\approx\,\,
    -25.4483 + {6} \tilde{c}_{FA}
  \,,\nonumber\\
  c^v_{FL} 
  &=& {6} \tilde{c}_{FL} 
    +\frac{445}{48} +  3\zeta_2
\,\,\approx\,\,
  + 14.2056 + {6} \tilde{c}_{FL}
\label{eq:cctilv}
  \,,\\
  c^s_{FF} 
  &=& 4\tilde{c}_{FF} 
    -\frac{257}{32}  
    + 6\ln2 
    - \frac{31}{2}\zeta_2 
    + 12\zeta_2\ln2 
    - 7\zeta_3
\,\,\approx\,\,
  -24.1011 + 4\tilde{c}_{FF}
  \,,\nonumber\\
  c^s_{FA} 
  &=& 4\tilde{c}_{FA} 
    -\frac{871}{96}  
    + \frac{5}{2}\zeta_2 
    - 6\zeta_2\ln2 
    + \frac{5}{2}\zeta_3
\,\,\approx\,\,
  -8.79653  + 4\tilde{c}_{FA}
  \,,\nonumber\\
  c^s_{FL} 
  &=& 4\tilde{c}_{FL} + 
    \frac{47}{24}  + 2\zeta_2
\,\,\approx\,\,
    5.2482 + 4\tilde{c}_{FL}
\,.
\label{eq:cctils}
\end{eqnarray}


\section{\label{sec:res}Spectral function in full QCD}

In this Section we discuss the computation of the polarization
function in full QCD. Explicit results are given for the imaginary
parts which constitute physical quantities. In a first step the input
quantities needed for the Pad\'e procedure are provided.

\begin{figure}[t]
  \begin{center}
    \begin{tabular}{c}
      \leavevmode
      \epsfxsize=14cm
      \epsffile[65 525 570 720]{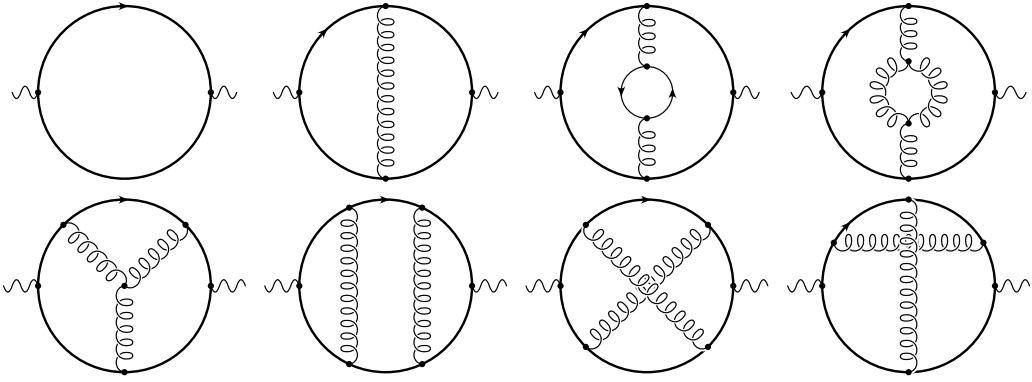}
    \end{tabular}
  \end{center}
  \vspace{-2.em}
  \caption{\label{fig:diags}Sample diagrams contributing to the
    current correlator of Eqs.~(\ref{eq:pivadef}) and~(\ref{eq:pispdef}).
    The straight and loopy lines represent quarks and gluons, respectively.
    One of the quarks connected to the external current carries mass
    $M$ whereas the other one is massless.
          }
\end{figure}

In Fig.~\ref{fig:diags} typical diagrams are pictured. Altogether
roughly 30 contributions have to be considered at three-loop order. 
Although their number is relatively
small we used {\tt GEFICOM}~\cite{geficom} for the automatic computation.
{\tt GEFICOM} uses {\tt QGRAF}~\cite{qgraf} for the generation of the
diagrams. In case an asymptotic expansion has to be applied
{\tt LMP}~\cite{Har:diss} or {\tt EXP}~\cite{Sei:dipl}
are used for the generation
of the sub-diagrams. The occurring vacuum diagrams are passed to 
{\tt MATAD}~\cite{matad} and the massless propagator type diagrams are
evaluated to {\tt MINCER}~\cite{mincer}. More details on the automatic
computation of Feynman diagrams can be found in~\cite{HarSte98}.

At three-loop order 
terms up to order $z^6$ ($z^7$) could be evaluated 
for the vector (scalar) correlator in the limit $z\to0$.
For convenience we list below the analytical results
for the one- and two-loop correlator up to order $z^7$ and 
present the results for the $\Pi^{(2),\delta}$ in the Appendix.
The analytical results can also be found under the URL
\verb|http://www-ttp.physik.uni-karlsruhe.de/Progdata/ttp01/ttp01-14|.
In the low-energy limit we obtain at order $\alpha_s^0$ and 
$\alpha_s^1$
\begin{eqnarray}
\Pi^{(0),v}&=& \frac{3}{16\pi^2}\Bigg[
+\frac{1}{2}\,z
+\frac{2}{15}\,z^2
+\frac{1}{18}\,z^3
+\frac{1}{35}\,z^4
+\frac{1}{60}\,z^5
+\frac{2}{189}\,z^6
+\frac{1}{140}\,z^7
\Bigg]+\ldots\,,
\nonumber\\
\Pi^{(1),v}
&=& \frac{3}{16\pi^2}\Bigg[
+\left(\frac{1}{2}\,\zeta_2 +
 \frac{25}{48}\right)\,z
+\left(\frac{2}{15}\,\zeta_2 +
 \frac{5}{18}\right)\,z^2
+\left(\frac{1}{18}\,\zeta_2 +
 \frac{503}{3240}\right)\,z^3
\nonumber\\&&\mbox{}
+\left(\frac{1}{35}\,\zeta_2 +
 \frac{1199}{12600}\right)\,z^4
+\left(\frac{1}{60}\,\zeta_2 +
 \frac{37883}{604800}\right)\,z^5
+\left(\frac{2}{189}\,\zeta_2 +
 \frac{23029}{529200}\right)\,z^6
\nonumber\\&&\mbox{}
+\left(\frac{1}{140}\,\zeta_2 +
 \frac{222433}{7056000}\right)\,z^7
\Bigg]+\ldots\,,
\\
\Pi^{(0),s}&=& \frac{3}{16\pi^2}\Bigg[
+\frac{2}{3}\,z
+\frac{1}{6}\,z^2
+\frac{1}{15}\,z^3
+\frac{1}{30}\,z^4
+\frac{2}{105}\,z^5
+\frac{1}{84}\,z^6
+\frac{1}{126}\,z^7
\Bigg]+\ldots\,,
\nonumber\\
\Pi^{(1),s}
&=& \frac{3}{16\pi^2}\Bigg[
+\left(\frac{2}{3}\,\zeta_2 +
 \frac{1}{12}\right)\,z
+\left(\frac{1}{6}\,\zeta_2 +
 \frac{19}{48}\right)\,z^2
+\left(\frac{1}{15}\,\zeta_2 +
 \frac{17}{72}\right)\,z^3
\nonumber\\&&\mbox{}
+\left(\frac{1}{30}\,\zeta_2 +
 \frac{31}{216}\right)\,z^4
+\left(\frac{2}{105}\,\zeta_2 +
 \frac{7001}{75600}\right)\,z^5
+\left(\frac{1}{84}\,\zeta_2 +
 \frac{38113}{604800}\right)\,z^6
\nonumber\\&&\mbox{}
+\left(\frac{1}{126}\,\zeta_2 +
 \frac{18961}{423360}\right)\,z^7
\Bigg]+\ldots\,,
\end{eqnarray}
where the on-shell quark mass $M$ has been used as a parameter.

In the high energy region eight expansion terms could be
obtained both in the vector and scalar case. 
Again we provide in the main text
the one- and two-loop results in analytic
form and refer for the analytic three-loop terms to the Appendix
and to the URL
\verb|http://www-ttp.physik.uni-karlsruhe.de/Progdata/ttp01/ttp01-14|.
Using again the on-shell mass the one-and two-loop results read
\begin{eqnarray}
\Pi^{(0),v}&=& \frac{3}{16\pi^2}\Bigg[
\frac{16}{9}
+\frac{8}{3}\,\lz
+\left(1 -
 4\,\lz\right)\,\frac{1}{z}
-
2\,\frac{1}{z^2}
+\left(\frac{4}{3}\,\lz -
 \frac{5}{9}\right)\,\frac{1}{z^3}
+\frac{1}{3}\,\frac{1}{z^4}
\nonumber\\&&\mbox{}
+\frac{1}{10}\,\frac{1}{z^5}
+\frac{2}{45}\,\frac{1}{z^6}
+\frac{1}{42}\,\frac{1}{z^7}
\Bigg]+\ldots\,,
\nonumber\\
\Pi^{(1),v}&=& \frac{3}{16\pi^2}\Bigg[
\frac{25}{18}
+2\,\lz
-
4\,\zeta_3
+\frac{4}{9}\,\zeta_2
+\left(3\,\lz -
 6\,\lz^2 -
 \zeta_2 +
 6\,\zeta_3 -
 \frac{23}{4}\right)\,\frac{1}{z}
\nonumber\\&&\mbox{}
+\left(2 -
 6\,\lz\right)\,\frac{1}{z^2}
+\left(-
\left(\frac{104}{27}\,\lz\right) +
 \frac{40}{9}\,\lz^2 -
 2\,\zeta_3 +
 \frac{28}{27}\right)\,\frac{1}{z^3}
\nonumber\\&&\mbox{}
+\left(\frac{31}{9}\,\lz -
 \frac{2}{3}\,\lz^2 -
 \frac{31}{16}\right)\,\frac{1}{z^4}
+\left(\frac{56}{75}\,\lz -
 \frac{1}{5}\,\lz^2 -
 \frac{1747}{36000}\right)\,\frac{1}{z^5}
\nonumber\\&&\mbox{}
+\left(\frac{7}{25}\,\lz -
 \frac{4}{45}\,\lz^2 +
 \frac{3833}{81000}\right)\,\frac{1}{z^6}
\nonumber\\&&\mbox{}
+\left(\frac{296}{2205}\,\lz -
 \frac{1}{21}\,\lz^2 +
 \frac{322099}{7408800}\right)\,\frac{1}{z^7}
\Bigg]+\ldots\,,
\\
\Pi^{(0),s}&=& \frac{3}{16\pi^2}\Bigg[
3
+4\,\lz
-
8\,\lz\,\frac{1}{z}
+\left(-
3 +
 4\,\lz\right)\,\frac{1}{z^2}
+\frac{2}{3}\,\frac{1}{z^3}
\nonumber\\&&\mbox{}
+\frac{1}{6}\,\frac{1}{z^4}
+\frac{1}{15}\,\frac{1}{z^5}
+\frac{1}{30}\,\frac{1}{z^6}
+\frac{2}{105}\,\frac{1}{z^7}
\Bigg]+\ldots\,,
\nonumber\\
\Pi^{(1),s}&=& \frac{3}{16\pi^2}\Bigg[
\frac{19}{4}
+6\,\lz^2
+9\,\lz
-
6\,\zeta_3
+\zeta_2
+\left(-
7 -
 24\,\lz^2 +
 12\,\zeta_3\right)\,\frac{1}{z}
\nonumber\\&&\mbox{}
+\left(8 -
 24\,\lz +
 24\,\lz^2 -
 6\,\zeta_3\right)\,\frac{1}{z^2}
+\left(\frac{128}{9}\,\lz -
 \frac{16}{3}\,\lz^2 -
 \frac{107}{18}\right)\,\frac{1}{z^3}
\nonumber\\&&\mbox{}
+\left(\frac{2}{9}\,\lz -
 \frac{1}{3}\,\lz^2 +
 \frac{263}{288}\right)\,\frac{1}{z^4}
+\left(\frac{47}{225}\,\lz -
 \frac{2}{15}\,\lz^2 +
 \frac{3521}{18000}\right)\,\frac{1}{z^5}
\nonumber\\&&\mbox{}
+\left(\frac{26}{225}\,\lz -
 \frac{1}{15}\,\lz^2 +
 \frac{10243}{108000}\right)\,\frac{1}{z^6}
\nonumber\\&&\mbox{}
+\left(\frac{743}{11025}\,\lz -
 \frac{4}{105}\,\lz^2 +
 \frac{532267}{9261000}\right)\,\frac{1}{z^7}
\Bigg]+\ldots\,,
\end{eqnarray}
with $\lz=-(\ln(-z))/2$.

In order to incorporate the available information at threshold one
has to perform an analytical continuation of the expressions in 
Eqs.~(\ref{eq:Rvthr}) and~(\ref{eq:Rsthr}).
Taking the logarithmic parts of Eqs.~(\ref{eq:Rvthr}) and~(\ref{eq:Rsthr})
one obtains the quadratic and cubic logarithms for the polarization functions
which read
\begin{eqnarray}
  \Pi^{(1),v,thr}_{log} &=&
    \frac{3}{16\pi^2}
      \frac{3}{2}\left(1-\frac{1}{z}\right)^2
      \ln^2\frac{1}{1-z} 
  \,,
  \nonumber\\
  \Pi^{(2),v,thr}_{FF,log} &=&
    \frac{3}{16\pi^2} 
    \left(1-\frac{1}{z}\right)^2
      \left[ 
      \left(\frac{49}{16}+5\zeta_2\right) \ln^2\frac{1}{1-z} 
                           + \frac{3}{4} \ln^3\frac{1}{1-z} 
      \right]
  \,,
  \nonumber\\
  \Pi^{(2),v,thr}_{FA,log} &=&
    \frac{3}{16\pi^2} 
    \left(1-\frac{1}{z}\right)^2
      \left[
         \left(\frac{423}{48}
              +\frac{19}{6}\zeta_2
              +\frac{11}{8}\ln\frac{\mu^2}{M^2}
         \right)\ln^2\frac{1}{1-z} 
         + \frac{11}{12} \ln^3\frac{1}{1-z}
    \right]
  \,,
  \nonumber\\
  \Pi^{(2),v,thr}_{FL,log} &=&
    \frac{3}{16\pi^2} 
    \left(1-\frac{1}{z}\right)^2
      \left[
         \left(
            -\frac{13}{4}
            -\frac{1}{2}\ln\frac{\mu^2}{M^2}
            -\frac{4}{3}\zeta_2
         \right)\ln^2\frac{1}{1-z} 
         - \frac{1}{3} \ln^3\frac{1}{1-z}
    \right]
  \,,
  \nonumber\\
  \Pi^{(2),v,thr}_{FH,log} &=& 0
  \,,
  \nonumber\\
  \Pi^{(1),s,thr}_{log} &=&
    \frac{3}{16\pi^2}
      \frac{3}{2}\left(1-\frac{1}{z}\right)^2
      \ln^2\frac{1}{1-z} 
  \,,
  \nonumber\\
  \Pi^{(2),s,thr}_{FF,log} &=&
    \frac{3}{16\pi^2} 
    \left(1-\frac{1}{z}\right)^2
      \left[ 
      \left(\frac{73}{16}+5\zeta_2\right) \ln^2\frac{1}{1-z} 
                           + \frac{3}{4} \ln^3\frac{1}{1-z} 
      \right]
  \,,
  \nonumber\\
  \Pi^{(2),s,thr}_{FA,log} &=&
    \frac{3}{16\pi^2} 
    \left(1-\frac{1}{z}\right)^2
      \left[
         \left(\frac{423}{48}
              +\frac{19}{6}\zeta_2
              +\frac{11}{8}\ln\frac{\mu^2}{M^2}
         \right)\ln^2\frac{1}{1-z} 
         + \frac{11}{12} \ln^3\frac{1}{1-z}
    \right]
  \,,
  \nonumber\\
  \Pi^{(2),s,thr}_{FL,log} &=&
    \frac{3}{16\pi^2} 
    \left(1-\frac{1}{z}\right)^2
      \left[
         \left(
            -\frac{13}{4}
            -\frac{1}{2}\ln\frac{\mu^2}{M^2}
            -\frac{4}{3}\zeta_2
         \right)\ln^2\frac{1}{1-z} 
         - \frac{1}{3} \ln^3\frac{1}{1-z}
    \right]
  \,,
  \nonumber\\
  \Pi^{(2),s,thr}_{FH,log} &=& 0
  \label{eq:Pithr}
  \,.
\end{eqnarray}
Note that, except for the coefficient of the quadratic
logarithm in the structure $FF$, there is agreement between the
coefficients of the vector and scalar correlator. 
The subscript ``{\it log}'' reminds that the imaginary parts of the
expressions in Eq.~(\ref{eq:Pithr}) only reproduce the logarithmic
parts of~(\ref{eq:Rvthr}) and~(\ref{eq:Rsthr}).
We want to mention that 
the polarization function itself also contains linear logarithmic terms 
which are not yet known at order $\alpha_s^2$. 
Below we will derive numerical estimates for them.
For the two-loop correlators the linear logarithms are known. They can
be obtained from the constant contributions in 
Eqs.~(\ref{eq:Rvthr}) and~(\ref{eq:Rsthr}).

Now the complete input is available and the individual steps described
in Section~\ref{sec:pade} can be performed leading to a large variety of
Pad\'e approximants. For the results we present
in the following those Pad\'e approximants are chosen which contain for
their construction at least terms of order $z^5$ and $1/z^5$ 
in the small and large momentum region, respectively.
Furthermore we demand that the difference of the degree in the
numerator and denominator in Eq.~(\ref{eq:padedef}) is less or equal to two.

\begin{figure}[t]
  \begin{center}
    \begin{tabular}{cc}
      \leavevmode
      \epsfxsize=7.cm
      \epsffile[110 280 460 560]{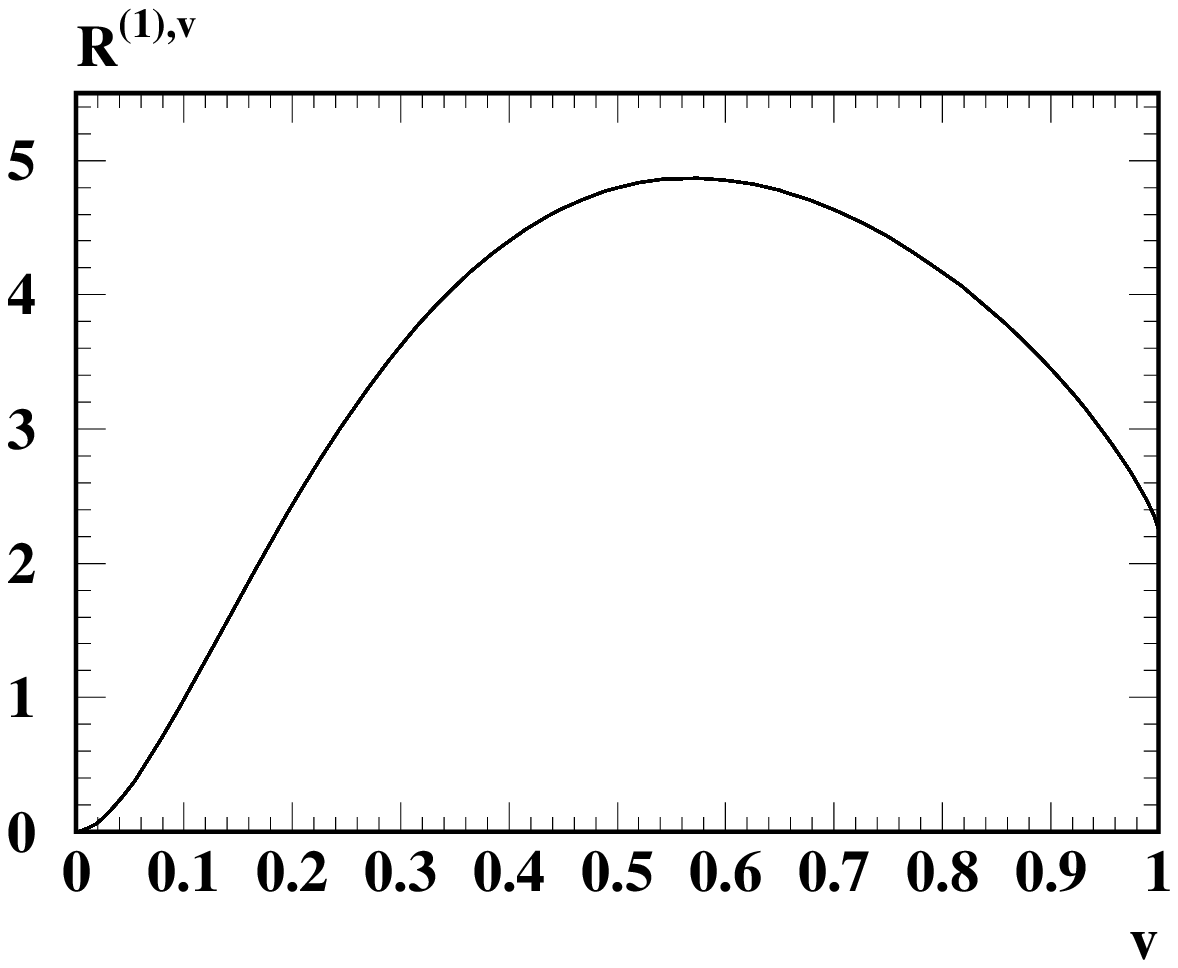}
      &
      \epsfxsize=7.cm
      \epsffile[110 280 460 560]{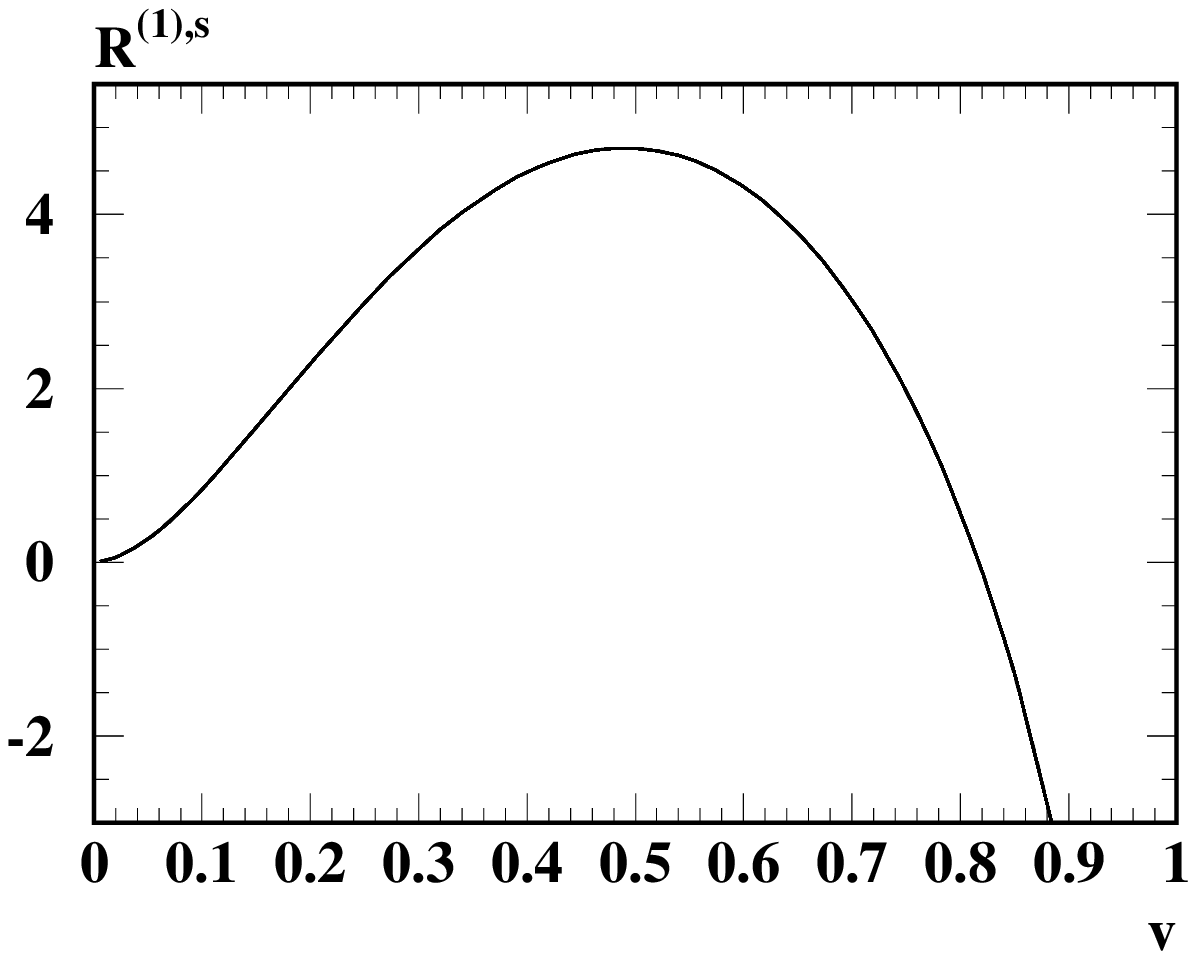}
    \end{tabular}
  \end{center}
  \vspace{-2.em}
  \caption{\label{fig:R2lv}$R^{(1),v}(s)$ and $R^{(1),s}(s)$ as a
    function of $v$.
          }
\end{figure}

\begin{figure}[t]
  \begin{center}
    \begin{tabular}{cc}
      \leavevmode
      \epsfxsize=7.cm
      \epsffile[110 280 460 560]{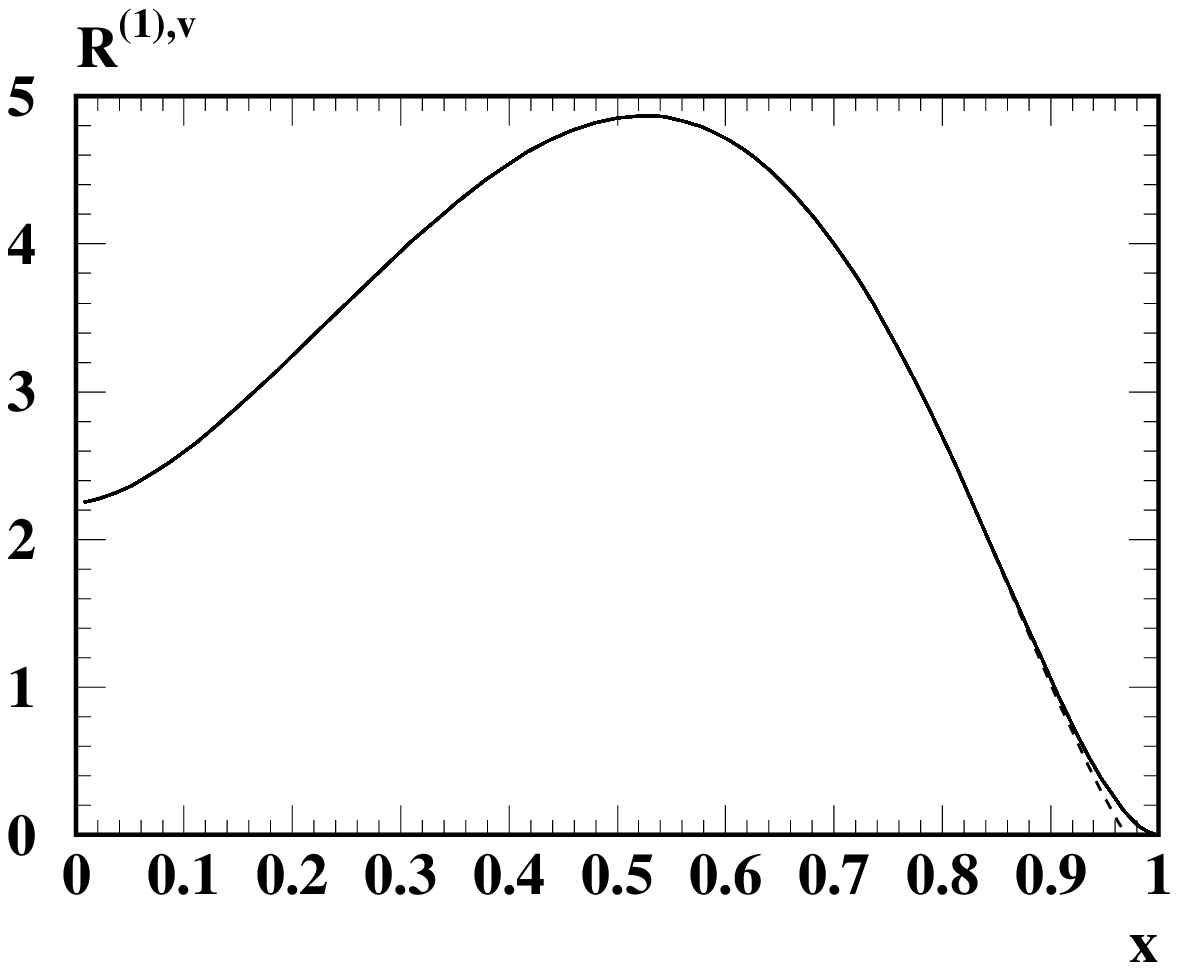}
      &
      \epsfxsize=7.cm
      \epsffile[110 280 460 560]{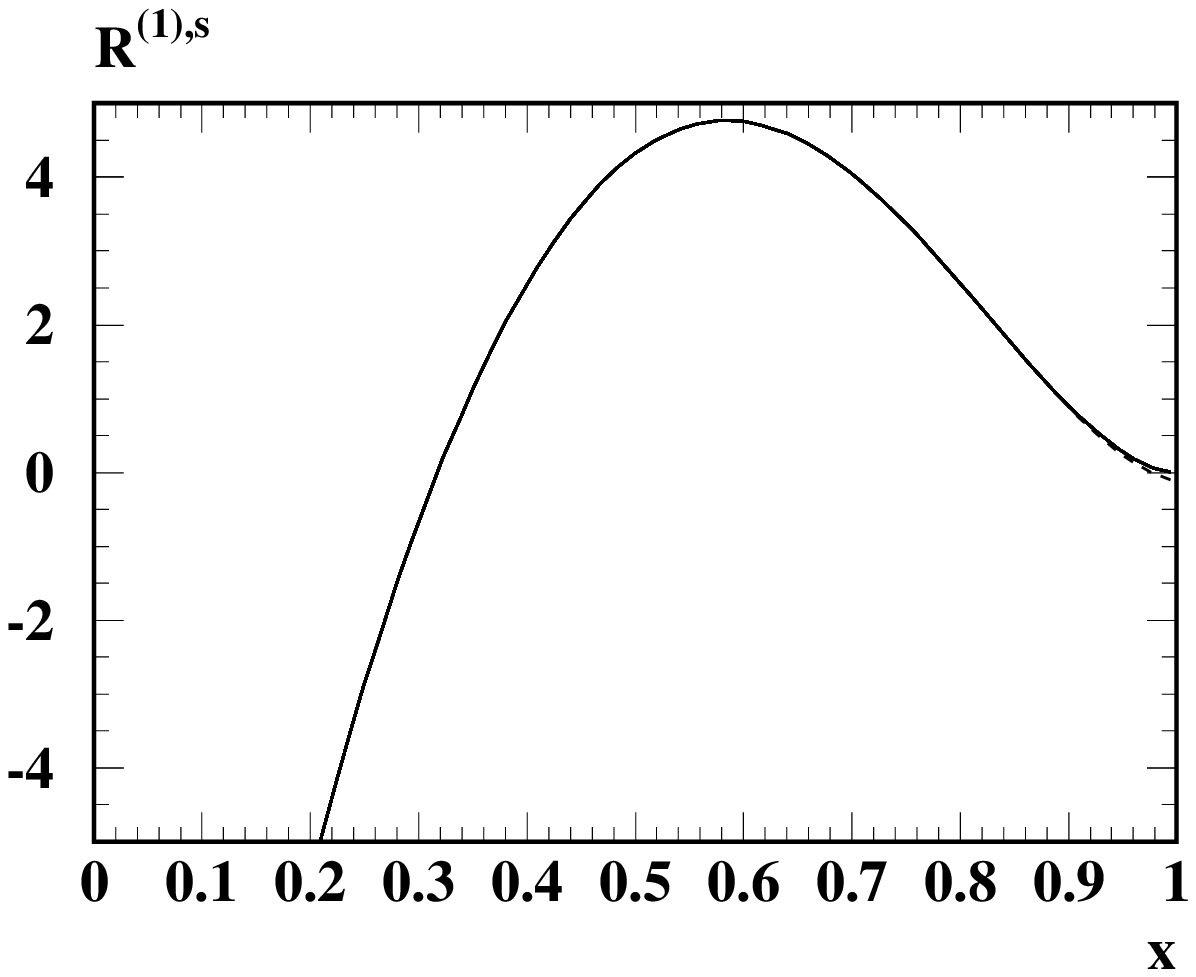}
    \end{tabular}
  \end{center}
  \vspace{-2.em}
  \caption{\label{fig:R2lx}$R^{(1),v}(s)$ and $R^{(1),s}(s)$ as a
    function of $x$.
          }
\end{figure}

In Figs.~\ref{fig:R2lv} and~\ref{fig:R2lx} the two-loop results are
plotted as a function of $v$ and $x$, respectively.
In addition to the roughly 15 Pad\'e results also the exact expression
is plotted. However, it is not possible to detect any difference ---
even close to the threshold at $v=0$.
The dashed lines in Fig.~\ref{fig:R2lx} correspond to the
results of the high-energy expansion including terms up to order
$1/z^7$. Although in these curves only the information from $x\to0$ is
incorporated one observes a perfect agreement with the exact results
up to $x\approx0.9$ ($v\approx0.10$).

\begin{figure}[t]
  \begin{center}
    \begin{tabular}{cc}
      \leavevmode
      \epsfxsize=7.cm
      \epsffile[110 280 460 560]{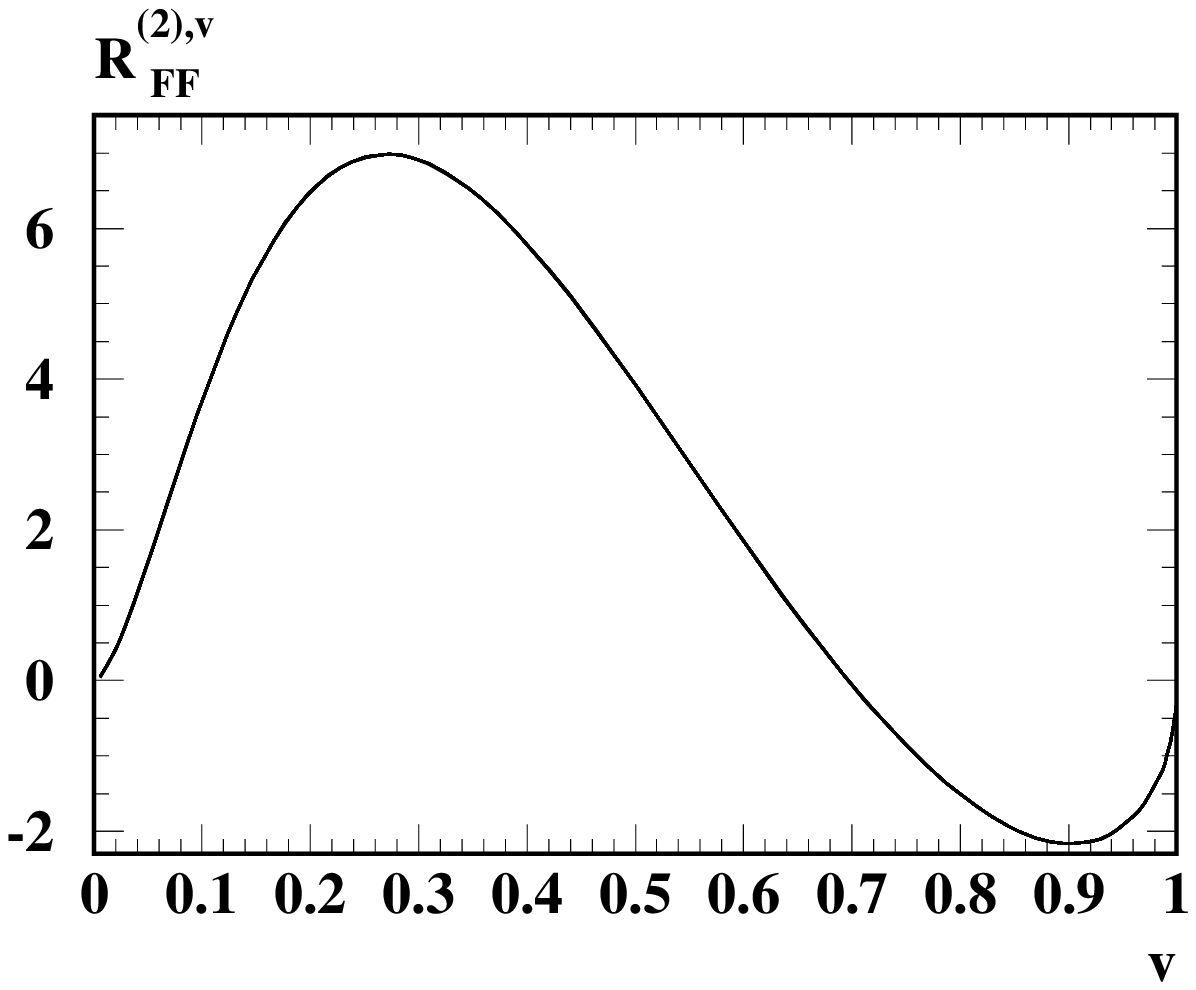}
      &
      \epsfxsize=7.cm
      \epsffile[110 280 460 560]{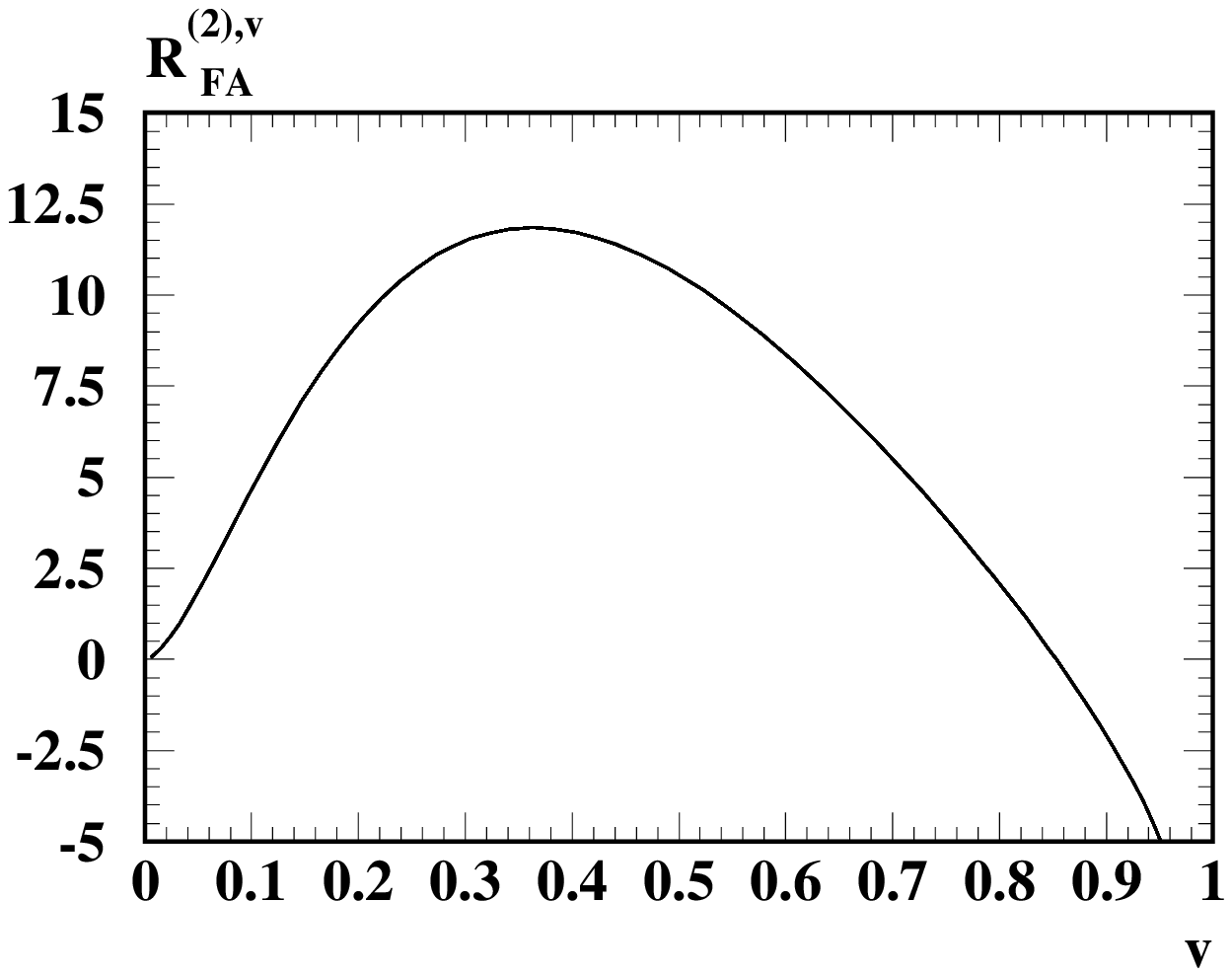}
      \\
      \epsfxsize=7.cm
      \epsffile[110 280 460 560]{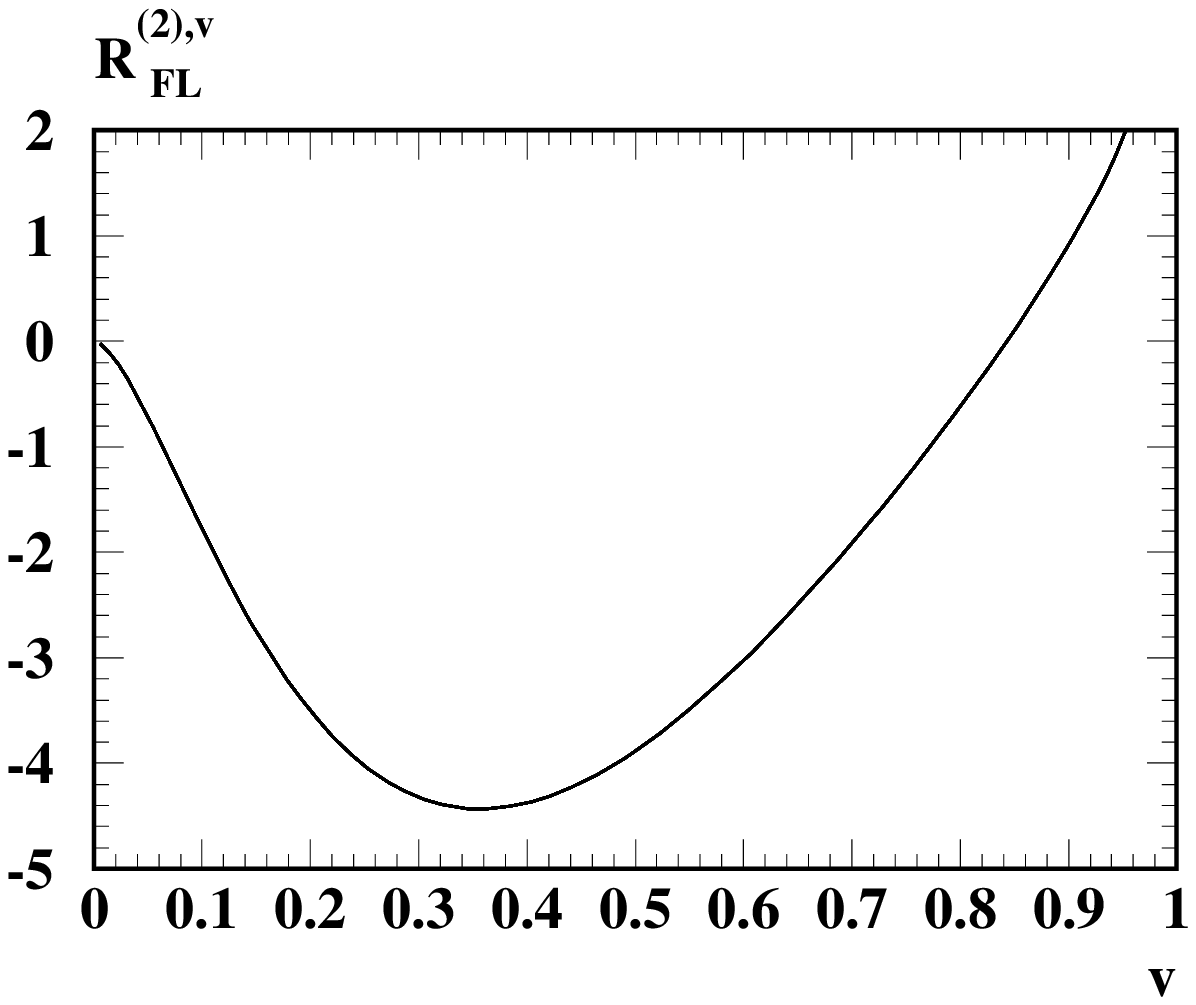}
      &
      \epsfxsize=7.cm
      \epsffile[110 280 460 560]{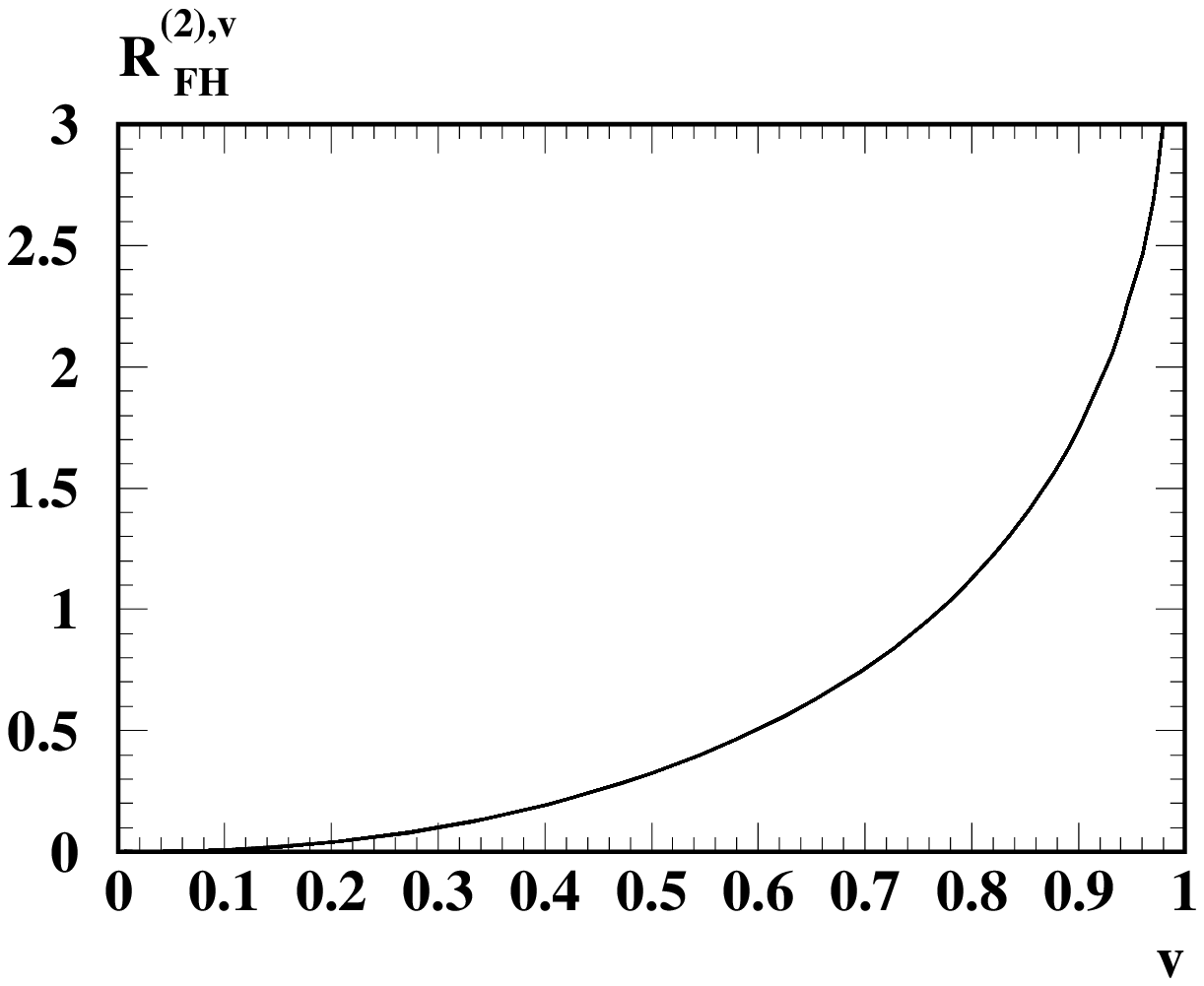}
    \end{tabular}
  \end{center}
  \vspace{-2.em}
  \caption{\label{fig:Rvv}$R^{(2),v}_{FF}(s)$, $R^{(2),v}_{FA}(s)$,
    $R^{(2),v}_{FL}(s)$  and $R^{(2),v}_{FH}(s)$ as a
    function of $v$.
          }
\end{figure}

\begin{figure}[t]
  \begin{center}
    \begin{tabular}{cc}
      \leavevmode
      \epsfxsize=7.cm
      \epsffile[110 280 460 560]{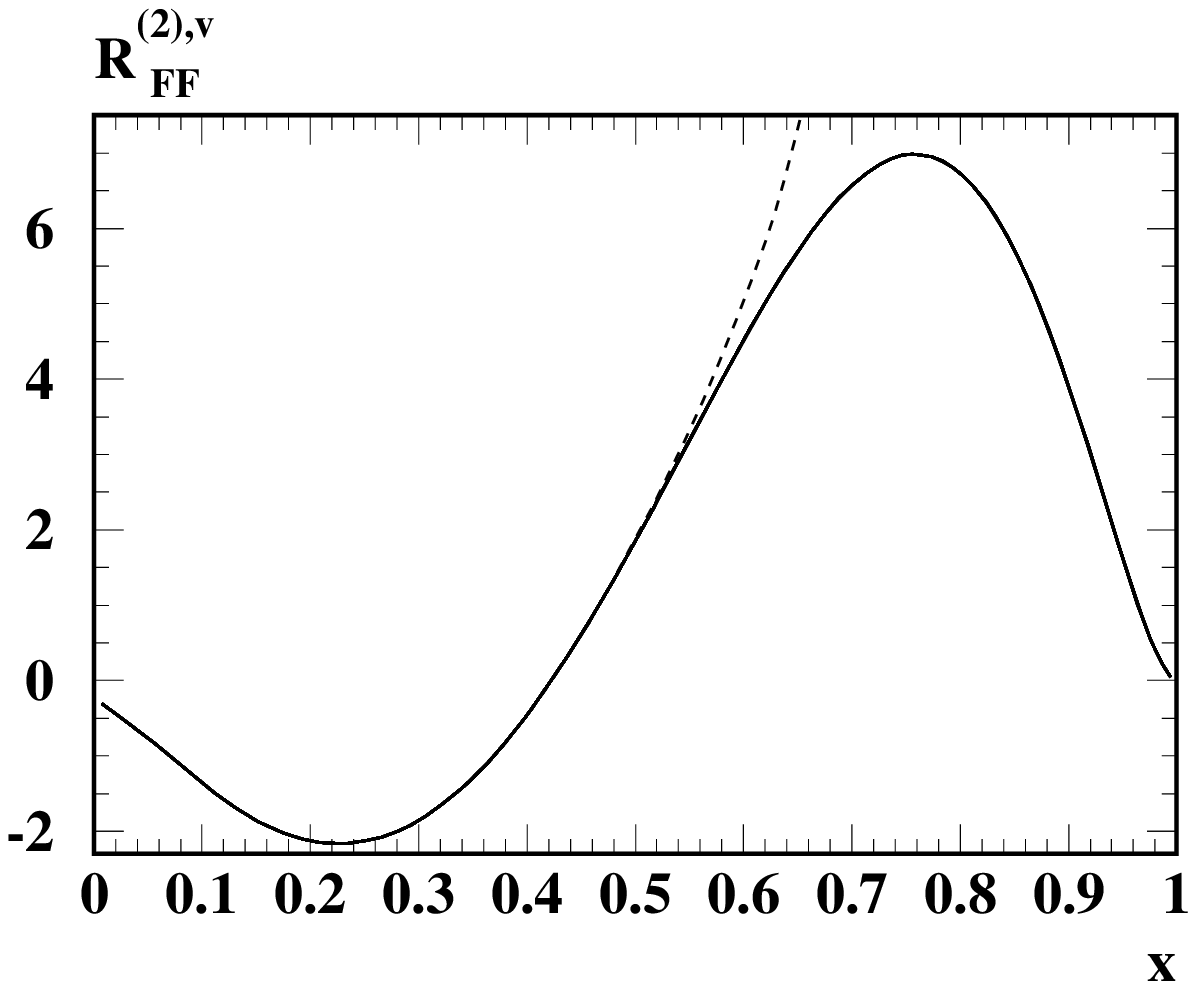}
      &
      \epsfxsize=7.cm
      \epsffile[110 280 460 560]{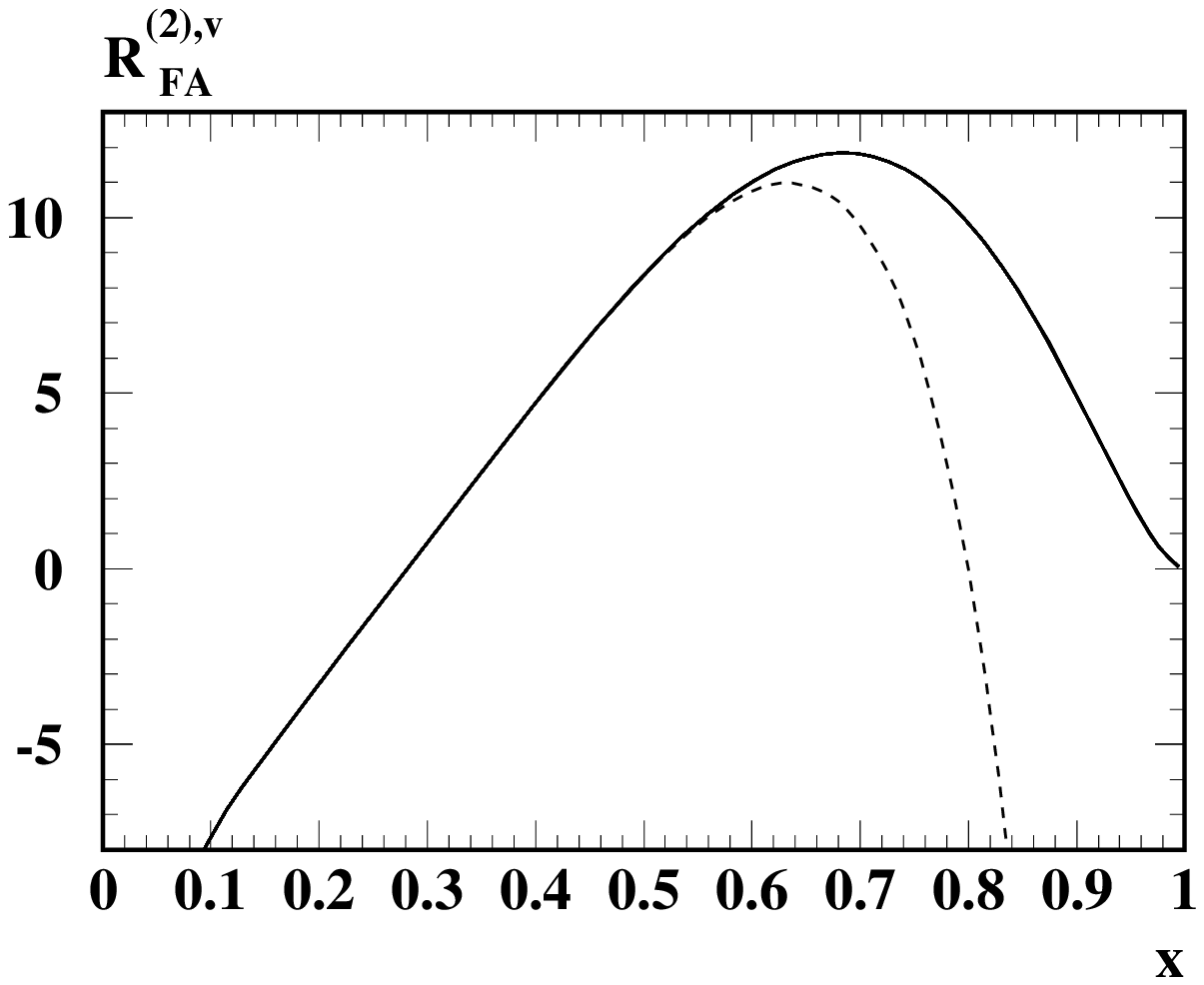}
      \\
      \epsfxsize=7.cm
      \epsffile[110 280 460 560]{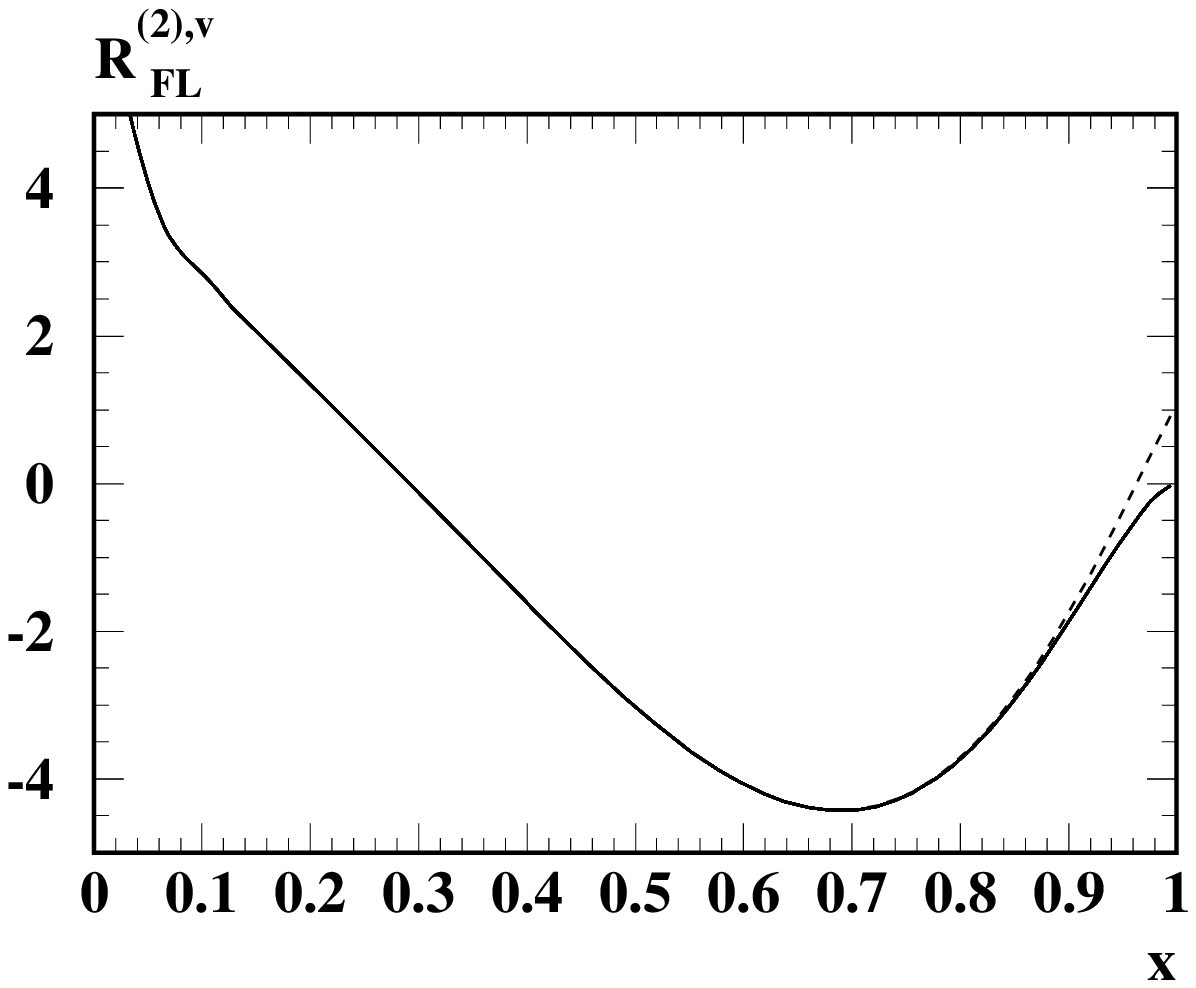}
      &
      \epsfxsize=7.cm
      \epsffile[110 280 460 560]{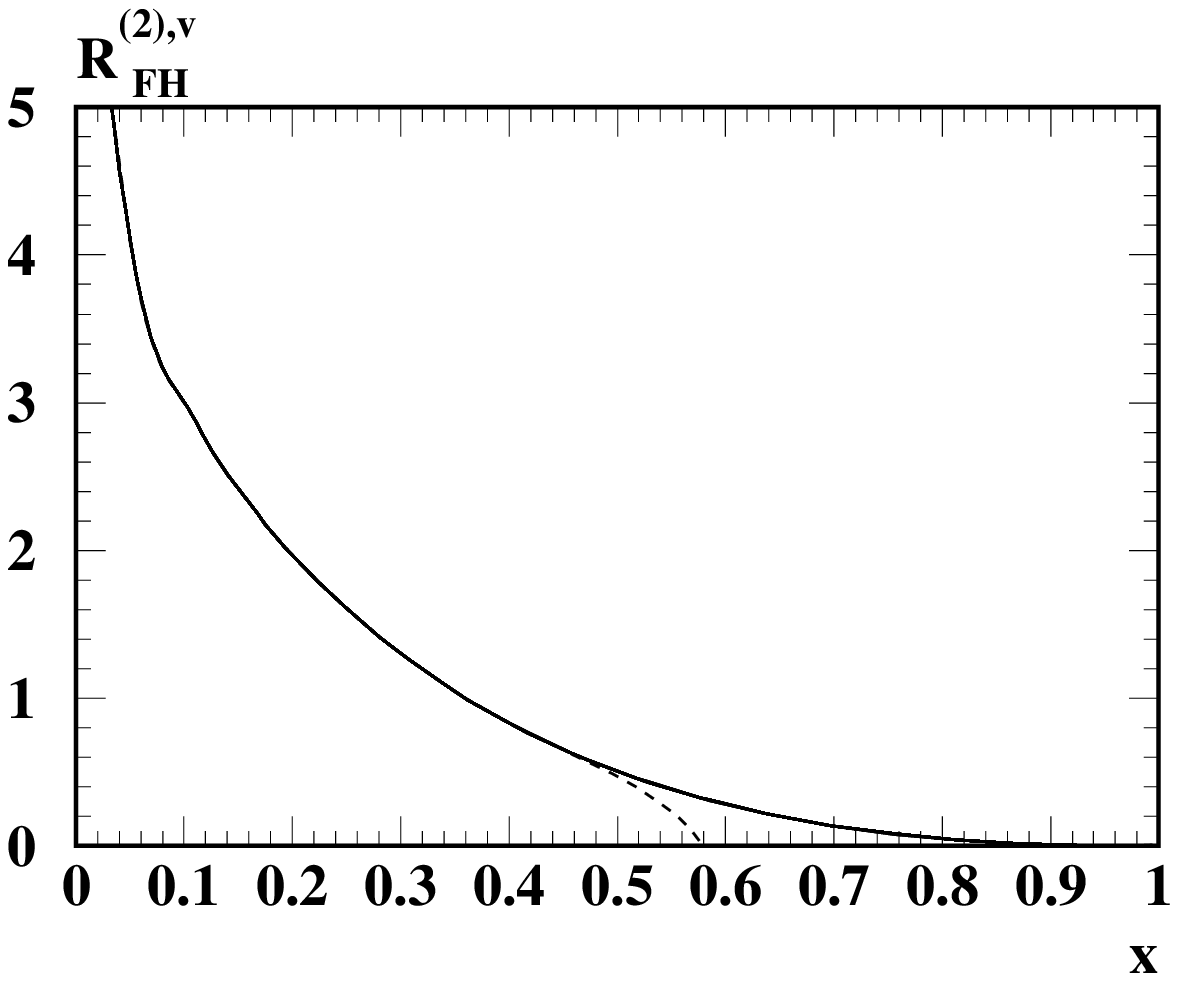}
    \end{tabular}
  \end{center}
  \vspace{-2.em}
  \caption{\label{fig:Rvx}$R^{(2),v}_{FF}(s)$, $R^{(2),v}_{FA}(s)$,
    $R^{(2),v}_{FL}(s)$  and $R^{(2),v}_{FF}(s)$ as a
    function of $x$.
          }
\end{figure}

\begin{figure}[t]
  \begin{center}
    \begin{tabular}{cc}
      \leavevmode
      \epsfxsize=7.cm
      \epsffile[110 280 460 560]{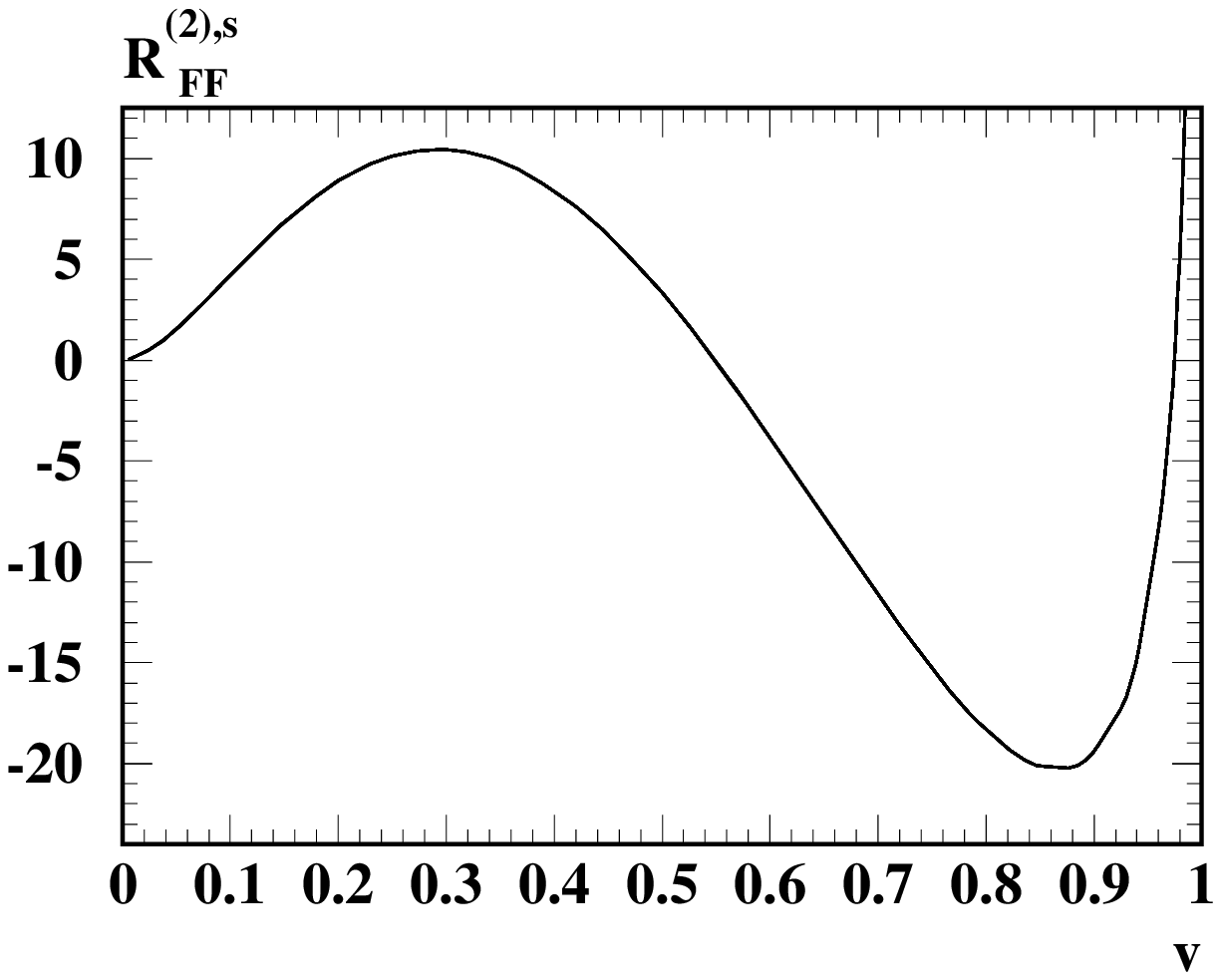}
      &
      \epsfxsize=7.cm
      \epsffile[110 280 460 560]{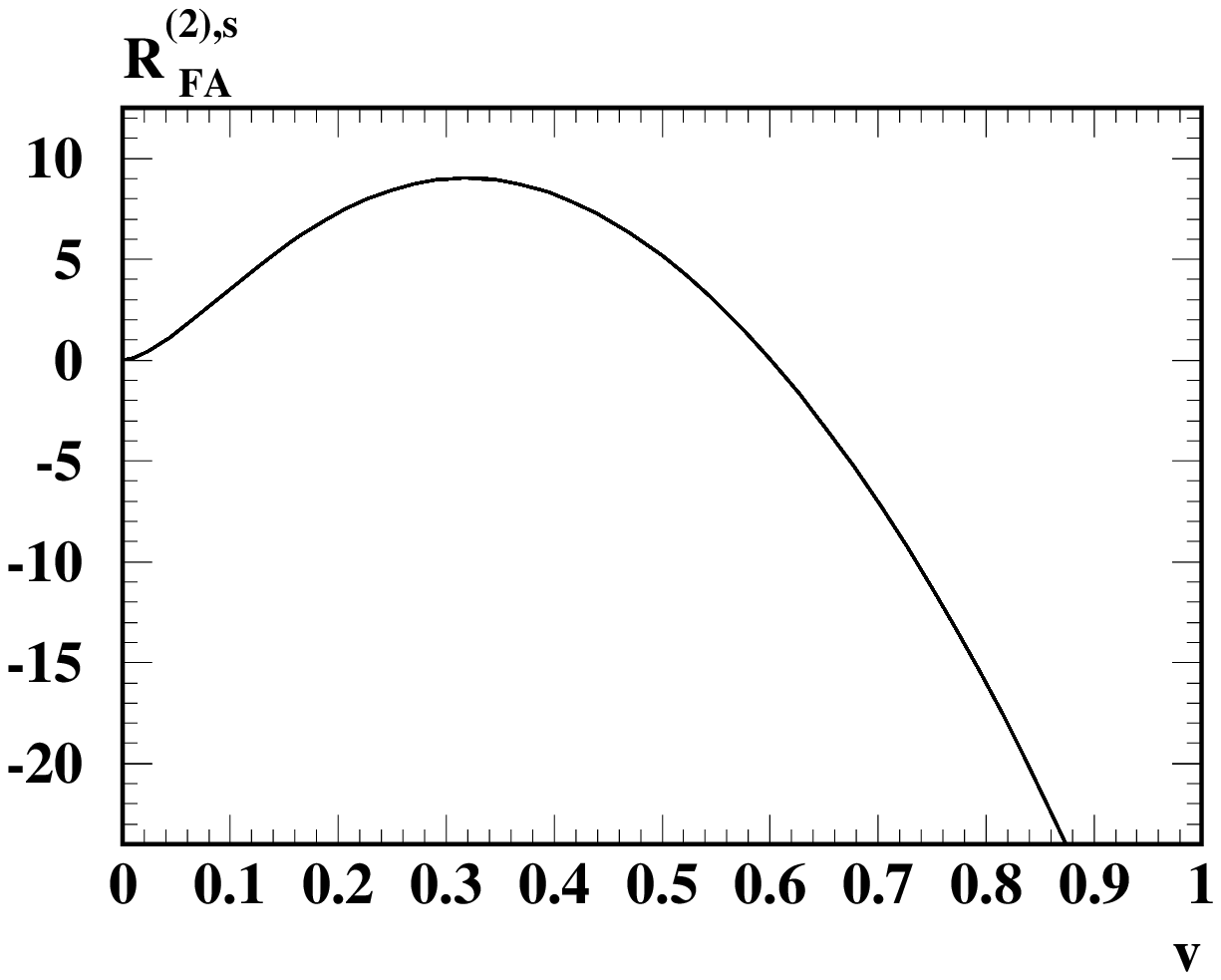}
      \\
      \epsfxsize=7.cm
      \epsffile[110 280 460 560]{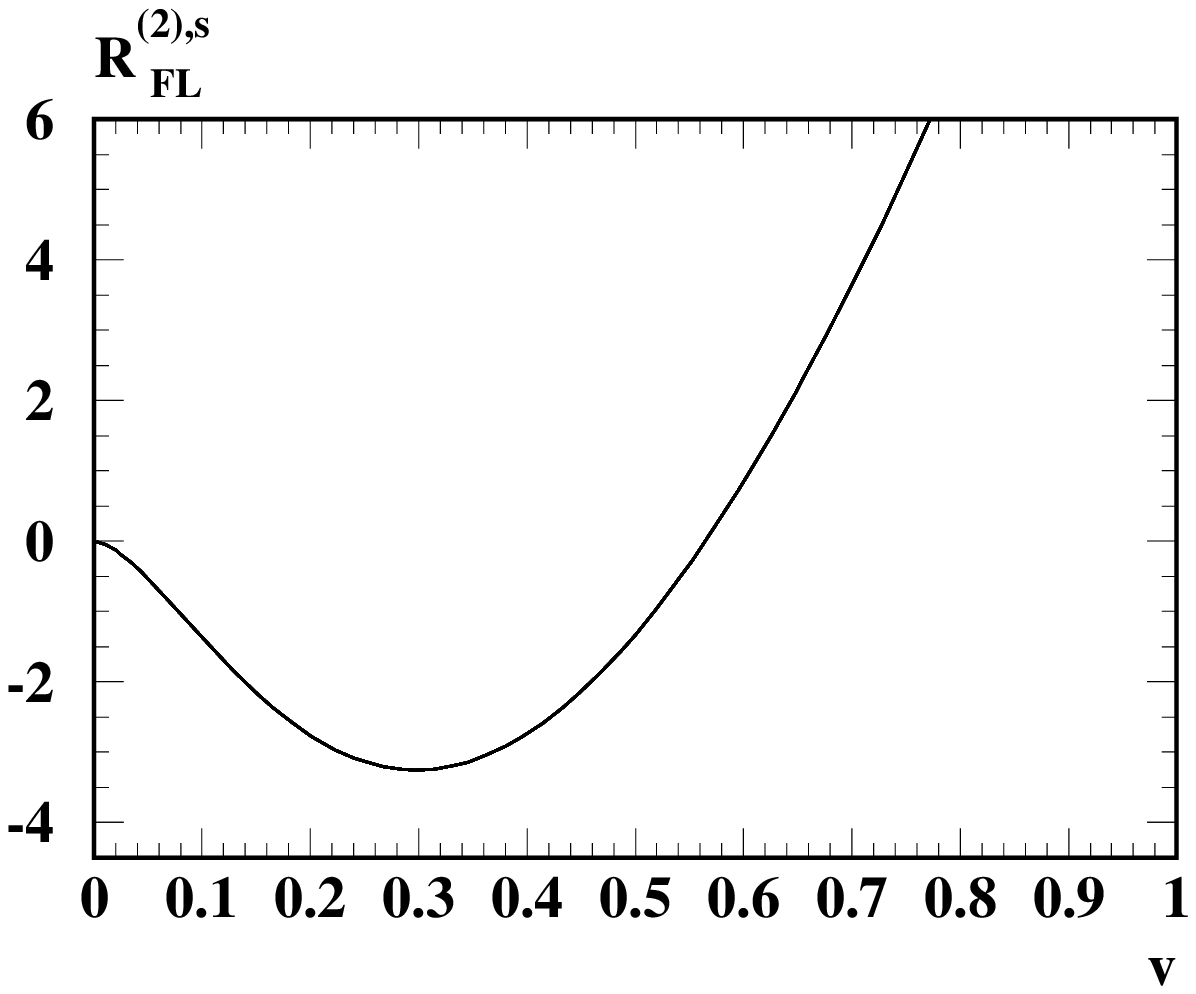}
      &
      \epsfxsize=7.cm
      \epsffile[110 280 460 560]{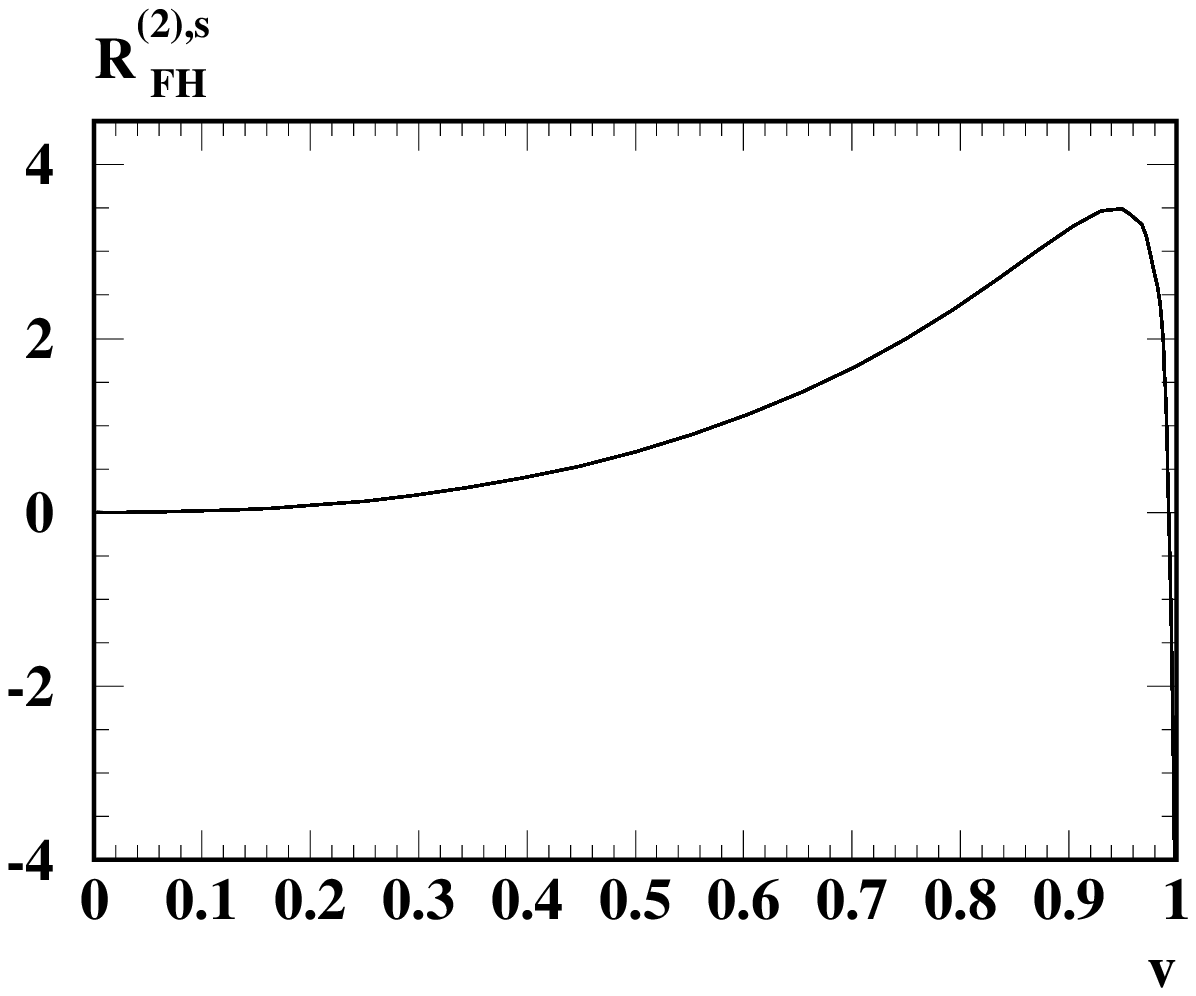}
    \end{tabular}
  \end{center}
  \vspace{-2.em}
  \caption{\label{fig:Rsv}$R^{(2),s}_{FF}(s)$, $R^{(2),s}_{FA}(s)$,
    $R^{(2),s}_{FL}(s)$  and $R^{(2),s}_{FH}(s)$ as a
    function of $v$.
          }
\end{figure}

\begin{figure}[t]
  \begin{center}
    \begin{tabular}{cc}
      \leavevmode
      \epsfxsize=7.cm
      \epsffile[110 280 460 560]{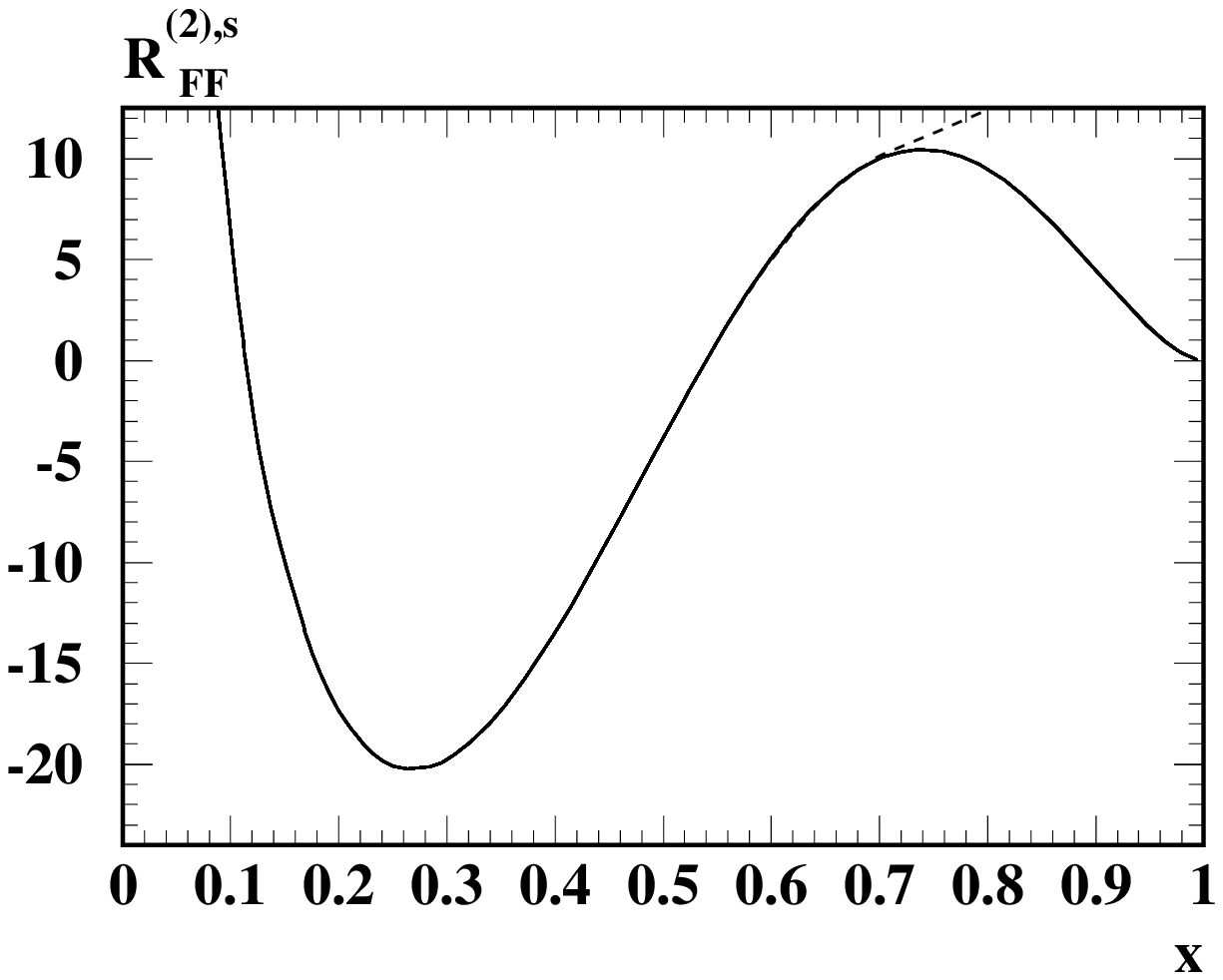}
      &
      \epsfxsize=7.cm
      \epsffile[110 280 460 560]{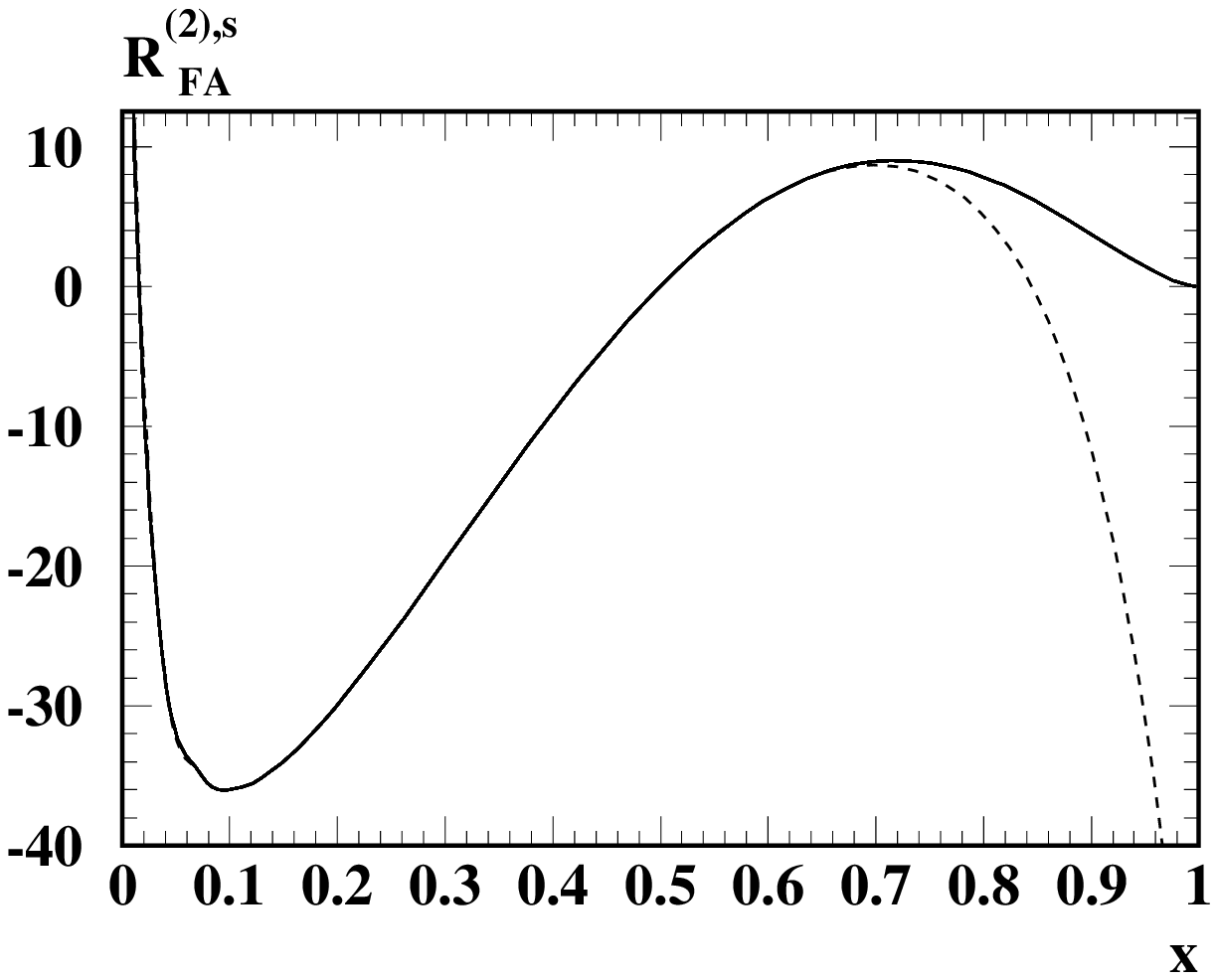}
      \\
      \epsfxsize=7.cm
      \epsffile[110 280 460 560]{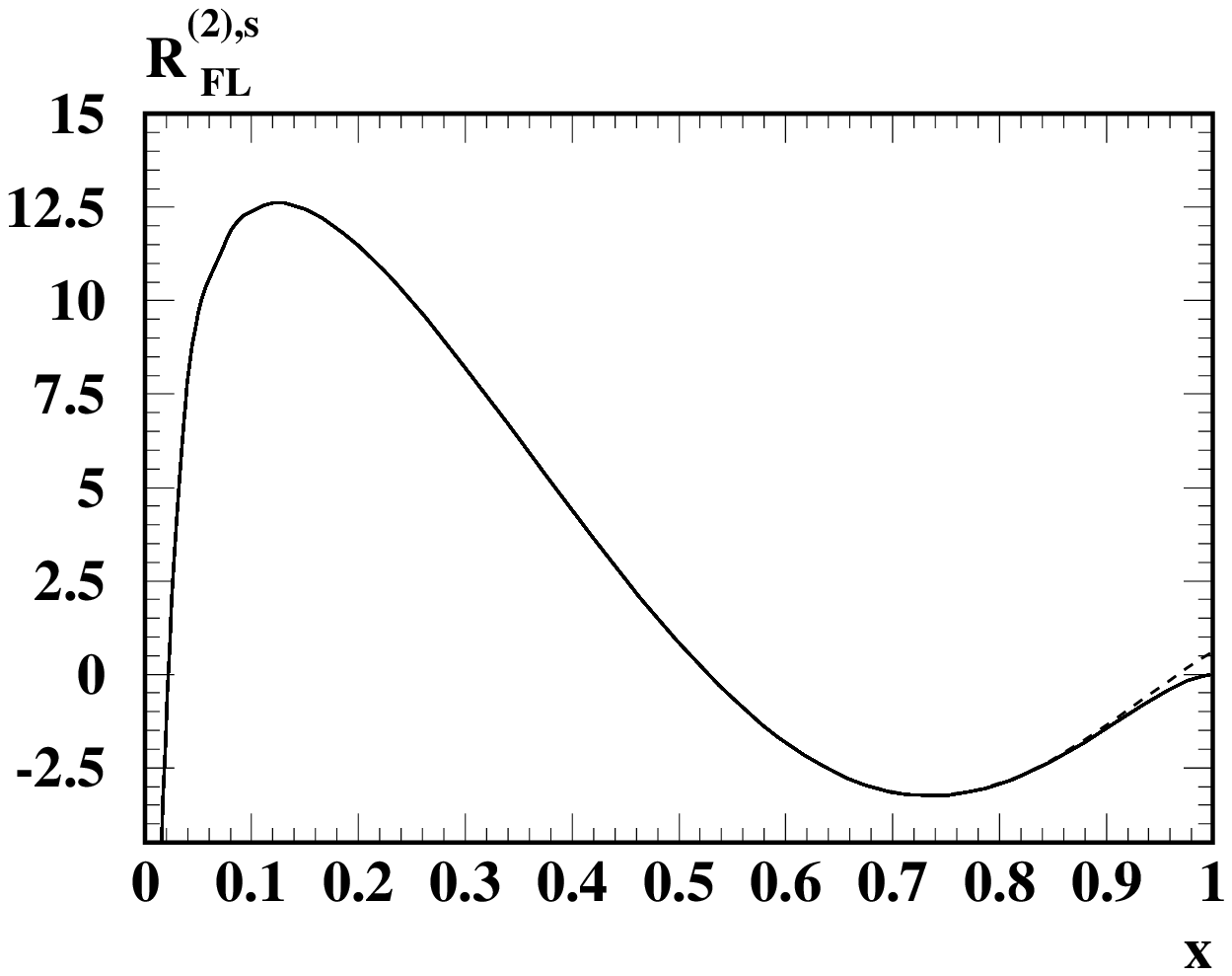}
      &
      \epsfxsize=7.cm
      \epsffile[110 280 460 560]{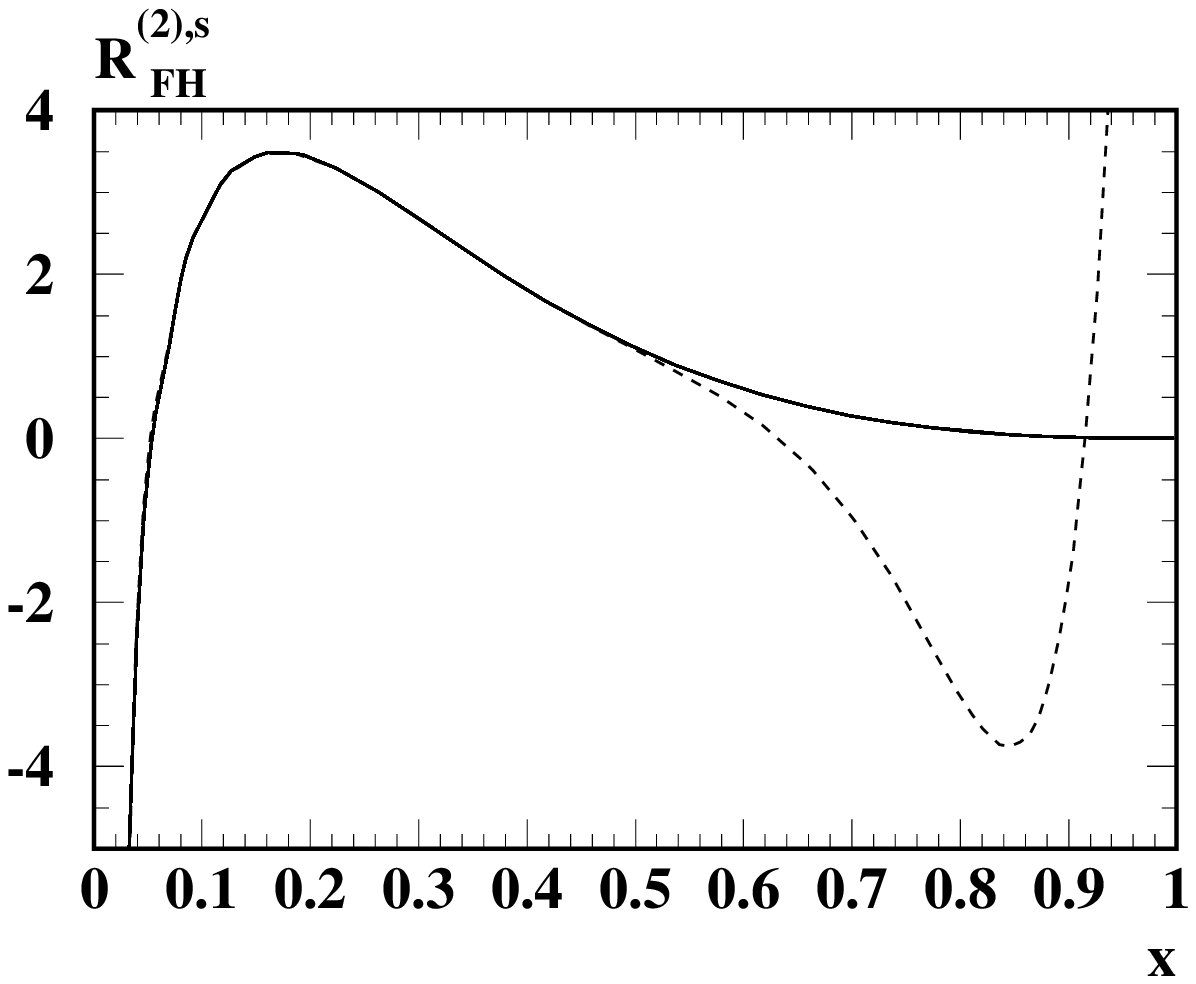}
    \end{tabular}
  \end{center}
  \vspace{-2.em}
  \caption{\label{fig:Rsx}$R^{(2),s}_{FF}(s)$, $R^{(2),s}_{FA}(s)$,
    $R^{(2),s}_{FL}(s)$  and $R^{(2),s}_{FH}(s)$ as a
    function of $x$.
          }
\end{figure}

The Pad\'e results for the individual 
colour structures at order $\alpha_s^2$
are plotted in Figs.~\ref{fig:Rvv},~\ref{fig:Rvx},~\ref{fig:Rsv}
and~\ref{fig:Rsx}.
Again, despite the fact that in each plot approximately 15 curves are
shown no difference between them can be observed.
In those plots where $x$ is chosen for the abscissa also the result of
the high-energy expansion containing terms up to order $1/z^7$ 
is plotted as a dashed line. There is excellent agreement with the 
semi-numerical result at least up to $x\approx0.5$, which corresponds
to $v\approx0.60$. In some cases even an agreement up to
$x\approx0.7$ is observed. This corresponds to $v\approx0.34$ which is
already quite close to the threshold.

A remark concerning the plots for $R_{FA}^{(2),s}$ and $R_{L}^{(2),s}$
are in order. In Fig.~\ref{fig:Rsx} it can be seen that for $x\to0$
they tend to $+\infty$ and $-\infty$, respectively.
However, the turn-over takes place in a small region of $x$ which is
beyond the resolution in Fig.~\ref{fig:Rsv} 
where the variable $v$ is used and thus the high-energy
region, i.e. the region for $v\to1$, gets squeezed.

The analytical formulae which result from the semi-numerical Pad\'e
procedure are quite long. Thus we refrain from listing them
explicitly. Instead, a typical representative for the two-loop results
and for each colour structure at three-loops can be found under the URL
\verb|http://www-ttp.physik.uni-karlsruhe.de/Progdata/ttp00/ttp00-25|.


\section{\label{sec:reseff}Spectral function in HQET}

In this Section we concentrate on the behaviour close to
the threshold. In particular we want to extract the leading
non-logarithmic term which is of order $v^2$ and afterwards
transform the result to HQET.

The expressions in Eq.~(\ref{eq:Pithr}) are constructed in such a way
that the combination 
$\mbox{Im}[\Pi(q^2)-\Pi^{thr}_{log}(q^2)]/v^2$ 
approaches a constant for $v\to0$.
However, in the method we use for the computation of $\Pi(q^2)$ it is not
possible to incorporate the constraint that $\mbox{Im}[\Pi(q^2)] \sim v^2$ in
analytical form. Our procedure provides rather a function which
numerically imitates the quadratic dependence. As a consequence 
the expression $\mbox{Im}[\Pi(q^2)-\Pi^{thr}_{log}(q^2)]/v^2$ diverges
for very small velocities and provides quite different numerical
values for the different Pad\'e approximants.
Nevertheless one could try to extract numerical approximations for the
non-logarithmic coefficient.

For this aim we consider the quantity
\begin{eqnarray}
  T^{(i),\delta}_X(v) &=&  \left[
  n_\delta\mbox{Im}[\Pi_X^{(i),\delta}(s+i\epsilon)]
  -R^{(i),\delta,thr}_{X,log}(s)
  \right]\Bigg|_{s=M^2(1+v)/(1-v)}
  \nonumber\\
  &=& N_c\, c^\delta_X v^2 + {\cal O}\left(v^3\right)
  \,,
  \label{eq:T}
\end{eqnarray}
where $n_v=12\pi$, $n_s=8\pi$ and $X\in\{F,FF,FA,FL,FH\}$. 
$R^{(i),\delta,thr}_{X,log}$ is obtained 
from Eqs.~(\ref{eq:Rvthr}) and~(\ref{eq:Rsthr}) 
by setting the unknown constants
$c^\delta_{X}$ to zero. 
We want to use $T^{(i),\delta}_X(v)$ in order to fit the
coefficient $c^\delta_X$ in Eq.~(\ref{eq:T}) 
for each individual Pad\'e approximant.
This requires small values of $v$.
However, due the very structure of the Pad\'e result the
ratio $T^{(i),\delta}_X(v)/v^2$ becomes unstable for $v\to0$.
Thus, in a first step we determine a minimal value $v_0$
for which $T^{(i),\delta}_X(v)/v^2$ evaluated for all Pad\'e results 
still show reasonable agreement. Furthermore we determine an
effective coefficient from the equation
\begin{eqnarray}
  T^{(i),\delta}_X(v_0) &=& N_c\, c^\delta_{X,{\rm eff}}\, v_0^2
  \,,
\end{eqnarray}
and require that the error on $c^\delta_{X,{\rm eff}}$ from the different
Pad\'e approximants is less than 10\%.
The results for $v_0$ and $c^\delta_{X,{\rm eff}}$ for each colour factor
both for the vector and scalar current can be found in Tab.~\ref{tab:c}.

\begin{table}[t]
  \begin{center}
    \begin{tabular}{|ll|r|r|r|r|r|r|}
    \hline
      $X$  & $\delta$ & $v_0$ & $c^\delta_{X, {\rm eff}}$ & 
                     $v_1$ & $c^\delta_X$ & $d^\delta_X$ & $\tilde{c}_X$
    \\
    \hline
        F & $v$ & $0.01$ & $  24.4(   0.1)$ & $0.02$ &
         $  26.3(   0.2)$ & $ -203.(    9.)$ & $   6.4$\\
        F & $s$ & $0.01$ & $  20.9(   0.1)$ & $0.02$ &
         $  21.8(   0.1)$ & $  -97.(    7.)$ & $   6.5$\\
        FF & $v$ & $0.03$ & $  -9.8(   0.9)$ & $0.05$ &
         $  12.1(   1.8)$ & $ -755.(   41.)$ & $  17.4$\\
        FF & $s$ & $0.02$ & $  47.9(   1.9)$ & $0.04$ &
         $  58.8(   2.9)$ & $ -567.(   70.)$ & $  20.7$\\
        FA & $v$ & $0.03$ & $ -30.5(   0.5)$ & $0.06$ &
         $ -18.2(   1.0)$ & $ -459.(   14.)$ & $   1.2$\\
        FA & $s$ & $0.045$ & $ -11.6(   0.4)$ & $0.09$ &
         $  -3.4(   0.6)$ & $ -201.(    7.)$ & $   1.3$\\
        FL & $v$ & $0.01$ & $   5.6(   0.1)$ & $0.02$ &
         $   2.0(   0.2)$ & $  380.(   11.)$ & $  -2.0$\\
        FL & $s$ & $0.01$ & $  -1.9(   0.1)$ & $0.02$ &
         $  -3.8(   0.1)$ & $  202.(    6.)$ & $  -2.3$\\
    \hline
    \end{tabular}
    \caption{\label{tab:c}Results from the fits 
    for the coefficients as described in the text
    separated according to the colour factors and the tensorial
    structure of the current correlator. The errors indicated in the
    brackets arise from the comparison of the different Pad\'e
    approximants. They are omitted in the case of $\tilde{c}_X$ as they
    are much smaller than 
    the systematic error of the extraction procedure.}
  \end{center}
\end{table}

In order to account also for higher order terms we choose in a next
step a value $v_1>v_0$ in such a way that,
first, $v_1-v_0 \approx v_0$ and, second, $T^{(i),\delta}_X(v)/v^2$ 
displays an approximately linear behaviour for $v_0\le v\le v_1$.
Then a two-parametric fit is performed in the interval $[v_0,v_1]$,
i.e. the coefficients $c^\delta_X$ and $d^\delta_X$ are determined from
the equation
\begin{eqnarray}
  T^{(i),\delta}_X(v) &=& N_c\,v^2\left(c_X^\delta + d_X^\delta v\right)
  \,.
\end{eqnarray}
Finally Eqs.~(\ref{eq:cctilv}) and~(\ref{eq:cctils}) are used to
convert the result for $c^\delta_X$ to a numerical estimate for
$\tilde{c}_X$. 
The corresponding results for the individual structures can again be
found in Tab.~\ref{tab:c}.
The corresponding errors are omitted in the case of $\tilde{c}_X$ 
since they are much smaller than 
the systematic error of the extraction procedure.

In order to get confidence in our 
prescription for the determination of $c^\delta_X$
let us have a look at the two-loop results. 
The values given in Tab.~\ref{tab:c} can be compared with 
the exact results which read
\begin{eqnarray}
  c^v_F &=& 27.00\,,
  \nonumber\\
  c^s_F &=& 22.00\,,
  \nonumber\\
  \tilde{c}_F &=& 6.50\,.
\end{eqnarray}
One observes a very good agreement of the approximated results for
both $\tilde{c}_F$ and $c_F^\delta$ with the exact 
ones\footnote{This is also the case for $v_0=0.03$ and $v_1=0.06$.}.
Although the error induced by the comparison of
the different Pad\'e approximations is slightly smaller than the
deviation from the exact result 
the considerations at order $\alpha_s$ are
quite promising for the determination of the coefficients at order 
$\alpha_s^2$.

As was already mentioned above the coefficients $c_{FH}^\delta$ are
completely determined. Thus, also they can be used in order to test
our prescription. However, both in the vector and scalar case
$c_{FH}^\delta$ is close to zero which, of course, leads to large
relative errors.
Nevertheless, let us present the results which 
read for $v_0=0.05$ and $v_1=0.1$\footnote{As can be seen in
  Figs.~\ref{fig:Rvv} and~\ref{fig:Rsv} 
  both $R_{FH}^{(2),v}$ and $R_{FH}^{(2),s}$
  are rather smooth around $v=0$ which allows for larger values of 
  $v_0$ and $v_1$.}
\begin{eqnarray}
  c_{FH,{\rm eff}}^v &=&  0.20(6)\,,
  \nonumber\\
  c_{FH}^v &=&  0.16(10)\,,
  \nonumber\\
  c_{FH,{\rm eff}}^s &=&  0.50(6)\,,
  \nonumber\\
  c_{FH}^s &=&  0.44(11)\,.
\end{eqnarray}
The comparison with the exact values obtained from 
Eqs.~(\ref{eq:Rvthr}) and~(\ref{eq:Rsthr})
\begin{eqnarray}
  c_{FH}^v &=&  0.175\ldots\,,
  \nonumber\\
  c_{FH}^s &=&  0.455\ldots\,,
\end{eqnarray}
shows that the agreement of the central values is within 10\%.

In order to obtain our final predictions for the coefficients
$\tilde{c}_X$ we proceed as follows:
the coefficients $\tilde{c}_X$ given in Tab.~\ref{tab:c} 
have to be independent from the tensorial structure of the current 
correlator from which they are determined.
On the other hand, guided by the numbers shown in Tab.~\ref{tab:c},
we can define criterions concerning the stability and reliability 
of the extraction procedure.
As a first criterion we require that $v_0$ should be as small as
possible. This clearly suppresses higher order terms in $v$.
Furthermore we consider the ratio $d^\delta_X/c^\delta_X$
and regard that colour structure as more reliable where the ratio 
is smallest.
As $d^\delta_X$ represents an effective constant containing the effects of
the higher orders in $v$ also this criterion selects the structure
where they are suppressed as compared to the $v^2$ terms.
Following these rules we choose the vector correlator in order to
obtain $\tilde{c}_{FA}$ and the scalar one for $\tilde{c}_{FF}$
and $\tilde{c}_{FL}$.
This leads to the following results
\begin{eqnarray}
  \tilde{c}_{FF} &=& 21(6)
  \,,
  \nonumber\\
  \tilde{c}_{FA} &=& 1.2(4)
  \,,
  \nonumber\\
  \tilde{c}_{FL} &=& -2.3(7)
  \,,
  \label{eq:ctil}
\end{eqnarray}
where we assigned a conservative error of 30\%.
Note that all numbers given in the last column of Tab.~\ref{tab:c} are
consistent with our final predictions of Eqs.~(\ref{eq:ctil}).

\begin{table}[t]
  \begin{center}
    \begin{tabular}{|ll|r|r|r|r|r|r|}
    \hline
       $n_l$  & $\delta$ & $v_0$ & $c^\delta_{n_l,{\rm eff}}$ & 
              $v_1$ & $c^\delta_{n_l}$ & $d^\delta_{n_l}$ & $\tilde{c}_{n_l}$
    \\
    \hline
        0 & $v$ & $0.03$ & $-143.0(   3.7)$ & $0.05$ &
         $ -46.9(   7.5)$ & $-3324.(  142.)$ & $  36.5$\\
        0 & $s$ & $0.03$ & $  46.3(   2.1)$ & $0.05$ &
         $ 106.8(   4.4)$ & $-2069.(   92.)$ & $  46.2$\\
        1 & $v$ & $0.03$ & $-134.7(   3.5)$ & $0.05$ &
         $ -42.5(   7.1)$ & $-3189.(  136.)$ & $  35.7$\\
        1 & $s$ & $0.03$ & $  47.5(   2.1)$ & $0.05$ &
         $ 105.6(   4.3)$ & $-1984.(   90.)$ & $  45.0$\\
        2 & $v$ & $0.03$ & $-126.4(   3.3)$ & $0.05$ &
         $ -38.1(   6.7)$ & $-3055.(  129.)$ & $  34.8$\\
        2 & $s$ & $0.03$ & $  48.8(   2.0)$ & $0.05$ &
         $ 104.4(   4.2)$ & $-1898.(   88.)$ & $  43.9$\\
        3 & $v$ & $0.03$ & $-118.0(   3.1)$ & $0.05$ &
         $ -33.6(   6.3)$ & $-2921.(  121.)$ & $  34.0$\\
        3 & $s$ & $0.03$ & $  50.1(   1.9)$ & $0.05$ &
         $ 103.2(   4.1)$ & $-1814.(   84.)$ & $  42.7$\\
        4 & $v$ & $0.03$ & $-109.6(   2.7)$ & $0.05$ &
         $ -28.9(   5.6)$ & $-2793.(  108.)$ & $  33.2$\\
        4 & $s$ & $0.03$ & $  51.4(   1.9)$ & $0.05$ &
         $ 102.0(   3.9)$ & $-1728.(   81.)$ & $  41.5$\\
        5 & $v$ & $0.03$ & $-101.4(   2.7)$ & $0.05$ &
         $ -24.8(   5.5)$ & $-2651.(  108.)$ & $  32.3$\\
        5 & $s$ & $0.03$ & $  52.6(   1.8)$ & $0.05$ &
         $ 100.7(   3.8)$ & $-1642.(   79.)$ & $  40.3$\\
    \hline
    \end{tabular}
    \caption{\label{tab:nl}Results from the fits 
    for the coefficients as described in the text
    separated for different values of $n_l$. 
    As far as the errors are concerned the same statements hold as in
    Tab.~\ref{tab:c}.}
  \end{center}
\end{table}

In practical applications it is often sufficient to know the result at
order $\alpha_s^2$ for the numerical value $N_c=3$ which 
motivates the following procedure:
in a first step the moments of the colour structures $FF$, $FA$
and $FL$ are added for fixed $n_l$ and the Pad\'e procedure is performed.
Afterwards the prescription described around Eq.~(\ref{eq:T}) is
applied.
The results of the corresponding analysis can be found in
Tab.~\ref{tab:nl} which allow to deduce 
the following compact expression for
$\tilde{c}_{n_l}$ which governs the dependence on $n_l$:
\begin{eqnarray}
  \tilde{c}_{n_l} &=& 46(15) - 1.2(4) \, n_l
  \,.
  \label{eq:ctilnl}
\end{eqnarray}
At first sight the errors appear quite large. However, one has to
recall that only the large- and small-$q^2$ behaviour of the
polarization function serves as input whereas the quantity 
in Eq.~(\ref{eq:ctilnl}) corresponds to the imaginary part at $q^2=M^2$.
Note also that at order $\alpha_s^2$ the numerical values for
$v_0$ have to be chosen
larger than at order $\alpha_s$ which increases the influence
of the higher order terms in $v$.

To summarize: the complicated  threshold structure of the 
Pad\'e approximants 
allows only a rather rough determination of the constants $c^v$ and $c^s$.
However, the peculiar structure of Eqs.~(\ref{eq:cctilv})
and~(\ref{eq:cctils}) connecting 
$c^\delta$ to $\tilde{c}$ makes it possible to extract the 
universal coefficient  $\tilde{c}$ with a reasonable accuracy.

Finally we are in the position to write down the expression for 
$\tilde{R}^\prime(\omega)$ up to order $\alpha_s^2$. Inserting the colour
factors into Eq.~(\ref{eq:Rtil}) and using the results from
Eq.~(\ref{eq:ctilnl}) one obtains
\begin{eqnarray}
  \tilde{R}^\prime(\omega) &=& N_c \omega^2 
    \Bigg[ 1 + \frac{\alpha_s^{(n_l)}(\mu)}{\pi}\left(8.667 +
               \Lw\right)
             + \left(\frac{\alpha_s^{(n_l)}(\mu)}{\pi}\right)^2
               \left(
                 46(15) + 35.54 \Lw 
\right.\nonumber\\&&\left.\mbox{}
                 + 1.875 \Lw^2
                 +n_l\left(-1.2(4)  - 1.583 \Lw - 0.08333
                           \Lw^2\right)
               \right)     
    \Bigg]
  \,.
  \label{eq:rtilfin}
\end{eqnarray}
We can conclude that the 
coefficient at order $\alpha_s^2$ is quite large and has a mild
dependence on $n_l$.


\section{\label{sec:con}Applications and conclusions}

\begin{figure}[t]
  \begin{center}
    \begin{tabular}{c}
      \leavevmode
      \epsfxsize=14.cm
      \epsffile[70 580 516 720]{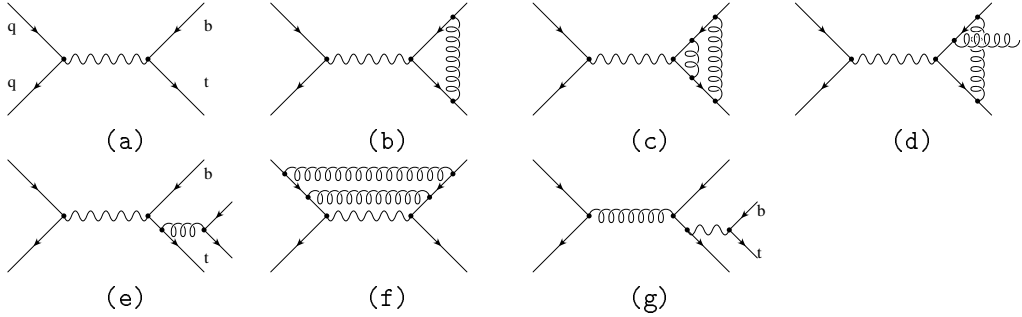}
    \end{tabular}
  \end{center}
  \vspace{-2.em}
  \caption{\label{fig:diags_qqtb}Sample diagrams contributing to the
    process $q\bar{q}\to t\bar{b}$. The wavy and loopy lines 
    represent $W$ bosons and gluons, respectively.
          }
\end{figure}

As an application of the vector and axial-vector current correlator
we want to discuss the single-top-quark production
via the process $q\bar{q}\to t\bar{b}$. Some sample diagrams
contributing to this process are shown in Fig.~\ref{fig:diags_qqtb}.

The corrections of order $\alpha_s$ to the (total) single-top-quark 
production rate are quite large. They amount to
about 54\% and 50\% for Tevatron and LHC energies,
respectively~\cite{SmithWillen96}, where 18\%, respectively, 17\% arise
from the final state corrections.
This makes it necessary to consider also the corrections of order 
$\alpha_s^2$. Due to the appearance of an interference between the
initial and final state (c.f. Fig.~\ref{fig:diags_qqtb}(f)), 
which for the first time happens at order
$\alpha_s^2$, the complete calculation is a non-trivial task.
However, with the results of this paper we are in the position to
perform a first step and consider the leading
term in the large-$N_c$ expansion.

One observes that the contributions of the diagrams where gluons
connect the initial and final states are
suppressed by at least a factor $1/N_c^2$ in the large $N_c$ limit
as compared to the diagrams in Fig.~\ref{fig:diags_qqtb}(c)--(e).
For the latter, together with the contributions of
Fig.~\ref{fig:diags_qqtb}(a) and~(b),
the differential cross section can be written in factorized form
\begin{eqnarray}
  \frac{{\rm d}\sigma}{{\rm d}q^2} (p\bar{p} \to t\bar{b}+X)
  &=& \sigma(p \bar{p} \to W^* +X) 
  \frac{\mathrm{Im}\, \Pi_W(q^2,M_t^2,M_b^2)}{\pi (q^2-M_W^2)^2}
  \label{eq:diff_cross}
  \,,
\end{eqnarray}
where $\Pi_W$ corresponds to the transversal part of the $W$ boson
self energy which is connected to the vector correlator of
Eq.~(\ref{eq:pivadef}) through
\begin{eqnarray}
  \Pi_W(q^2) &=& \sqrt{2} G_F M_W^2 |V_{tb}|^2 q^2 \Pi^v(q^2)
  \,.
  \label{eq:piw}
\end{eqnarray}

At order $\alpha_s^2$ there are also diagrams like the one in
Fig.~\ref{fig:diags_qqtb}(g) which appear for the first time.
In principle they also lead to the same final state as 
the diagram in Fig.~\ref{fig:diags_qqtb}(e). However, one 
has to note that the $W$ boson generating the top and bottom quark
is radiated from a light quark flavour. This suggests that their
contribution is small although there is only a suppression
by a factor $1/N_c$ as compared to Fig.~\ref{fig:diags_qqtb}(c)
and not by $1/N_c^2$ like for the diagram in Fig.~\ref{fig:diags_qqtb}(f).

Thus, if we restrict ourselves to the leading term in $1/N_c$
it is possible to use the results for $R^v$ 
obtained above in combination
with Eq.~(\ref{eq:diff_cross}) to perform a theoretical analysis
at order $\alpha_s^2$ to the single-top-quark production in the
large-$N_c$ limit. In order to obtain the total cross section the 
corresponding parton distribution functions would be needed to the
same order.

\begin{figure}[t]
  \begin{center}
    \begin{tabular}{c}
      \leavevmode
      \epsfxsize=14.cm
      \epsffile[110 280 470 560]{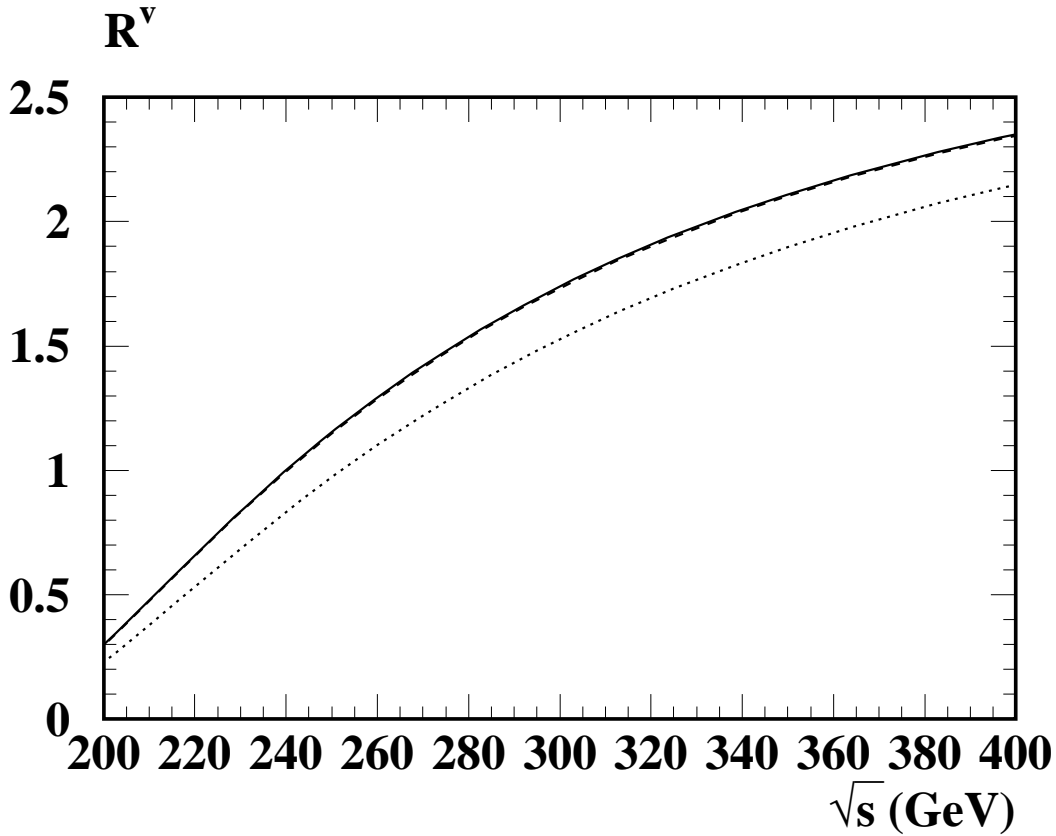}
    \end{tabular}
  \end{center}  
  \vspace{-2.em}
  \caption{\label{fig:dsigma_qqtb}LO (dotted), 
    NLO (dashed) and NNLO (solid) results of $R^v(s)$.
          }
\end{figure}

The production cross section of the virtual $W^*$ boson is identical to
that of the Drell-Yan process $q\bar{q}\to e\bar{\nu}_e$.
The latter is known to
${\cal{O}}(\alpha^2_s)$ from Ref.~\cite{Drell_Yan}.
Thus we can take the proper ratios to make predictions 
in the large-$N_c$ limit 
at NNLO free from any  dependence on parton distribution functions.  
As an example, we can consider the following expression
\begin{eqnarray}
  \frac{\frac{{\rm d}\sigma}{{\rm d}q^2}\left(pp\to W^*\to tb\right)}
       {\frac{{\rm d}\sigma}{{\rm d}q^2}\left(pp\to W^*\to
       e\nu_e\right)}
  &=& \frac{\mbox{Im}\left[\Pi_{tb}(q^2)\right]}
           {\mbox{Im}\left[\Pi_{e\nu}(q^2)\right]} 
  \nonumber\\
  &=& N_c |V_{tb}|^2 R^v(s)
  \,.
  \label{eq:ratio}
\end{eqnarray}
In Fig.~\ref{fig:dsigma_qqtb} the LO, NLO and NNLO results
of $R^v(s)$ are plotted in the range $\sqrt{s} = 200 \ldots 400$ GeV
where $M_t=175$~GeV and $\alpha_s(M_Z)=0.118$ has been chosen for the
numerical analysis.
Whereas the ${\cal O}(\alpha_s)$ corrections are significant
there is only a moderate contribution from the order $\alpha_s^2$
terms. In the range in $q^2$ shown in Fig.~\ref{fig:dsigma_qqtb} they are
below 1\% of the Born result.
Note that the NNLO correction to the Drell-Yan process are also 
small and amount to at most a few percent (see e.g.~\cite{Mar00}).
Thus, in case there is no kinematical magnification for the 
diagrams in Fig.~\ref{fig:diags_qqtb}(f) and~(g) we can conclude
that the radiative corrections to the single-top-quark production via the 
process $q\bar{q}\to t\bar{b}$ are well under control.

The scalar and pseudo-scalar correlator covers properties connected to
a charged Higgs boson.
The latter appear in theories beyond the SM which are usually 
characterized by an enlarged Higgs
sector containing Higgs bosons with different quantum numbers.
For example, one of the most appealing extensions of the SM,
the Minimal Supersymmetric Standard Model (MSSM), contains two complex
iso-doublets with opposite hyper-charge (see
e.g.~\cite{GunHabKanDaw90}), resulting in five mass eigenstates of
(pseudo-)scalar physical Higgs fields: 
two neutral CP-even ($H^0$ and $h^0$),
one neutral CP-odd ($A$)
and two charged ($H^{\pm}$) Higgs bosons.

Let us consider a generic charged Higgs boson coupled to 
fermions through
\begin{eqnarray}
  {\cal L}_{H^+D\bar{U}} &=&
  \left(\sqrt{2} G_F\right)^{1/2} \, H^+\, J_{H^+}
  \,,
\end{eqnarray}
where the corresponding quark current is given by
\begin{eqnarray}
  J_{H^+} &=& 
  \frac{m_U}{\sqrt{2}} \,  
  \bar{U} \left[ a \, (1-\gamma_5) + b \,  (1+\gamma_5) \right] D
  \label{eq:Higgs_current} 
  \,.
\end{eqnarray}
Here $U$ and $D$ represent generic up- and down-type quarks,
respectively, with $\overline{\rm MS}$ masses $m_U$ and $m_D=0$.
Eq.~(\ref{eq:Higgs_current}) only covers the contributions from 
a $H^+$ boson; the formulae for a Higgs boson with negative charge are
analogous. The parameters $a$ and $b$ are model dependent and are left
unspecified.

\begin{figure}[t]
  \begin{center}
    \begin{tabular}{c}
      \leavevmode
      \epsfxsize=14.cm
      \epsffile[110 280 470 560]{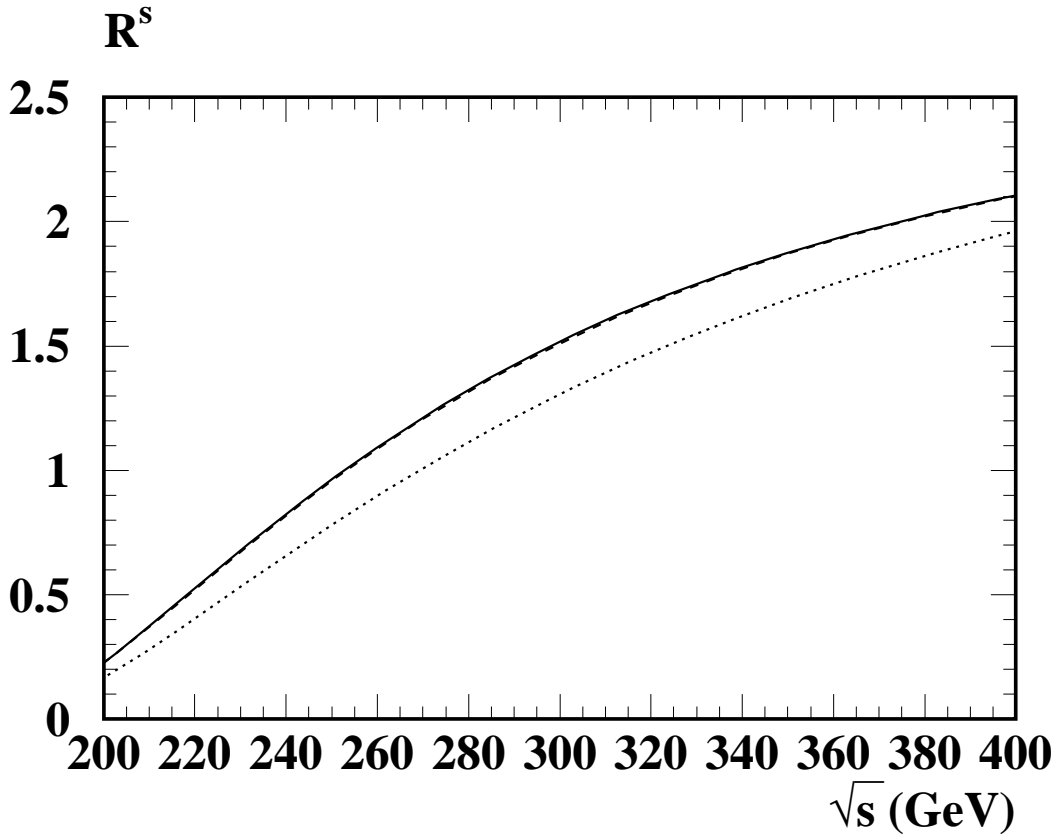}
    \end{tabular}
  \end{center}
  \vspace{-2.em}
  \caption{\label{fig:higgs}LO (dotted), 
    NLO (dashed) and NNLO (solid) results of $R^s(s)$, $M=M_t=175$~GeV.
          }
\end{figure}

The decay rate of the boson $H^+$ into quarks and gluons
can be written in the form 
\begin{eqnarray}
  \Gamma(H^+ \to U \bar{D})
  &=& 
  \sqrt{2} G_F M_{H^+} \, \mathrm{Im}\, \left[ \Pi_H(M_{H^+}^2) \right]
  \,,
\end{eqnarray}
where $M_U$ is the pole quark mass and $\Pi_H(q^2)$ is given by
\begin{eqnarray}
  q^2 \Pi_H(q^2) &=& \int {\rm d}x \, e^{iqx} 
  \langle     
  T J_H{^+}(x) J_{H^-}(0)
  \rangle
  \,\,=\,\,(a^2+b^2) q^2 \Pi^s(q^2) 
  \label{eq:Higgs_corr}
  \,,
\end{eqnarray}
Thus, we arrive at the following expression for
the hadronic decay rate of the charged Higgs boson
\begin{eqnarray}
  \Gamma(H^+ \to U \bar{D})
  & = &\frac{\sqrt{2}G_F}{8\pi}M_{H^+} (a^2+b^2) R^s(M_{H^+}^2)
  \,.
\end{eqnarray}
In Fig.~\ref{fig:higgs} $R^s(M_{H^+}^2)$ is plotted at LO, NLO and NNLO.
Again it turns out that the radiative corrections are well under
control as order $\alpha_s^2$ terms contribute at most of the order of 1\%.

As an application of the correlator in the effective theory
we want to mention the determination of the meson decay constants 
via QCD sum rules
where the Borel transform of
$\tilde{R}^\prime$ as 
given in Eq.~(\ref{eq:rtilfin}) enters as a building block.
Besides the perturbative part the sum rules also obtain contributions
from non-perturbative condensates which, however, are numerical less
important~\cite{BroGro92}.
The typical scale which has to be used in Eq.~(\ref{eq:rtilfin}) 
is of the order of 1~GeV~\cite{BroGro92,Bagan92} 
which leads to sizeable corrections both at
order $\alpha_s$ and at order $\alpha_s^2$. 

To be more precise let us choose $\alpha_s^{(5)}(M_Z)=0.118$ which,
using two-loop accuracy,
leads to $\alpha_s^{(4)}(1.3~\mbox{GeV})\approx0.37$~\cite{rundec}.
For $\omega=1.3$~GeV and $n_l=4$ the order $\alpha_s$ corrections in
Eq.~(\ref{eq:rtilfin}) amount to about 100\%.
The terms at ${\cal O}(\alpha_s^2)$ contribute with additional 60(20)\%
where the sign is the same as for the LO correction.
A careful analysis is
necessary in order to decide whether perturbation theory can safely be
applied in this case.

\vspace{1em}

To conclude, in this paper the non-diagonal current correlator formed
by a massive and massless quark has been considered. Moments in the
low and high energy region have been computed analytically in full
QCD. Furthermore the leading logarithmic contributions arising at the
quark threshold $q^2=M^2$ have been obtained from the reconstruction
of the logarithmic terms of the spectral function in the effective
theory. This information is combined with the help of conformal
mapping and Pad\'e approximation to obtain semi-numerical results for
the vector and scalar correlator and in particular their imaginary
parts valid for all values of $q$ and $M$.
In a next step the various Pad\'e results are used in order to obtain
the leading non-logarithmic coefficient at threshold which is
in turn transformed to the effective theory leading to a prediction
of the spectral function up to order $\alpha_s^2$ 
(see Eq.~(\ref{eq:rtilfin})).
As applications we considered the effect of the $\alpha_s^2$
correction to the single-top quark production and the decay of a
charged Higgs boson.


\section*{Acknowledgments}

The authors are grateful to A. Grozin, J.H. K\"uhn and V.A. Smirnov 
for useful discussions and advice.  
This work was supported in part by the {\it DFG-Forschergruppe
``Quantenfeldtheorie, Computeralgebra und Monte-Carlo-Simulation''} 
(contract FOR 264/2-1), by SUN Microsystems through Academic
Equipment Grant No.~14WU0148 and by
the European Union under contract
HPRN-CT-2000-00149.


\begin{appendix}

\section*{Appendix:\\ Analytical results for the moments at order
$\alpha_s^2$}

In the limit $z\to0$ the results for $\Pi^{(2),v}$ and $\Pi^{(2),s}$
parameterized in terms of the on-shell mass read
{\scriptsize
\begin{eqnarray}
\Pi^{(2),v}_{FF}
&=& \frac{3}{16\pi^2}\Bigg[
+\left(-
\left(\frac{245}{48}\,\frac{1}{\sqrtthree}\,\cl\right) +
 \frac{3}{2}\,\cl^2 +
 \left(-
3\,\ln2 +
 \frac{80579}{25920}\right)\,\zeta_2 +
 \frac{179}{144}\,\zeta_3 -
 \frac{5}{16}\,\zeta_4 +
 \frac{649}{41472}\right)\,z
\nonumber\\&&\mbox{}
+\left(-
\left(\frac{6187}{1800}\,\frac{1}{\sqrtthree}\,\cl\right) +
 \frac{2}{5}\,\cl^2 +
 \left(-
\left(\frac{8}{5}\,\ln2\right) +
 \frac{54089}{40500}\right)\,\zeta_2 +
 \frac{4261}{2700}\,\zeta_3 -
 \frac{1}{12}\,\zeta_4 +
 \frac{15229}{51840}\right)\,z^2
\nonumber\\&&\mbox{}
+\left(-
\left(\frac{26869}{364500}\,\frac{1}{\sqrtthree}\,\cl\right) +
 \frac{1}{6}\,\cl^2 +
 \left(-
\ln2 +
 \frac{1637813}{2268000}\right)\,\zeta_2 +
 \frac{409}{2700}\,\zeta_3 -
 \frac{5}{144}\,\zeta_4 +
 \frac{5303639}{23328000}\right)\,z^3
\nonumber\\&&\mbox{}
+\left(-
\left(\frac{31433}{79380}\,\frac{1}{\sqrtthree}\,\cl\right) +
 \frac{3}{35}\,\cl^2 +
 \left(-
\left(\frac{24}{35}\,\ln2\right) +
 \frac{455767}{1029000}\right)\,\zeta_2 +
 \frac{155243}{396900}\,\zeta_3 -
 \frac{1}{56}\,\zeta_4 +
 \frac{7315733}{65318400}\right)\,z^4
\nonumber\\&&\mbox{}
+\left(\frac{552604783}{800150400}\,\frac{1}{\sqrtthree}\,\cl +
 \frac{1}{20}\,\cl^2 +
 \left(-
\left(\frac{1}{2}\,\ln2\right) +
 \frac{523855033}{1778112000}\right)\,\zeta_2 -
 \frac{59293}{423360}\,\zeta_3 -
 \frac{1}{96}\,\zeta_4 +
 \frac{2245278797}{34139750400}\right)\,z^5
\nonumber\\&&\mbox{}
+\left(\frac{1448186461}{5952139200}\,\frac{1}{\sqrtthree}\,\cl +
 \frac{2}{63}\,\cl^2 +
 \left(-
\left(\frac{8}{21}\,\ln2\right) +
 \frac{82974271}{400075200}\right)\,\zeta_2 +
 \frac{26561}{317520}\,\zeta_3 -
 \frac{5}{756}\,\zeta_4
\right.\nonumber\\&&\left.\mbox{}
+ \frac{23465474813}{1904684544000}\right)\,z^6
\Bigg]+\ldots\,,
\nonumber\\
\Pi^{(2),v}_{FA}&=& \frac{3}{16\pi^2}\Bigg[
+\left(\frac{245}{96}\,\frac{1}{\sqrtthree}\,\cl -
 \frac{3}{4}\,\cl^2 +
 \left(\frac{3}{2}\,\ln2 +
 \frac{6757}{10368}\right)\,\zeta_2 -
 \frac{1093}{864}\,\zeta_3 -
 \frac{27}{32}\,\zeta_4 +
 \frac{147835}{82944}\right)\,z
\nonumber\\&&\mbox{}
+\left(\frac{6187}{3600}\,\frac{1}{\sqrtthree}\,\cl -
 \frac{1}{5}\,\cl^2 +
 \left(\frac{4}{5}\,\ln2 +
 \frac{66997}{324000}\right)\,\zeta_2 -
 \frac{2711}{2700}\,\zeta_3 -
 \frac{9}{40}\,\zeta_4 +
 \frac{112693}{129600}\right)\,z^2
\nonumber\\&&\mbox{}
+\left(\frac{26869}{729000}\,\frac{1}{\sqrtthree}\,\cl -
 \frac{1}{12}\,\cl^2 +
 \left(\frac{1}{2}\,\ln2 +
 \frac{4573}{54000}\right)\,\zeta_2 -
 \frac{69}{400}\,\zeta_3 -
 \frac{3}{32}\,\zeta_4 +
 \frac{11769743}{23328000}\right)\,z^3
\nonumber\\&&\mbox{}
+\left(\frac{31433}{158760}\,\frac{1}{\sqrtthree}\,\cl -
 \frac{3}{70}\,\cl^2 +
 \left(\frac{12}{35}\,\ln2 +
 \frac{144869}{3472875}\right)\,\zeta_2 -
 \frac{24547}{99225}\,\zeta_3 -
 \frac{27}{560}\,\zeta_4 +
 \frac{39697543}{114307200}\right)\,z^4
\nonumber\\&&\mbox{}
+\left(-
\left(\frac{552604783}{1600300800}\,\frac{1}{\sqrtthree}\,\cl\right) -
 \frac{1}{40}\,\cl^2 +
 \left(\frac{1}{4}\,\ln2 +
 \frac{16650583}{711244800}\right)\,\zeta_2 +
 \frac{23603}{604800}\,\zeta_3 -
 \frac{9}{320}\,\zeta_4 +
 \frac{16737914551}{68279500800}\right)\,z^5
\nonumber\\&&\mbox{}
+\left(-
\left(\frac{1448186461}{11904278400}\,\frac{1}{\sqrtthree}\,\cl\right) -
 \frac{1}{63}\,\cl^2 +
 \left(\frac{4}{21}\,\ln2 +
 \frac{69820073}{4800902400}\right)\,\zeta_2 -
 \frac{8417}{136080}\,\zeta_3 -
 \frac{1}{56}\,\zeta_4
\right.\nonumber\\&&\left.\mbox{}
+ \frac{10342199087}{53331167232}\right)\,z^6
\Bigg]+\ldots\,,
\nonumber\\
\Pi^{(2),v}_{FL}&=& \frac{3}{16\pi^2}\Bigg[
+\left(-
\left(\frac{125}{216}\,\zeta_2\right) +
 \frac{1}{3}\,\zeta_3 -
 \frac{871}{1728}\right)\,z
+\left(-
\left(\frac{367}{1350}\,\zeta_2\right) +
 \frac{4}{45}\,\zeta_3 -
 \frac{1373}{6480}\right)\,z^2
\nonumber\\&&\mbox{}
+\left(-
\left(\frac{11}{72}\,\zeta_2\right) +
 \frac{1}{27}\,\zeta_3 -
 \frac{25771}{233280}\right)\,z^3
+\left(-
\left(\frac{12857}{132300}\,\zeta_2\right) +
 \frac{2}{105}\,\zeta_3 -
 \frac{43117}{680400}\right)\,z^4
\nonumber\\&&\mbox{}
+\left(-
\left(\frac{10133}{151200}\,\zeta_2\right) +
 \frac{1}{90}\,\zeta_3 -
 \frac{112051}{2903040}\right)\,z^5
+\left(-
\left(\frac{34933}{714420}\,\zeta_2\right) +
 \frac{4}{567}\,\zeta_3 -
 \frac{718927}{29393280}\right)\,z^6
\Bigg]+\ldots\,,
\nonumber\\
\Pi^{(2),v}_{FH}&=& \frac{3}{16\pi^2}\Bigg[
+\left(\frac{139}{12}\,\frac{1}{\sqrtthree}\,\cl +
 \zeta_2 -
 \frac{194}{27}\,\zeta_3 +
 \frac{665}{1728}\right)\,z
+\left(\frac{382}{25}\,\frac{1}{\sqrtthree}\,\cl +
 \frac{8}{15}\,\zeta_2 -
 \frac{1168}{135}\,\zeta_3 +
 \frac{9661}{16200}\right)\,z^2
\nonumber\\&&\mbox{}
+\left(\frac{1400461}{72900}\,\frac{1}{\sqrtthree}\,\cl +
 \frac{1}{3}\,\zeta_2 -
 \frac{4102}{405}\,\zeta_3 +
 \frac{440741}{1166400}\right)\,z^3
\nonumber\\&&\mbox{}
+\left(\frac{13738391}{595350}\,\frac{1}{\sqrtthree}\,\cl +
 \frac{8}{35}\,\zeta_2 -
 \frac{3644}{315}\,\zeta_3 +
 \frac{52709}{4762800}\right)\,z^4
\nonumber\\&&\mbox{}
+\left(\frac{128030963}{4762800}\,\frac{1}{\sqrtthree}\,\cl +
 \frac{1}{6}\,\zeta_2 -
 \frac{1363}{105}\,\zeta_3 -
 \frac{8545133}{20321280}\right)\,z^5
\nonumber\\&&\mbox{}
+\left(\frac{15959744567}{520812180}\,\frac{1}{\sqrtthree}\,\cl +
 \frac{8}{63}\,\zeta_2 -
 \frac{24452}{1701}\,\zeta_3 -
 \frac{147478621319}{166659897600}\right)\,z^6
\Bigg]+\ldots\,,
\nonumber\\
\Pi^{(2),s}_{FF}&=& \frac{3}{16\pi^2}\Bigg[
+\left(\frac{7}{2}\,\sqrtthree\,\cl +
 2\,\cl^2 +
 \frac{121}{144}\,\zeta_2 -
 \frac{85}{12}\,\zeta_3 -
 \frac{5}{12}\,\zeta_4 +
 \frac{103}{384}\right)\,z
\nonumber\\&&\mbox{}
+\left(-
\left(\frac{283}{80}\,\sqrtthree\,\cl\right) +
 \frac{1}{2}\,\cl^2 +
 \left(-
\ln2 +
 \frac{59623}{43200}\right)\,\zeta_2 +
 \frac{3461}{720}\,\zeta_3 -
 \frac{5}{48}\,\zeta_4 +
 \frac{23477}{69120}\right)\,z^2
\nonumber\\&&\mbox{}
+\left(\frac{1049}{1800}\,\frac{1}{\sqrtthree}\,\cl +
 \frac{1}{5}\,\cl^2 +
 \left(-
\left(\frac{4}{5}\,\ln2\right) +
 \frac{92717}{108000}\right)\,\zeta_2 -
 \frac{923}{1800}\,\zeta_3 -
 \frac{1}{24}\,\zeta_4 +
 \frac{65329}{103680}\right)\,z^3
\nonumber\\&&\mbox{}
+\left(-
\left(\frac{324853}{85050}\,\frac{1}{\sqrtthree}\,\cl\right) +
 \frac{1}{10}\,\cl^2 +
 \left(-
\left(\frac{3}{5}\,\ln2\right) +
 \frac{2892733}{5292000}\right)\,\zeta_2 +
 \frac{967}{525}\,\zeta_3 -
 \frac{1}{48}\,\zeta_4 +
 \frac{4524487}{10886400}\right)\,z^4
\nonumber\\&&\mbox{}
+\left(\frac{442717}{1587600}\,\frac{1}{\sqrtthree}\,\cl +
 \frac{2}{35}\,\cl^2 +
 \left(-
\left(\frac{16}{35}\,\ln2\right) +
 \frac{27193603}{74088000}\right)\,\zeta_2 -
 \frac{599}{3528}\,\zeta_3 -
 \frac{1}{84}\,\zeta_4 +
 \frac{33815837}{87091200}\right)\,z^5
\nonumber\\&&\mbox{}
+\left(-
\left(\frac{219606437}{114307200}\,\frac{1}{\sqrtthree}\,\cl\right) +
 \frac{1}{28}\,\cl^2 +
 \left(-
\left(\frac{5}{14}\,\ln2\right) +
 \frac{91843391}{355622400}\right)\,\zeta_2 +
 \frac{137203}{141120}\,\zeta_3 -
 \frac{5}{672}\,\zeta_4 
\right.\nonumber\\&&\left.\mbox{}
+
 \frac{3685964659}{14631321600}\right)\,z^6
\nonumber\\&&\mbox{}
+\left(\frac{2861558023}{13888324800}\,\frac{1}{\sqrtthree}\,\cl +
 \frac{1}{42}\,\cl^2 +
 \left(-
\left(\frac{2}{7}\,\ln2\right) +
 \frac{100704841}{533433600}\right)\,\zeta_2 -
 \frac{18371}{211680}\,\zeta_3 -
 \frac{5}{1008}\,\zeta_4
\right.\nonumber\\&&\left.\mbox{}
+
 \frac{10307800934177}{44442639360000}\right)\,z^7
\Bigg]+\ldots\,,
\nonumber\\
\Pi^{(2),s}_{FA}&=& \frac{3}{16\pi^2}\Bigg[
+\left(-
\left(\frac{7}{4}\,\sqrtthree\,\cl\right) -
 \cl^2 +
 \frac{277}{216}\,\zeta_2 +
 \frac{8}{3}\,\zeta_3 -
 \frac{9}{8}\,\zeta_4 +
 \frac{175}{1728}\right)\,z
\nonumber\\&&\mbox{}
+\left(\frac{283}{160}\,\sqrtthree\,\cl -
 \frac{1}{4}\,\cl^2 +
 \left(\frac{1}{2}\,\ln2 +
 \frac{1723}{5760}\right)\,\zeta_2 -
 \frac{3857}{1440}\,\zeta_3 -
 \frac{9}{32}\,\zeta_4 +
 \frac{953}{1024}\right)\,z^2
\nonumber\\&&\mbox{}
+\left(-
\left(\frac{1049}{3600}\,\frac{1}{\sqrtthree}\,\cl\right) -
 \frac{1}{10}\,\cl^2 +
 \left(\frac{2}{5}\,\ln2 +
 \frac{1153}{12000}\right)\,\zeta_2 +
 \frac{31}{225}\,\zeta_3 -
 \frac{9}{80}\,\zeta_4 +
 \frac{33307}{64800}\right)\,z^3
\nonumber\\&&\mbox{}
+\left(\frac{324853}{170100}\,\frac{1}{\sqrtthree}\,\cl -
 \frac{1}{20}\,\cl^2 +
 \left(\frac{3}{10}\,\ln2 +
 \frac{313}{8400}\right)\,\zeta_2 -
 \frac{24761}{25200}\,\zeta_3 -
 \frac{9}{160}\,\zeta_4 +
 \frac{3961231}{10886400}\right)\,z^4
\nonumber\\&&\mbox{}
+\left(-
\left(\frac{442717}{3175200}\,\frac{1}{\sqrtthree}\,\cl\right) -
 \frac{1}{35}\,\cl^2 +
 \left(\frac{8}{35}\,\ln2 +
 \frac{243241}{14817600}\right)\,\zeta_2 +
 \frac{359}{7350}\,\zeta_3 -
 \frac{9}{280}\,\zeta_4 +
 \frac{31355501}{152409600}\right)\,z^5
\nonumber\\&&\mbox{}
+\left(\frac{219606437}{228614400}\,\frac{1}{\sqrtthree}\,\cl -
 \frac{1}{56}\,\cl^2 +
 \left(\frac{5}{28}\,\ln2 +
 \frac{272033}{33868800}\right)\,\zeta_2 -
 \frac{143663}{282240}\,\zeta_3 -
 \frac{9}{448}\,\zeta_4
\right.\nonumber\\&&\left.\mbox{}
 +
 \frac{4739201353}{29262643200}\right)\,z^6
\nonumber\\&&\mbox{}
+\left(-
\left(\frac{2861558023}{27776649600}\,\frac{1}{\sqrtthree}\,\cl\right) -
 \frac{1}{84}\,\cl^2 +
 \left(\frac{1}{7}\,\ln2 +
 \frac{28543}{6531840}\right)\,\zeta_2 +
 \frac{4439}{158760}\,\zeta_3 -
 \frac{3}{224}\,\zeta_4
\right.\nonumber\\&&\left.\mbox{}
+
 \frac{108189401963}{1111065984000}\right)\,z^7
\Bigg]+\ldots\,,
\nonumber\\
\Pi^{(2),s}_{FL}&=& \frac{3}{16\pi^2}\Bigg[
+\left(-
\left(\frac{1}{6}\,\zeta_2\right) +
 \frac{4}{9}\,\zeta_3 -
 \frac{5}{48}\right)\,z
+\left(-
\left(\frac{73}{360}\,\zeta_2\right) +
 \frac{1}{9}\,\zeta_3 -
 \frac{143}{576}\right)\,z^2
\nonumber\\&&\mbox{}
+\left(-
\left(\frac{119}{900}\,\zeta_2\right) +
 \frac{2}{45}\,\zeta_3 -
 \frac{2083}{12960}\right)\,z^3
+\left(-
\left(\frac{1123}{12600}\,\zeta_2\right) +
 \frac{1}{45}\,\zeta_3 -
 \frac{7759}{77760}\right)\,z^4
\nonumber\\&&\mbox{}
+\left(-
\left(\frac{2791}{44100}\,\zeta_2\right) +
 \frac{4}{315}\,\zeta_3 -
 \frac{346519}{5443200}\right)\,z^5
+\left(-
\left(\frac{9941}{211680}\,\zeta_2\right) +
 \frac{1}{126}\,\zeta_3 -
 \frac{608813}{14515200}\right)\,z^6
\nonumber\\&&\mbox{}
+\left(-
\left(\frac{34387}{952560}\,\zeta_2\right) +
 \frac{1}{189}\,\zeta_3 -
 \frac{27828673}{979776000}\right)\,z^7
\Bigg]+\ldots\,,
\nonumber\\
\Pi^{(2),s}_{FH}&=& \frac{3}{16\pi^2}\Bigg[
+\left(7\,\sqrtthree\,\cl -
 \frac{104}{9}\,\zeta_3 +
 \frac{33}{16}\right)\,z
+\left(\frac{479}{20}\,\frac{1}{\sqrtthree}\,\cl +
 \frac{1}{3}\,\zeta_2 -
 \frac{38}{3}\,\zeta_3 +
 \frac{2101}{2880}\right)\,z^2
\nonumber\\&&\mbox{}
+\left(\frac{12473}{450}\,\frac{1}{\sqrtthree}\,\cl +
 \frac{4}{15}\,\zeta_2 -
 \frac{628}{45}\,\zeta_3 +
 \frac{7837}{64800}\right)\,z^3
\nonumber\\&&\mbox{}
+\left(\frac{5363167}{170100}\,\frac{1}{\sqrtthree}\,\cl +
 \frac{1}{5}\,\zeta_2 -
 \frac{2062}{135}\,\zeta_3 -
 \frac{1179673}{2721600}\right)\,z^4
\nonumber\\&&\mbox{}
+\left(\frac{21024149}{595350}\,\frac{1}{\sqrtthree}\,\cl +
 \frac{16}{105}\,\zeta_2 -
 \frac{15688}{945}\,\zeta_3 -
 \frac{37456457}{38102400}\right)\,z^5
\nonumber\\&&\mbox{}
+\left(\frac{334931357}{8573040}\,\frac{1}{\sqrtthree}\,\cl +
 \frac{5}{42}\,\zeta_2 -
 \frac{3389}{189}\,\zeta_3 -
 \frac{4201269473}{2743372800}\right)\,z^6
\nonumber\\&&\mbox{}
+\left(\frac{14859593107}{347208120}\,\frac{1}{\sqrtthree}\,\cl +
 \frac{2}{21}\,\zeta_2 -
 \frac{10922}{567}\,\zeta_3 -
 \frac{1154471458391}{555532992000}\right)\,z^7
\Bigg]+\ldots\,,
\nonumber
\end{eqnarray}
}

\noindent
where $\mu^2=M^2$ has been adopted.
In the high-energy limit the corresponding results read
{\scriptsize
\begin{eqnarray}
\Pi^{(2),v}_{FF}&=& \frac{3}{16\pi^2}\Bigg[
\frac{71}{576}
-
\left(\frac{8}{3}\,\cl^2\right)
+\frac{11}{2}\,\frac{1}{\sqrtthree}\,\cl
-
\left(\frac{64}{3}\,\lifour\right)
-
\left(\frac{8}{9}\,\ln^4 2\right)
-
\left(\frac{1}{4}\,\lz\right)
\nonumber\\&&\mbox{}
+\left(8\,\ln2 +
 \frac{16}{3}\,\ln^2 2 -
 \frac{713}{216}\right)\,\zeta_2
-
\left(\frac{158}{9}\,\zeta_3\right)
+\frac{107}{9}\,\zeta_4
+10\,\zeta_5
\nonumber\\&&\mbox{}
+\left(-
\left(\frac{39}{4}\,\sqrtthree\,\cl\right) -
 \frac{3}{2}\,\cl^2 +
 32\,\lifour +
 \frac{4}{3}\,\ln^4 2 +
 \left(\left(15 -
 24\,\ln2\right)\,\zeta_2 +
 15\,\zeta_3 -
 \frac{121}{8}\right)\,\lz +
 \frac{15}{4}\,\lz^2 -
 6\,\lz^3
\right.\nonumber\\&&\left.\mbox{}
 +
 \left(-
6\,\ln2 -
 8\,\ln^2 2 +
 \frac{155}{72}\right)\,\zeta_2 +
 \frac{491}{12}\,\zeta_3 -
 \frac{363}{16}\,\zeta_4 -
 \frac{85}{3}\,\zeta_5 +
 \frac{3859}{288}\right)\,\frac{1}{z}
\nonumber\\&&\mbox{}
+\left(\frac{21}{4}\,\lz -
 18\,\lz^2 +
 \left(15 -
 24\,\ln2\right)\,\zeta_2 +
 \frac{68}{3}\,\zeta_3 -
 \frac{80}{3}\,\zeta_5 -
 \frac{31}{24}\right)\,\frac{1}{z^2}
\nonumber\\&&\mbox{}
+\left(-
\left(\frac{32}{3}\,\lifour\right) -
 \frac{4}{9}\,\ln^4 2 +
 \left(\left(-
15 +
 24\,\ln2\right)\,\zeta_2 -
 \frac{101}{3}\,\zeta_3 +
 \frac{4337}{324}\right)\,\lz -
 \frac{85}{6}\,\lz^2 +
 \frac{890}{81}\,\lz^3
\right.\nonumber\\&&\left.\mbox{}
 +
 \left(-
6\,\ln2 +
 \frac{8}{3}\,\ln^2 2 +
 \frac{15}{4}\right)\,\zeta_2 -
 \frac{47}{6}\,\zeta_3 +
 \frac{25}{3}\,\zeta_4 +
 \frac{5}{3}\,\zeta_5 -
 \frac{16883}{2916}\right)\,\frac{1}{z^3}
\nonumber\\&&\mbox{}
+\left(\left(-
2\,\zeta_3 +
 \frac{757}{972}\right)\,\lz +
 \frac{19855}{648}\,\lz^2 -
 \frac{1345}{81}\,\lz^3 +
 \left(-
5 +
 8\,\ln2\right)\,\zeta_2 +
 \frac{373}{36}\,\zeta_3 -
 \frac{1760405}{93312}\right)\,\frac{1}{z^4}
\nonumber\\&&\mbox{}
+\left(\left(-
\left(\frac{3}{5}\,\zeta_3\right) +
 \frac{27867253}{1620000}\right)\,\lz -
 \frac{617069}{54000}\,\lz^2 -
 \frac{64663}{4050}\,\lz^3 +
 \left(3\,\ln2 -
 \frac{15}{8}\right)\,\zeta_2 +
 \frac{1306}{675}\,\zeta_3 +
 \frac{804239879}{116640000}\right)\,\frac{1}{z^5}
\nonumber\\&&\mbox{}
+\left(\left(-
\left(\frac{4}{15}\,\zeta_3\right) +
 \frac{36020231}{2430000}\right)\,\lz -
 \frac{4796371}{81000}\,\lz^2 -
 \frac{50866}{2025}\,\lz^3 +
 \left(-
1 +
 \frac{8}{5}\,\ln2\right)\,\zeta_2 +
 \frac{10987}{900}\,\zeta_3
\right.\nonumber\\&&\left.\mbox{}
 +
 \frac{700685473}{29160000}\right)\,\frac{1}{z^6}
\nonumber\\&&\mbox{}
+\left(\left(-
\left(\frac{1}{7}\,\zeta_3\right) -
 \frac{7266775483}{121550625}\right)\,\lz -
 \frac{20158677919}{111132000}\,\lz^2 -
 \frac{3780751}{99225}\,\lz^3 +
 \left(\ln2 -
 \frac{5}{8}\right)\,\zeta_2 +
 \frac{2673157}{176400}\,\zeta_3
\right.\nonumber\\&&\left.\mbox{}
 +
 \frac{44535682588013}{490092120000}\right)\,\frac{1}{z^7}
\Bigg]+\ldots\,,
\nonumber\\
\Pi^{(2),v}_{FA}&=& \frac{3}{16\pi^2}\Bigg[
\frac{18433}{2592}
+\frac{4}{3}\,\cl^2
-
\left(\frac{11}{4}\,\frac{1}{\sqrtthree}\,\cl\right)
+\frac{32}{3}\,\lifour
+\frac{4}{9}\,\ln^4 2
+\left(-
\left(\frac{22}{3}\,\zeta_3\right) +
 \frac{41}{4}\right)\,\lz
\nonumber\\&&\mbox{}
+\left(-
4\,\ln2 -
 \frac{8}{3}\,\ln^2 2 +
 \frac{1985}{648}\right)\,\zeta_2
-
\left(\frac{17}{2}\,\zeta_3\right)
-
\left(\frac{41}{6}\,\zeta_4\right)
-
\left(\frac{5}{3}\,\zeta_5\right)
+\frac{11}{6}\,\lz^2
\nonumber\\&&\mbox{}
+\left(\frac{39}{8}\,\sqrtthree\,\cl +
 \frac{3}{4}\,\cl^2 -
 16\,\lifour -
 \frac{2}{3}\,\ln^4 2 +
 \left(\left(-
4 +
 12\,\ln2\right)\,\zeta_2 +
 \frac{25}{2}\,\zeta_3 -
 \frac{11}{3}\right)\,\lz -
 \frac{38}{3}\,\lz^2 -
 \frac{11}{3}\,\lz^3
\right.\nonumber\\&&\left.\mbox{}
 +
 \left(3\,\ln2 +
 4\,\ln^2 2 -
 \frac{97}{24}\right)\,\zeta_2 +
 12\,\zeta_3 +
 \frac{427}{32}\,\zeta_4 -
 \frac{65}{6}\,\zeta_5 -
 \frac{4015}{288}\right)\,\frac{1}{z}
\nonumber\\&&\mbox{}
+\left(-
\left(\frac{91}{6}\,\lz\right) +
 \left(-
4 +
 12\,\ln2\right)\,\zeta_2 +
 \frac{52}{3}\,\zeta_3 -
 20\,\zeta_5 +
 \frac{1561}{144}\right)\,\frac{1}{z^2}
\nonumber\\&&\mbox{}
+\left(\frac{16}{3}\,\lifour +
 \frac{2}{9}\,\ln^4 2 +
 \left(\left(4 -
 12\,\ln2\right)\,\zeta_2 +
 \frac{13}{2}\,\zeta_3 -
 \frac{22153}{1944}\right)\,\lz +
 \frac{3409}{324}\,\lz^2 +
 \frac{143}{81}\,\lz^3
\right.\nonumber\\&&\left.\mbox{}
 +
 \left(-
1 +
 3\,\ln2 -
 \frac{4}{3}\,\ln^2 2\right)\,\zeta_2 +
 \frac{293}{72}\,\zeta_3 -
 \frac{25}{6}\,\zeta_4 -
 \frac{5}{2}\,\zeta_5 +
 \frac{108761}{23328}\right)\,\frac{1}{z^3}
\nonumber\\&&\mbox{}
+\left(\left(2\,\zeta_3 -
 \frac{11267}{3888}\right)\,\lz -
 \frac{1789}{324}\,\lz^2 +
 \frac{491}{81}\,\lz^3 +
 \left(-
4\,\ln2 +
 \frac{4}{3}\right)\,\zeta_2 -
 \frac{80}{9}\,\zeta_3 +
 \frac{1658717}{186624}\right)\,\frac{1}{z^4}
\nonumber\\&&\mbox{}
+\left(\left(\frac{3}{5}\,\zeta_3 -
 \frac{10253071}{1944000}\right)\,\lz +
 \frac{43999}{6480}\,\lz^2 +
 \frac{331}{45}\,\lz^3 +
 \left(-
\left(\frac{3}{2}\,\ln2\right) +
 \frac{1}{2}\right)\,\zeta_2 -
 \frac{20327}{10800}\,\zeta_3 -
 \frac{69281173}{14580000}\right)\,\frac{1}{z^5}
\nonumber\\&&\mbox{}
+\left(\left(\frac{4}{15}\,\zeta_3 -
 \frac{13535191}{2430000}\right)\,\lz +
 \frac{40111}{1350}\,\lz^2 +
 \frac{2767}{225}\,\lz^3 +
 \left(-
\left(\frac{4}{5}\,\ln2\right) +
 \frac{4}{15}\right)\,\zeta_2 -
 \frac{5837}{900}\,\zeta_3
\right.\nonumber\\&&\left.\mbox{}
 -
 \frac{120018929}{9720000}\right)\,\frac{1}{z^6}
\nonumber\\&&\mbox{}
+\left(\left(\frac{1}{7}\,\zeta_3 +
 \frac{11463009101}{370440000}\right)\,\lz +
 \frac{2877743137}{31752000}\,\lz^2 +
 \frac{1072571}{56700}\,\lz^3 +
 \left(-
\left(\frac{1}{2}\,\ln2\right) +
 \frac{1}{6}\right)\,\zeta_2 -
 \frac{2742577}{352800}\,\zeta_3
\right.\nonumber\\&&\left.\mbox{}
 -
 \frac{104055061483}{2286144000}\right)\,\frac{1}{z^7}
\Bigg]+\ldots\,,
\nonumber\\
\Pi^{(2),v}_{FL}&=& \frac{3}{16\pi^2}\Bigg[
-\frac{589}{216}
+\frac{122}{27}\,\zeta_3
+\frac{2}{9}\,\zeta_2
-
\left(\frac{2}{3}\,\lz^2\right)
+\left(\frac{8}{3}\,\zeta_3 -
 \frac{11}{3}\right)\,\lz
\nonumber\\&&\mbox{}
+\left(\left(-
4\,\zeta_2 -
 4\,\zeta_3 +
 \frac{7}{3}\right)\,\lz +
 \frac{10}{3}\,\lz^2 +
 \frac{4}{3}\,\lz^3 +
 \frac{1}{6}\,\zeta_2 -
 \frac{22}{3}\,\zeta_3 +
 \frac{89}{18}\right)\,\frac{1}{z}
\nonumber\\&&\mbox{}
+\left(\frac{16}{3}\,\lz -
 4\,\zeta_2 -
 \frac{2}{3}\,\zeta_3 -
 \frac{131}{36}\right)\,\frac{1}{z^2}
\nonumber\\&&\mbox{}
+\left(\left(4\,\zeta_2 +
 \frac{4}{3}\,\zeta_3 +
 \frac{2183}{486}\right)\,\lz -
 \frac{299}{81}\,\lz^2 -
 \frac{52}{81}\,\lz^3 -
 \zeta_2 +
 2\,\zeta_3 -
 \frac{11491}{5832}\right)\,\frac{1}{z^3}
\nonumber\\&&\mbox{}
+\left(-
\left(\frac{1625}{648}\,\lz\right) -
 \frac{23}{54}\,\lz^2 +
 \frac{8}{27}\,\lz^3 +
 \frac{4}{3}\,\zeta_2 +
 \frac{2}{3}\,\zeta_3 +
 \frac{15455}{7776}\right)\,\frac{1}{z^4}
\nonumber\\&&\mbox{}
+\left(-
\left(\frac{199679}{162000}\,\lz\right) +
 \frac{497}{2700}\,\lz^2 +
 \frac{4}{45}\,\lz^3 +
 \frac{1}{2}\,\zeta_2 +
 \frac{1}{5}\,\zeta_3 +
 \frac{233149}{405000}\right)\,\frac{1}{z^5}
\nonumber\\&&\mbox{}
+\left(-
\left(\frac{19067}{30375}\,\lz\right) +
 \frac{41}{225}\,\lz^2 +
 \frac{16}{405}\,\lz^3 +
 \frac{4}{15}\,\zeta_2 +
 \frac{4}{45}\,\zeta_3 +
 \frac{3236209}{14580000}\right)\,\frac{1}{z^6}
\nonumber\\&&\mbox{}
+\left(-
\left(\frac{3920897}{11113200}\,\lz\right) +
 \frac{247}{1764}\,\lz^2 +
 \frac{4}{189}\,\lz^3 +
 \frac{1}{6}\,\zeta_2 +
 \frac{1}{21}\,\zeta_3 +
 \frac{245309459}{2333772000}\right)\,\frac{1}{z^7}
\Bigg]+\ldots\,,
\nonumber\\
\Pi^{(2),v}_{FH}&=& \frac{3}{16\pi^2}\Bigg[
-\frac{1141}{216}
+\frac{410}{27}\,\zeta_3
-
\left(\frac{8}{3}\,\zeta_2\right)
-
\left(\frac{2}{3}\,\lz^2\right)
+\left(\frac{8}{3}\,\zeta_3 -
 \frac{11}{3}\right)\,\lz
-
16\,\frac{1}{\sqrtthree}\,\cl
\nonumber\\&&\mbox{}
+\left(9\,\sqrtthree\,\cl +
 \left(8\,\zeta_2 -
 4\,\zeta_3 -
 \frac{11}{3}\right)\,\lz +
 \frac{10}{3}\,\lz^2 +
 \frac{4}{3}\,\lz^3 +
 2\,\zeta_2 -
 \frac{20}{3}\,\zeta_3 -
 \frac{95}{9}\right)\,\frac{1}{z}
\nonumber\\&&\mbox{}
+\left(\left(-
16\,\zeta_3 +
 \frac{68}{3}\right)\,\lz -
 4\,\lz^2 +
 8\,\zeta_2 +
 \frac{22}{3}\,\zeta_3 -
 \frac{551}{36}\right)\,\frac{1}{z^2}
\nonumber\\&&\mbox{}
+\left(\left(-
8\,\zeta_2 +
 \frac{4}{3}\,\zeta_3 +
 \frac{1111}{486}\right)\,\lz +
 \frac{173}{27}\,\lz^2 -
 \frac{20}{9}\,\lz^3 +
 2\,\zeta_2 -
 \frac{32}{27}\,\zeta_3 +
 \frac{2129}{5832}\right)\,\frac{1}{z^3}
\nonumber\\&&\mbox{}
+\left(\frac{1177}{216}\,\lz -
 \frac{31}{6}\,\lz^2 +
 \frac{176}{27}\,\lz^3 -
 \frac{8}{3}\,\zeta_2 -
 \frac{4}{3}\,\zeta_3 -
 \frac{3191}{7776}\right)\,\frac{1}{z^4}
\nonumber\\&&\mbox{}
+\left(-
\left(\frac{136549}{54000}\,\lz\right) +
 \frac{6679}{900}\,\lz^2 +
 \frac{1864}{135}\,\lz^3 -
 \zeta_2 -
 \frac{34}{5}\,\zeta_3 -
 \frac{3922271}{1215000}\right)\,\frac{1}{z^5}
\nonumber\\&&\mbox{}
+\left(\frac{824581}{121500}\,\lz +
 \frac{11756}{225}\,\lz^2 +
 \frac{8192}{405}\,\lz^3 -
 \frac{8}{15}\,\zeta_2 -
 \frac{1568}{135}\,\zeta_3 -
 \frac{337954307}{14580000}\right)\,\frac{1}{z^6}
\nonumber\\&&\mbox{}
+\left(\frac{1750009337}{18522000}\,\lz +
 \frac{2488393}{14700}\,\lz^2 +
 \frac{8264}{315}\,\lz^3 -
 \frac{1}{3}\,\zeta_2 -
 \frac{5078}{315}\,\zeta_3 -
 \frac{912529125371}{11668860000}\right)\,\frac{1}{z^7}
\Bigg]+\ldots\,,
\nonumber\\
\Pi^{(2),s}_{FF}&=& \frac{3}{16\pi^2}\Bigg[
\frac{1961}{288}
+\frac{3}{2}\,\cl^2
+\frac{39}{4}\,\sqrtthree\,\cl
-
32\,\lifour
-
\left(\frac{4}{3}\,\ln^4 2\right)
+6\,\lz^3
+\frac{57}{4}\,\lz^2
\nonumber\\&&\mbox{}
+\left(\left(-
15 +
 24\,\ln2\right)\,\zeta_2 -
 15\,\zeta_3 +
 \frac{109}{8}\right)\,\lz
+\left(30\,\ln2 +
 8\,\ln^2 2 -
 \frac{1235}{72}\right)\,\zeta_2
-
\left(\frac{631}{12}\,\zeta_3\right)
+\frac{363}{16}\,\zeta_4
+15\,\zeta_5
\nonumber\\&&\mbox{}
+\left(64\,\lifour +
 \frac{8}{3}\,\ln^4 2 +
 \left(-
42 +
 \left(60 -
 96\,\ln2\right)\,\zeta_2 +
 72\,\zeta_3\right)\,\lz -
 3\,\lz^2 -
 48\,\lz^3
\right.\nonumber\\&&\left.\mbox{}
 +
 \left(15 -
 24\,\ln2 -
 16\,\ln^2 2\right)\,\zeta_2 +
 \frac{137}{2}\,\zeta_3 -
 50\,\zeta_4 -
 50\,\zeta_5 +
 \frac{3}{8}\right)\,\frac{1}{z}
\nonumber\\&&\mbox{}
+\left(-
32\,\lifour -
 \frac{4}{3}\,\ln^4 2 +
 \left(\left(-
45 +
 72\,\ln2\right)\,\zeta_2 -
 69\,\zeta_3 +
 \frac{299}{4}\right)\,\lz -
 \frac{201}{2}\,\lz^2 +
 78\,\lz^3
\right.\nonumber\\&&\left.\mbox{}
 +
 \left(-
42\,\ln2 +
 8\,\ln^2 2 +
 \frac{105}{4}\right)\,\zeta_2 +
 \frac{51}{2}\,\zeta_3 +
 25\,\zeta_4 +
 5\,\zeta_5 -
 \frac{237}{8}\right)\,\frac{1}{z^2}
\nonumber\\&&\mbox{}
+\left(\left(-
16\,\zeta_3 -
 \frac{2351}{27}\right)\,\lz +
 96\,\lz^2 -
 \frac{896}{27}\,\lz^3 +
 \left(-
10 +
 16\,\ln2\right)\,\zeta_2 -
 41\,\zeta_3 +
 \frac{137243}{3888}\right)\,\frac{1}{z^3}
\nonumber\\&&\mbox{}
+\left(\left(-
\zeta_3 +
 \frac{139555}{2592}\right)\,\lz +
 \frac{2719}{432}\,\lz^2 -
 \frac{575}{54}\,\lz^3 +
 \left(5\,\ln2 -
 \frac{25}{8}\right)\,\zeta_2 +
 \frac{113}{4}\,\zeta_3 -
 \frac{1271839}{20736}\right)\,\frac{1}{z^4}
\nonumber\\&&\mbox{}
+\left(\left(-
\left(\frac{2}{5}\,\zeta_3\right) +
 \frac{17774837}{1620000}\right)\,\lz +
 \frac{63049}{54000}\,\lz^2 -
 \frac{1514}{225}\,\lz^3 +
 \left(\frac{12}{5}\,\ln2 -
 \frac{3}{2}\right)\,\zeta_2 -
 \frac{809}{200}\,\zeta_3 +
 \frac{175044221}{19440000}\right)\,\frac{1}{z^5}
\nonumber\\&&\mbox{}
+\left(\left(-
\left(\frac{1}{5}\,\zeta_3\right) +
 \frac{5650789}{162000}\right)\,\lz -
 \frac{78107}{36000}\,\lz^2 -
 \frac{7247}{675}\,\lz^3 +
 \left(\frac{7}{5}\,\ln2 -
 \frac{7}{8}\right)\,\zeta_2 +
 \frac{11261}{1200}\,\zeta_3 -
 \frac{2411369}{405000}\right)\,\frac{1}{z^6}
\nonumber\\&&\mbox{}
+\left(\left(-
\left(\frac{4}{35}\,\zeta_3\right) +
 \frac{215940050209}{3889620000}\right)\,\lz +
 \frac{18459061}{9261000}\,\lz^2 -
 \frac{70724}{6615}\,\lz^3 +
 \left(\frac{32}{35}\,\ln2 -
 \frac{4}{7}\right)\,\zeta_2 +
 \frac{1261}{588}\,\zeta_3
\right.\nonumber\\&&\left.\mbox{}
 +
 \frac{339666207683}{32672808000}\right)\,\frac{1}{z^7}
\Bigg]+\ldots\,,
\nonumber\\
\Pi^{(2),s}_{FA}&=& \frac{3}{16\pi^2}\Bigg[
\frac{4783}{288}
-
\left(\frac{3}{4}\,\cl^2\right)
-
\left(\frac{39}{8}\,\sqrtthree\,\cl\right)
+16\,\lifour
+\frac{2}{3}\,\ln^4 2
+\frac{11}{3}\,\lz^3
+\frac{71}{3}\,\lz^2
\nonumber\\&&\mbox{}
+\left(\left(4 -
 12\,\ln2\right)\,\zeta_2 -
 \frac{25}{2}\,\zeta_3 +
 \frac{98}{3}\right)\,\lz
+\left(-
15\,\ln2 -
 4\,\ln^2 2 +
 \frac{193}{24}\right)\,\zeta_2
-
\left(\frac{5}{2}\,\zeta_3\right)
-
\left(\frac{427}{32}\,\zeta_4\right)
-
\left(\frac{5}{2}\,\zeta_5\right)
\nonumber\\&&\mbox{}
+\left(-
32\,\lifour -
 \frac{4}{3}\,\ln^4 2 +
 \left(\left(-
16 +
 48\,\ln2\right)\,\zeta_2 +
 22\,\zeta_3 -
 \frac{77}{6}\right)\,\lz -
 \frac{185}{3}\,\lz^2 -
 \frac{44}{3}\,\lz^3
\right.\nonumber\\&&\left.\mbox{}
 +
 \left(-
4 +
 12\,\ln2 +
 8\,\ln^2 2\right)\,\zeta_2 +
 \frac{41}{6}\,\zeta_3 +
 25\,\zeta_4 +
 5\,\zeta_5 -
 \frac{821}{72}\right)\,\frac{1}{z}
\nonumber\\&&\mbox{}
+\left(16\,\lifour +
 \frac{2}{3}\,\ln^4 2 +
 \left(\left(12 -
 36\,\ln2\right)\,\zeta_2 -
 \frac{25}{2}\,\zeta_3 -
 \frac{405}{8}\right)\,\lz +
 \frac{509}{12}\,\lz^2 +
 \frac{55}{3}\,\lz^3
\right.\nonumber\\&&\left.\mbox{}
 +
 \left(-
7 +
 21\,\ln2 -
 4\,\ln^2 2\right)\,\zeta_2 +
 \frac{17}{24}\,\zeta_3 -
 \frac{25}{2}\,\zeta_4 -
 \frac{15}{2}\,\zeta_5 +
 \frac{5239}{288}\right)\,\frac{1}{z^2}
\nonumber\\&&\mbox{}
+\left(\left(16\,\zeta_3 +
 \frac{9395}{324}\right)\,\lz +
 \frac{209}{27}\,\lz^2 -
 \frac{248}{27}\,\lz^3 +
 \left(-
8\,\ln2 +
 \frac{8}{3}\right)\,\zeta_2 +
 8\,\zeta_3 -
 \frac{16153}{972}\right)\,\frac{1}{z^3}
\nonumber\\&&\mbox{}
+\left(\left(\zeta_3 -
 \frac{42247}{5184}\right)\,\lz -
 \frac{6847}{432}\,\lz^2 +
 \frac{193}{27}\,\lz^3 +
 \left(-
\left(\frac{5}{2}\,\ln2\right) +
 \frac{5}{6}\right)\,\zeta_2 -
 \frac{725}{48}\,\zeta_3 +
 \frac{420179}{15552}\right)\,\frac{1}{z^4}
\nonumber\\&&\mbox{}
+\left(\left(\frac{2}{5}\,\zeta_3 -
 \frac{650497}{162000}\right)\,\lz -
 \frac{47}{40}\,\lz^2 +
 \frac{427}{135}\,\lz^3 +
 \left(-
\left(\frac{6}{5}\,\ln2\right) +
 \frac{2}{5}\right)\,\zeta_2 +
 \frac{118}{75}\,\zeta_3 -
 \frac{15978853}{4320000}\right)\,\frac{1}{z^5}
\nonumber\\&&\mbox{}
+\left(\left(\frac{1}{5}\,\zeta_3 -
 \frac{10567931}{648000}\right)\,\lz +
 \frac{1979}{2880}\,\lz^2 +
 \frac{2833}{540}\,\lz^3 +
 \left(-
\left(\frac{7}{10}\,\ln2\right) +
 \frac{7}{30}\right)\,\zeta_2 -
 \frac{3939}{800}\,\zeta_3 +
 \frac{16344943}{5184000}\right)\,\frac{1}{z^6}
\nonumber\\&&\mbox{}
+\left(\left(\frac{4}{35}\,\zeta_3 -
 \frac{3758939963}{138915000}\right)\,\lz -
 \frac{13834}{11025}\,\lz^2 +
 \frac{24902}{4725}\,\lz^3 +
 \left(-
\left(\frac{16}{35}\,\ln2\right) +
 \frac{16}{105}\right)\,\zeta_2 -
 \frac{591}{490}\,\zeta_3
\right.\nonumber\\&&\left.\mbox{}
 -
 \frac{471073727}{92610000}\right)\,\frac{1}{z^7}
\Bigg]+\ldots\,,
\nonumber\\
\Pi^{(2),s}_{FL}&=& \frac{3}{16\pi^2}\Bigg[
-\frac{95}{18}
+6\,\zeta_3
+\frac{23}{6}\,\zeta_2
-
\left(\frac{4}{3}\,\lz^3\right)
-
\left(\frac{22}{3}\,\lz^2\right)
+\left(4\,\zeta_2 +
 4\,\zeta_3 -
 \frac{31}{3}\right)\,\lz
\nonumber\\&&\mbox{}
+\left(\left(-
16\,\zeta_2 -
 8\,\zeta_3 +
 \frac{14}{3}\right)\,\lz +
 \frac{52}{3}\,\lz^2 +
 \frac{16}{3}\,\lz^3 -
 4\,\zeta_2 -
 \frac{26}{3}\,\zeta_3 +
 \frac{61}{18}\right)\,\frac{1}{z}
\nonumber\\&&\mbox{}
+\left(\left(12\,\zeta_2 +
 4\,\zeta_3 +
 \frac{27}{2}\right)\,\lz -
 \frac{31}{3}\,\lz^2 -
 \frac{20}{3}\,\lz^3 -
 7\,\zeta_2 -
 \frac{2}{3}\,\zeta_3 -
 \frac{341}{72}\right)\,\frac{1}{z^2}
\nonumber\\&&\mbox{}
+\left(-
\left(\frac{118}{81}\,\lz\right) -
 \frac{136}{27}\,\lz^2 +
 \frac{64}{27}\,\lz^3 +
 \frac{8}{3}\,\zeta_2 +
 \frac{10}{3}\,\zeta_3 -
 \frac{191}{486}\right)\,\frac{1}{z^3}
\nonumber\\&&\mbox{}
+\left(-
\left(\frac{4981}{1296}\,\lz\right) +
 \frac{143}{108}\,\lz^2 +
 \frac{4}{27}\,\lz^3 +
 \frac{5}{6}\,\zeta_2 +
 \frac{1}{3}\,\zeta_3 +
 \frac{6755}{3888}\right)\,\frac{1}{z^4}
\nonumber\\&&\mbox{}
+\left(-
\left(\frac{85273}{81000}\,\lz\right) +
 \frac{649}{1350}\,\lz^2 +
 \frac{8}{135}\,\lz^3 +
 \frac{2}{5}\,\zeta_2 +
 \frac{2}{15}\,\zeta_3 +
 \frac{1217977}{4860000}\right)\,\frac{1}{z^5}
\nonumber\\&&\mbox{}
+\left(-
\left(\frac{26531}{54000}\,\lz\right) +
 \frac{719}{2700}\,\lz^2 +
 \frac{4}{135}\,\lz^3 +
 \frac{7}{30}\,\zeta_2 +
 \frac{1}{15}\,\zeta_3 +
 \frac{54839}{540000}\right)\,\frac{1}{z^6}
\nonumber\\&&\mbox{}
+\left(-
\left(\frac{320456}{1157625}\,\lz\right) +
 \frac{5693}{33075}\,\lz^2 +
 \frac{16}{945}\,\lz^3 +
 \frac{16}{105}\,\zeta_2 +
 \frac{4}{105}\,\zeta_3 +
 \frac{218460799}{3889620000}\right)\,\frac{1}{z^7}
\Bigg]+\ldots\,,
\nonumber\\
\Pi^{(2),s}_{FH}&=& \frac{3}{16\pi^2}\Bigg[
-\frac{46}{9}
+\frac{64}{3}\,\zeta_3
-
10\,\zeta_2
-
\left(\frac{4}{3}\,\lz^3\right)
-
\left(\frac{22}{3}\,\lz^2\right)
+\left(-
8\,\zeta_2 +
 4\,\zeta_3 -
 \frac{13}{3}\right)\,\lz
-
9\,\sqrtthree\,\cl
\nonumber\\&&\mbox{}
+\left(\left(32\,\zeta_2 -
 8\,\zeta_3 +
 \frac{14}{3}\right)\,\lz +
 \frac{52}{3}\,\lz^2 +
 \frac{16}{3}\,\lz^3 +
 8\,\zeta_2 +
 \frac{46}{3}\,\zeta_3 -
 \frac{191}{18}\right)\,\frac{1}{z}
\nonumber\\&&\mbox{}
+\left(\left(-
24\,\zeta_2 +
 4\,\zeta_3 +
 \frac{3}{2}\right)\,\lz -
 \frac{13}{3}\,\lz^2 -
 \frac{20}{3}\,\lz^3 +
 14\,\zeta_2 -
 \frac{20}{3}\,\zeta_3 +
 \frac{1639}{72}\right)\,\frac{1}{z^2}
\nonumber\\&&\mbox{}
+\left(-
\left(\frac{1826}{81}\,\lz\right) +
 \frac{124}{9}\,\lz^2 -
 \frac{16}{3}\,\zeta_2 +
 \frac{14}{9}\,\zeta_3 +
 \frac{3685}{486}\right)\,\frac{1}{z^3}
\nonumber\\&&\mbox{}
+\left(\frac{8347}{1296}\,\lz +
 \frac{217}{36}\,\lz^2 +
 \frac{8}{9}\,\lz^3 -
 \frac{5}{3}\,\zeta_2 +
 \frac{26}{9}\,\zeta_3 -
 \frac{34093}{3888}\right)\,\frac{1}{z^4}
\nonumber\\&&\mbox{}
+\left(\frac{752611}{81000}\,\lz +
 \frac{3683}{450}\,\lz^2 +
 \frac{272}{45}\,\lz^3 -
 \frac{4}{5}\,\zeta_2 -
 \frac{44}{45}\,\zeta_3 -
 \frac{34816841}{4860000}\right)\,\frac{1}{z^5}
\nonumber\\&&\mbox{}
+\left(\frac{3698791}{162000}\,\lz +
 \frac{34153}{900}\,\lz^2 +
 \frac{1528}{135}\,\lz^3 -
 \frac{7}{15}\,\zeta_2 -
 \frac{74}{15}\,\zeta_3 -
 \frac{29545141}{1620000}\right)\,\frac{1}{z^6}
\nonumber\\&&\mbox{}
+\left(\frac{378496261}{3472875}\,\lz +
 \frac{491947}{3675}\,\lz^2 +
 \frac{15712}{945}\,\lz^3 -
 \frac{32}{105}\,\zeta_2 -
 \frac{312}{35}\,\zeta_3 -
 \frac{74992868899}{1296540000}\right)\,\frac{1}{z^7}
\Bigg]+\ldots\,.
\nonumber
\end{eqnarray}
}

\noindent
In the above expressions the following symbols have been used:
$\lifour = \mbox{Li}_4(1/2)$,
$\cl = \mbox{Cl}_2(\pi/3) = \mbox{Im}[\mbox{Li}_2(\exp(i\pi/3))]$,
$\zeta_2=\pi^2/6$,
$\zeta_3\approx1.202\,057$,
$\zeta_4=\pi^4/90$,
$\zeta_5\approx1.036\,928$
and $\lz=-(\ln(-z))/2$.
A {\tt Mathematica} input can be found under the URL
\verb|http://www-ttp.physik.uni-karlsruhe.de/Progdata/ttp01/ttp01-14|.

\end{appendix}


\end{document}